\documentclass[a4paper,11pt]{article}
\pdfoutput=1
\usepackage{jheppub}

\usepackage[utf8]{inputenc}

\usepackage{xcolor}

\usepackage{amsmath}
\usepackage{amssymb}
\usepackage{tikz}
\usepackage{siunitx}
\usepackage{physics}
\ExplSyntaxOn
\msg_redirect_name:nnn{siunitx}{physics-pkg}{none}
\ExplSyntaxOff

\usepackage{slashed}

\usepackage{overpic}

\usepackage{graphicx}

\usepackage{mathtools}

\newcommand{\Lt}{\mathbf{L}}
\newcommand{\Pt}{\mathbf{P}}

\newcommand{\Kt}{\mathbf{K}}
\newcommand{\rt}{\mathbf{r}}
\newcommand{\xt}{\mathbf{x}}

\newcommand{\kt}{\mathbf{k}}
\newcommand{\qt}{\mathbf{q}}

\newcommand{\Kb}{K}
\newcommand{\li}{\text{Li}_2}

\newcommand{\kmin}{k_\text{min}}
\newcommand{\deltas}{\delta_s}
\newcommand{\msbar}{\overline{\text{MS}}}

\newcommand{\lo}{\mathrm{LO}}

\newcommand{\cf}{C_F}
\newcommand{\as}{\alpha_s}
\newcommand{\aem}{\alpha_\text{em}}

\newcommand{\dipole}{\mathcal{S}_{01}}
\newcommand{\tripole}{\mathcal{S}_{012}}

\renewcommand{\order}[1]{O \qty( #1 )} 
\renewcommand{\vdot}{\cdot} 

\title{Exploring rapidity regularization schemes at low $x$ with the DIS longitudinal structure function}

\author[a]{Tolga Altinoluk,}
\affiliation[a]{Theoretical Physics Division, National Centre for Nuclear Research,
Pasteura 7, Warsaw 02-093, Poland}
\emailAdd{tolga.altinoluk@ncbj.gov.pl}

\author[a]{Guillaume Beuf,}
\emailAdd{guillaume.beuf@ncbj.gov.pl}

\author[b,c]{Jani Penttala}
\affiliation[b]{
Department of Physics and Astronomy, University of California, Los Angeles, CA 90095, USA
}
\affiliation[c]{Mani L. Bhaumik Institute for Theoretical Physics, University of California, Los Angeles, California 90095, USA}
\emailAdd{janipenttala@physics.ucla.edu}

\abstract{
We propose three possible rapidity regulators for higher-order calculations in low $x$ QCD with gluon saturation, as  alternatives to the usual lower cut-off for the integrals over the light-cone momentum $k^+$. These rapidity regulators are closely related to the $\eta$ regulator and to the pure rapidity regulator, which have been used primarily in studies of transverse-momentum-dependent (TMD) factorization within the soft-collinear effective theory (SCET).
By choosing one of the three rapidity regulators that we propose, formulated in terms of $k^+$, $k^-$ or rapidity respectively, one can set from the start of the calculation in which of these three variables one wishes to formulate the low $x$ evolution equations, which is one of the main advantages of our approach. 

As a test of the viability of these rapidity regulators and of their practical implementation in higher order calculations with gluon saturation effects, we use them to revisit the calculation of the NLO corrections to the dipole factorization of the $F_L$ structure function in inclusive DIS at low $x$.
}

\begin{document}

\maketitle


\section{Introduction}

For scattering processes in Quantum Chromodynamics (QCD) in the Regge-Gribov limit, in which the energy of the collision is parametrically larger than any other scale, large logarithms of the energy arise in the perturbative expansion of observables, which need to be resummed. If the two colliding particles are dilute, that resummation can be performed thanks to the Balitsky-Fadin-Kuraev-Lipatov (BFKL) evolution equation~\cite{Lipatov:1976zz,Kuraev:1977fs,Balitsky:1978ic}. However, the BFKL evolution results in a rapid increase of the density of partons of low momentum fraction $x$ resolved in the colliding hadrons, eventually leading to a breakdown of the dilute approximation, or equivalently of the parton picture, with the onset of the nonlinear phenomenon of gluon saturation~\cite{Gribov:1983ivg,McLerran:1993ni,McLerran:1993ka,McLerran:1994vd}, which can be studied using semi-classical techniques, in particular in the Color Glass Condensate formalism (see \cite{Gelis:2010nm,Albacete:2014fwa,Blaizot:2016qgz} for reviews and references therein).   
Then, when gluon saturation is taken into account in the resummation of high-energy logarithms, the BFKL evolution is replaced by the Balitsky and Jalilian-Marian–Iancu–McLerran–Weigert–Leonidov–Kovner (B-JIMWLK) evolution~\cite{Balitsky:1995ub,Jalilian-Marian:1996mkd,Jalilian-Marian:1997qno,Jalilian-Marian:1997jhx,Jalilian-Marian:1997ubg,Kovner:2000pt,Weigert:2000gi,Iancu:2000hn,Iancu:2001ad,Ferreiro:2001qy}, or in a mean field approximation by the Balitsky-Kovchegov (BK) equation~\cite{Balitsky:1995ub,Kovchegov:1999yj,Kovchegov:1999ua}. This approach has been successful in providing a unified description of observed phenomena at low $x$ in deep inelastic scattering (DIS) at the Hadron-Electron Ring Accelerator (HERA) (using the dipole factorization~\cite{Bjorken:1970ah,Nikolaev:1990ja,Nikolaev:1991et}) and in proton-proton, proton-nucleus and nucleus-nucleus collisions at the Relativistic Heavy Ion Collider (RHIC) and at the Large Hadron
Collider (LHC). 

As usual in QCD, in order to obtain precise and reliable theoretical results, it is necessary to include the next-to-leading order (NLO) corrections in the 
QCD coupling $\as$. Their derivation in the presence of gluon saturation effects has been the subject of intense studies in recent years. 
On the one hand, NLO corrections have been derived for the BK equation~\cite{Balitsky:2007feb,Balitsky:2009xg} and for the B-JIMWLK evolution~\cite{Balitsky:2013fea,Kovner:2013ona,Kovner:2014lca,Lublinsky:2016meo}. Partial results have also been obtained for the NNLO corrections to the BK equation~\cite{Caron-Huot:2016tzz,Brunello:2025rhh}.
On the other hand, the fixed-order NLO corrections to many  observables sensitive to gluon saturation at low $x$ have been calculated, with a particular emphasis on DIS observables, in preparation for the future Electron-Ion Collider (EIC). The inclusive DIS structure functions have been calculated for both massless quarks~\cite{Balitsky:2010ze,Balitsky:2012bs,Beuf:2011xd,Beuf:2016wdz,Beuf:2017bpd,Hanninen:2017ddy}
and massive quarks~\cite{Beuf:2021qqa,Beuf:2021srj,Beuf:2022ndu}. Using these results in the dipole factorization approach, it has been possible to fit~\cite{Beuf:2020dxl,Hanninen:2022gje,Casuga:2025etc} the dipole-proton scattering amplitude to the HERA data on inclusive structure function and on their heavy quark contribution, at NLO accuracy.
Other DIS observables that have been derived at NLO in this context include diffractive structure functions~\cite{Beuf:2022kyp,Beuf:2024msh}, exclusive production of a light~\cite{Boussarie:2016bkq,Mantysaari:2022bsp} or heavy~\cite{Mantysaari:2021ryb,Mantysaari:2022kdm} vector meson, inclusive production of dijet~\cite{Caucal:2021ent,Taels:2022tza,Caucal:2022ulg,Caucal:2023nci,Caucal:2023fsf}, dihadron~\cite{Bergabo:2023wed,Bergabo:2022tcu,Iancu:2022gpw,Caucal:2024nsb}, photon plus dijet~\cite{Roy:2019hwr}, single jet or hadron~\cite{Bergabo:2022zhe,Bergabo:2024ivx,Caucal:2024cdq,Altinoluk:2025dwd}, and diffractive production of dijet~\cite{Boussarie:2016ogo,Boussarie:2019ero}, dihadron~\cite{Fucilla:2022wcg} or single hadron~\cite{Fucilla:2023mkl}.
In the context of proton-nucleus or proton-proton collisions, the NLO corrections have been derived for the forward inclusive production of single hadron~\cite{Chirilli:2012jd,Altinoluk:2014eka,Liu:2019iml,Liu:2020mpy,Shi:2021hwx,Altinoluk:2023hfz},
single jet~\cite{Liu:2022ijp}
and Drell-Yan pair plus jet~\cite{Taels:2023czt}.

In all of these NLO calculations, various types of divergences are encountered. Transverse integrations lead to UV and  collinear divergences, usually handled with dimensional regularization, evaluating the transverse integrals in $D\!-\!2 = 2\!-\!2\epsilon$ dimensions. In addition, divergences arise from integrations over light-cone momenta\footnote{In this article, we always choose a frame in which the dense target is left-moving, with a large momentum component $P^-$, and the dilute projectile (such as the virtual photon in DIS) is right-moving, with a large momentum component $q^+$.} $k^+$, when the $k^+$ of gluon line goes to zero. Either soft or rapidity divergences appear in such a way. Since rapidity divergences cannot be regularized by dimensional regularization, a second regulator is needed. In the vast majority of the above mentioned NLO calculations including gluon saturation, that second regulator is a lower cut-off in $k^+$
\begin{equation}
    \int_0^\infty \frac{\dd{k_2^+}}{k_2^+}\, (\cdots) \mapsto 
    \int_{\kmin^+}^\infty \frac{\dd{k_2^+}}{k_2^+}\, (\cdots),
    \label{k_plus_cutoff}
\end{equation}
with the limit $\kmin^+ \to 0$ taken after the $\epsilon\to 0$ limit from dimensional regularization. Although very convenient for practical calculations in this context, that regularization has several shortcomings. The first issue is that, due to this ordering of limits for the regulators, not only rapidity divergences but also soft divergences are regularized by the cut-off, even though the latter could be instead regularized by dimensional regularization. Then, it can be difficult to disentangle rapidity from soft divergences.  

Another issue is related to a scheme ambiguity in the formulation of low $x$ evolution equations. In the case of the BFKL equation, it is well known that several choices of evolution variable are possible, and that they correspond to different versions of the BFKL equation starting at NLO~\cite{Fadin:1998py}. More precisely, large collinear  or anti-collinear logarithms (respectively collinear to the target or to the projectile), which question the validity of the perturbative expansion, appear in the BFKL equation at NLO and in the NLO impact factors for any observable, in a pattern that depends on the choice of evolution variable. Choosing the rapidity as the evolution variable, in order to treat the projectile and target on equal footing, one obtains such large corrections both in the collinear and anticollinear regimes. If instead one formulates the BFKL evolution in terms of momentum fraction with respect to the target (or $k^-$), the largest higher order corrections disappear in the collinear regime, allowing a smooth matching with the DGLAP evolution of the target~\cite{Fadin:1998py}. Similarly, formulating the BFKL evolution in terms of momentum fraction with respect to the projectile  (or $k^+$) removes the largest corrections in the anticollinear regime, allowing a smooth matching with the DGLAP evolution of the projectile. 
As shown in Ref.~\cite{Salam:1998tj}, the largest NLO corrections in both the collinear and the anticollinear regime can be simultaneously resummed by imposing a kinematical constraint in the LO evolution. That constraint enforces a simultaneous ordering in $k^+$ and in $k^-$ (in opposite directions) along the evolution, which corresponds to the consistent phase-space contributing to high-energy logarithms.  
Only the ordering along the chosen evolution variable is automatically obtained, whereas the other one is missed in the standard implementation of the Regge-Gribov high-energy limit, and can be restored a posteriori thanks to such kinematical constraint, in order to restore the internal consistency of the high-energy resummation and prevent the appearance of pathologically large collinear or anticollinear logs at higher orders.

These issues of scheme ambiguity related to the choice of evolution variable, appearance of large collinear or anticollinear logs, and the need to resum them by improving the treatment of kinematics, first encountered in the dilute BFKL case, remain present and essentially unchanged in the presence of nonlinear gluon saturation effects~\cite{Motyka:2009gi,Avsar:2011ds,Beuf:2014uia,Lappi:2015fma,Lappi:2016fmu}.  
In the widely used $k^+$ cut-off regularization scheme~\eqref{k_plus_cutoff} for NLO calculations with gluon saturation, one obtains the high-energy evolution, BK or B-JIMWLK, with $k^+$ as evolution variable, so that the problem of transverse large logs appears predominantly in the regime collinear to the target. However, in application to DIS processes, the collinear regime is more relevant than the anticollinear regime, due to the asymmetry of the process, with the hard scale $Q^2$ associated to the photon projectile side. Thus, using the $k^+$ cut-off scheme for DIS processes leads to pathologically large corrections specifically in the important collinear regime, spoiling the matching with the DGLAP evolution of the target. In order to address this issue, several schemes have been proposed for the resummation of the large collinear logs in the high-energy evolution by imposing a posteriori the ordering in $k^-$ in the BK~\cite{Motyka:2009gi,Beuf:2014uia,Iancu:2015vea,Iancu:2015joa} or JIMWLK~\cite{Hatta:2016ujq} evolution in $k^+$, obtained by using the  $k^+$ cut-off~\eqref{k_plus_cutoff}.
Due to the importance of the collinear regime in DIS processes, it would however be more appropriate to use $k^-$ as the high-energy evolution variable. With this motivation in mind, it has been shown in Ref.~\cite{Ducloue:2019ezk} how  perturbative results obtained in the $k^+$ cut-off scheme can be approximately translated a posteriori into a scheme using $k^-$ as evolution variable. Then, an implementation of the kinematical constraint was also proposed in that case, in order to enforce this time the missing $k^+$ ordering, in order to resum large anticollinear  logs. Very recently~\cite{Boussarie:2025mzh}, a more general and systematic method has been presented in order to perform such a posteriori translation of the high-energy evolution, from an evolution in $k^+$ obtained from the cut-off regularization~\eqref{k_plus_cutoff} to an evolution in $k^-$ suitable for DIS processes. That method is similar to the one used in Ref.~\cite{Balitsky:2009xg} in order to transform the NLO BK equation obtained using the $k^+$ cut-off \eqref{k_plus_cutoff} into a scheme in which conformal symmetry is restored (in ${\cal N}=4$ sYM) and which essentially corresponds to a scheme in which the rapidity plays the role of evolution variable.  

Many of these complications are thus related to the widespread use of the $k^+$ cut-off \eqref{k_plus_cutoff} as the regulator for rapidity (and soft) divergences in higher order calculations involving gluon saturation. In the present study, we explore  alternative regularization procedures for rapidity divergences, in order to alleviate these issues. 
In the transverse-momentum-dependent (TMD) factorization  
for processes involving a hard and a not-so-hard scales  (see \cite{Collins:2011zzd,Boussarie:2023izj} for reviews), rapidity divergences are also crucial, and multiple ways to regularize them have been suggested in the literature: non-light-like slope for the Wilson lines~\cite{Collins:2011zzd},
the $\eta$ regulator~\cite{Chiu:2011qc,Chiu:2012ir},
the $\delta$ regulator~\cite{Echevarria:2011epo,Echevarria:2012js,Echevarria:2015usa,Echevarria:2015byo},
the analytic regulator~\cite{Becher:2011dz},
the exponential regulator~\cite{Li:2016axz} and the pure rapidity regulator~\cite{Ebert:2018gsn}.
We are then considering several rapidity regulators for low $x$ QCD inspired by some of these rapidity regulators for TMDs.
In particular, we show that by choosing a specific rapidity regulator, one chooses, from the start of the calculation, which is the evolution variable for the high-energy evolution equation, $k^+$, $k^-$ or rapidity. Moreover, by reversing the order of limit for the regulators compared to the usual cut-off scheme \eqref{k_plus_cutoff}, we ensure that the rapidity regulator acts only on divergences which cannot be regularized by dimensional regularization, following the standard practices from the TMD community, so that a clean separation of soft and rapidity divergences is obtained. 

As a test for the proposed rapidity regularization procedures in the low $x$ QCD context, we use them to revisit the calculation of the NLO corrections to the dipole factorization formula for the $F_L$ inclusive DIS structure function at low $x$, performed in Refs.~\cite{Beuf:2016wdz,Beuf:2017bpd} with the $k^+$ cut-off~\eqref{k_plus_cutoff}. Indeed, despite being a relatively simple observable, $F_L$ features at NLO nontrivial diagrams with multiple divergences (UV, soft and rapidity). All of the divergences are expected to cancel in the sum of diagrams except for rapidity divergences associated with the B-JIMWLK evolution. This provides an ideal testing ground for the implementation of new regulation procedures in low $x$ QCD.

The plan of the present paper is then as follows. In Sec.~\ref{sec:framework}, we remind the setup for the calculation of $F_L$ at NLO at low $x$ in dipole factorization, and we define the rapidity regularization schemes that we consider. In Sec.~\ref{sec:gamma_L_qqbar}, we calculate the loop correction to the $\gamma_L^* \to q  \bar q$ light-front wave function, which is an important building block for the NLO corrections to $F_L$ (and to other DIS observables) at low $x$. 
In Sec.~\ref{sec:FL_qqbarg}, we calculate the NLO corrections due to $q\bar q g$ Fock states scattering on the target, and assemble the final results for $F_L$ at NLO, in each of the considered rapidity regularization schemes. Conclusions are provided in Sec.~\ref{sec:conclu}. 
Technical details concerning the calculation of some of the loop integrals or Fourier transforms are provided in Appendices~\ref{app:details}
and~\ref{app:FT} respectively.


\section{Theoretical framework\label{sec:framework}}

\subsection{Scattering in the high-energy limit}

In the high-energy limit of QCD, 
the lifetime of quantum fluctuations for the projectile is much longer than the interaction with the target.
This allows us 
to factorize the process $\gamma^* A \to \gamma^* A$ into three different parts:
First, the photon fluctuates into a quantum state, such as a quark--antiquark dipole $q \bar q$ at leading order, which can be computed perturbatively in terms of light-front wave functions (LFWFs) using light-front perturbation theory (LFPT).
This quantum state then scatters off the target eikonally,
leading to Wilson lines that rotate the particles in the color space.
Finally, the scattered state forms the final state, which can again be described in terms of perturbative LFWFs.
The computation of the scattering amplitude then reduces to calculating the relevant LFWFs at the desired order and convoluting them with the Wilson lines.

For inclusive DIS, we  can use the optical theorem to relate the imaginary part of the forward scattering amplitude $\gamma^* A \to \gamma^* A$ to the total $\gamma^* A$ cross section.
This allows us to write \cite{Beuf:2017bpd}:
\begin{equation}
\begin{split}
    &\sigma_\lambda = 2 \Im \mathcal{M}(\gamma_\lambda^* A \to \gamma_\lambda^* A)
\\
    =& 
2 N_c  \widetilde{ \sum_{q \bar q}}
   2 q^+ 2\pi \delta(k_0^+ + k_1^+ - q^+)
   \abs{\widetilde \psi_{\gamma_\lambda^* \to q_0 + \bar q_1}}^2
    \Re[ 1 - \dipole ]
    \\
    &+
 2 N_c C_F \widetilde{ \sum_{q \bar q g}}
   2 q^+ 2\pi \delta(k_0^+ + k_1^+ + k_2^+ - q^+)
   \abs{\widetilde \psi_{\gamma_\lambda^* \to q_0 + \bar q_1 + g_2}}^2
    \Re[ 1 - \tripole ]
    +O(\alpha_{\textrm{em}} \alpha_s^2)
\end{split}
\label{eq:sigma_NLO}
\end{equation}
where $\lambda$ is the polarization of the photon, and $\widetilde \psi$ are reduced LFWFs in the mixed space of coordinates where the phase space of the Fock state $n$ is defined as~\footnote{This expression is suitable in the present study, because all the partons in the two considered Fock states, $q\bar q$ and $q \bar q g$, are of different nature, and thus discernible. In the case of more general Fock states, which contain, for example, several gluons or several quarks, symmetry factors should be included in Eq.~\eqref{Fock_state_phase_space}.  Moreover, note that the summation over color degrees of freedom is not included in the expression \eqref{Fock_state_phase_space}, because it has already been performed explicitly in Eq.~\eqref{eq:sigma_NLO}.}
\begin{equation}
    \widetilde{\sum_n}
    = 
    \sum_\text{helicities}
    \int
    \prod_{i \in n}
    \qty[ \frac{\dd[D-2]{\xt_i} \dd{k_i^+}}{2k_i^+ (2\pi)} ].
    \label{Fock_state_phase_space}
\end{equation}
In Eq.~\eqref{eq:sigma_NLO}, we have only included the intermediate states $q\bar q$ and $q \bar q g$ that are relevant at NLO. At LO, only the $q\bar q$ state contributes, which leads to the dipole factorization formula for inclusive DIS~\cite{Bjorken:1970ah,Nikolaev:1990ja}.

This inclusive $\gamma^* A$ cross section can then be related to the DIS structure functions $F_2$ and $F_L$ by
\begin{align}
    F_\lambda &=  \frac{Q^2}{(2\pi)^2 \aem} \sigma_\lambda ,
    \\
    F_2 &= F_L + F_T,
\end{align}
where the transverse cross section is defined as the average of the transverse polarizations
$\sigma_T = \frac{1}{2} \sum_{\lambda = \pm 1} \sigma_\lambda$.
In this work, we will only focus on the longitudinally polarized case $\sigma_L$, and the transverse polarization $\sigma_T$ is left for future work.

\subsection{Regularization scheme}
\label{sec:regulators}

As discussed in the introduction, we would like to study various options for rapidity regularization, supplementing dimensional regularization in $D = 4-2\epsilon$ dimensions for the transverse degrees of freedom. 
Moreover, by contrast to the implementation of the $k^+$ cut-off \eqref{k_plus_cutoff}, we will consider only regularization procedures in which the rapidity regulator is removed first, at finite $\epsilon$, before taking the $\epsilon\to 0$ limit.
In this way, dimensional regularization is utilized as much as possible, for UV, collinear and soft divergences, whereas the rapidity regulator acts only on genuine rapidity divergences.
This also allows for a more direct comparison to results using only dimensional regularization.

The choice of the rapidity regulator is not unique.
In this work, we will consider three different rapidity regulators that are related, but not identical, to the $\eta$ regulator~\cite{Chiu:2011qc,Chiu:2012ir}, the analytic regulator~\cite{Becher:2011dz} or to the pure rapidity regulator~\cite{Ebert:2018gsn},
used to regulate rapidity divergences in many TMD calculations (see also Refs.~\cite{Kang:2019ysm,Liu:2020mpy,Liu:2022ijp,Mukherjee:2023snp} for applications of the $\eta$ regulator in small-$x$ calculations). 
Specifically, the regulators we consider are the following:\footnote{As a remark, the $\eta$ regulator from Refs.~\cite{Chiu:2011qc,Chiu:2012ir}, is formulated within soft-collinear effective theory (SCET) \cite{Bauer:2000ew,Bauer:2000yr,Bauer:2001ct,Bauer:2001yt}, in which modes are sorted according to their scaling in terms of hard and soft scales of the process. Then, the $\eta$ regulator from Refs.~\cite{Chiu:2011qc,Chiu:2012ir} amounts to use our $\eta^-$ regulator \eqref{def_etaminus_reg} for the modes collinear to the target, our $\eta^+$ regulator \eqref{def_etaplus_reg} (up to the change $\eta\mapsto -\eta$) for the modes collinear to the projectile, and a regulator factor $(|k^+\!-\!k^-|/\nu)^{-\frac{\eta}{2}}$ for the soft modes.
}

\begin{enumerate}
    \item $\eta^+$ \textit{regulator}: Integrals over the gluons' longitudinal momenta are replaced by
\begin{equation}
    \int_0^\infty\frac{\dd{k_2^+}}{k_2^+}\, (\cdots) \mapsto 
    \int_{0}^\infty \frac{\dd{k_2^+}}{k_2^+}\, 
    \qty(
    \frac{|k_2^+|}{\nu_B^+}
    )^\eta\, (\cdots).
    \label{def_etaplus_reg}
\end{equation}
    \item $\eta^-$ \textit{regulator}: Integrals over the gluons' longitudinal momenta are replaced by
\begin{equation}
    \int_0^\infty \frac{\dd{k_2^+}}{k_2^+}\, (\cdots) \mapsto 
    \int_{0}^\infty \frac{\dd{k_2^+}}{k_2^+}\, 
    \qty(
    \frac{\nu_B^-}{|k_2^-|}
    )^\eta\,
    (\cdots)
    =
    \int_{0}^\infty \frac{\dd{k_2^+}}{k_2^+}\, 
    \qty(
    \frac{2k_2^+\nu_B^-}{\kt_2^2}
    )^\eta\,
    (\cdots).
    \label{def_etaminus_reg}
\end{equation}
Here, the minus-momentum $k_2^- = \kt_2^2 / (2k_2^+)$ is taken at its on-shell value.

    \item \textit{Pure rapidity regulator}:  Integrals over the gluons' longitudinal momenta are replaced by
\begin{equation}
    \int_0^\infty \frac{\dd{k_2^+}}{k_2^+}\, (\cdots) \mapsto 
    \int_{0}^\infty \frac{\dd{k_2^+}}{k_2^+}\, 
    \qty(
    \frac{|k_2^+| \nu_B^-}{|k_2^-| \nu_B^+}
    )^{\frac{\eta}{2}}
    (\cdots)
    =
    \int_{0}^\infty \frac{\dd{k_2^+}}{k_2^+}\, 
    \qty(
    \frac{2 (k_2^+)^2 \nu_B^-}{\kt_2^2 \nu_B^+}
    )^{\frac{\eta}{2}}
    (\cdots).
    \label{def_pure_rap_reg}
\end{equation}
This regulator corresponds to the one introduced in Ref.~\cite{Ebert:2018gsn}. It can be understood as interpolating between the $\eta^+$ and $\eta^-$ regulators.
\end{enumerate}
In all of these regulators, $\eta$ plays a similar role to that of $\epsilon$ in dimensional regularization.
\footnote{A further comment is in order concerning our notations in the implementation of the rapidity regulators by comparison to Refs.~\cite{Chiu:2011qc,Chiu:2012ir,Ebert:2018gsn}. In these references the so-called bookkeeping parameter $w$ is introduced together with the rapidity regularization. For example, for the modes collinear to the target, the factor compensating the change of dimension of the integral is $w^2 \nu^{\eta}$, which corresponds to $(\nu_B^-)^{\eta}$ for our $\eta^-$ regulator  \eqref{def_etaminus_reg}. Indeed, the bookkeeping parameter is defined by its properties $\lim_{\eta\rightarrow 0} w =1$ and $\nu \partial_{\nu} w = -\eta\, w/2$, so that $\nu \partial_{\nu} (w^2 \nu^{\eta}) = 0$. In the analogy with dimensional regularization, $\nu$ would corresponds to $\mu$, $w^2$ would play a role of renormalized coupling, and $w^2 \nu^{\eta}$ or our $(\nu_B^-)^{\eta}$ would play a role of bare coupling, dimensionful but independent of the scale $\nu$ (hence the subscript $B$ for bare in the scales $\nu_B^{\pm}$ in Eqs.~\eqref{def_etaplus_reg}. \eqref{def_etaminus_reg} and \eqref{def_pure_rap_reg}). }
As a remark, in Ref.~\cite{Liu:2019iml}, another rapidity regularization was used in a NLO calculation at low $x$ in LFPT, by modifying the power of each energy denominator. This procedure leads to evolution equations formulated with a $k^-$ evolution variable, like our $\eta^-$ regulator. However, by experience, the regularization from \cite{Liu:2019iml} does not seem suitable for generic processes, because it would spoil the simplifications between numerators and denominators typical from gauge theories.

In addition to these different rapidity regulators, we will also consider two slightly different versions of dimensional regularization.
These are called the \textit{conventional dimensional regularization} (CDR) and
\textit{four-dimensional helicity} (FDH) schemes.
In our case,
the difference in these two schemes corresponds to whether the transverse polarizations of the gluon are computed in $D-2$ (CDR) or $2$ (FDH) dimensions.
This can be taken into account by introducing the factor
\begin{equation}
    \deltas 
    =
    \begin{cases}
        1 \text{ for CDR}, \\
        0 \text{ for FDH},
    \end{cases}
\end{equation}
in the computations. In particular, in the evaluation of a NLO diagram, when the trace of the metric is obtained purely from the numerator algebra, it should be evaluated as
\begin{equation}
{g^{\mu}}_{\mu} \mapsto D_s = 4-2 \deltas \epsilon
\, .
\end{equation}
%
In $F_L$ at NLO at low $x$, the only leftover divergence expected in the final result is the rapidity divergence associated with the low-$x$ evolution of the dipole operator. Hence, poles at $\epsilon =0$ should cancel in the final result, as well as the dependence on $\deltas$. This 
 serves as a cross-check of our calculations.

Finally, let us discuss how the expansion around $\eta=0$ at fixed $\epsilon$ is performed in practice. In NLO contributions with a rapidity divergence, after applying one of the three rapidity regulators defined in Eqs.~\eqref{def_etaplus_reg} \eqref{def_etaminus_reg} and \eqref{def_pure_rap_reg}, and changing the variable to light-cone momentum fraction, one typically arrives at an expression of the type
\begin{equation}
I(\eta,\epsilon)
\equiv
\int_0^1 d\xi\: \xi^{-1+\eta}\: F(\xi,\eta,\epsilon)\, , 
\label{split_etapole_plusdistr_0}
\end{equation}
with $F(\xi,\eta,\epsilon)$ regular around $\xi=0$. Adding and subtracting its value at $\xi=0$, one has
\begin{equation}
I(\eta,\epsilon)
=
F(0,\eta,\epsilon)\int_0^1 d\xi\: \xi^{-1+\eta}\:
+\int_0^1 d\xi\: \xi^{-1+\eta}\: \left[F(\xi,\eta,\epsilon)\,
-F(0,\eta,\epsilon)\right]\, . 
\end{equation}
In the first term, the integration over $\xi$ can now be performed. In the second term, the subtraction removes the potential divergence at $\xi=0$, making the rapidity regulator unnecessary, so that the $\eta =0$ limit can be taken.
Then, one obtains
\begin{equation}
I(\eta,\epsilon)
=
\frac{1}{\eta}\, F(0,\eta,\epsilon)\:
+\int_0^1 \frac{d\xi}{(\xi)_+}\:  F(\xi,0,\epsilon)
+O(\eta)
\, . 
\label{split_etapole_plusdistr}
\end{equation}
Now, the pole at $\eta=0$ in the first term signals the rapidity divergence, regularized in one of the three schemes \eqref{def_etaplus_reg} \eqref{def_etaminus_reg} or \eqref{def_pure_rap_reg}. 
The second term, rapidity safe, contains the plus-distribution defined in the standard way:
\begin{equation}
    \int_0^1 \dd{x} \frac{f(x)}{(x)_+}
    =
    \int_0^1 \frac{\dd{x}}{x} \Big[f(x) - f(0)\Big].
    \label{def_plus_distr}
\end{equation}
By construction, the same second term (with the plus-distribution) in Eq.\eqref{split_etapole_plusdistr} is obtained for any choice of rapidity regulator. By contrast, the rapidity regularization scheme dependence is carried by the $\eta$ dependence of $F(0,\eta,\epsilon)$ in the first term of Eq.\eqref{split_etapole_plusdistr}.  
Further expanding that term at $\eta=0$, one finds from Eq.~\eqref{split_etapole_plusdistr}
\begin{equation}
I(\eta,\epsilon)
=
\frac{1}{\eta}\, F(0,0,\epsilon)\:
+ \Big[\partial_{\eta}F\Big](0,\eta=0,\epsilon)\:
+\int_0^1 \frac{d\xi}{(\xi)_+}\:  F(\xi,0,\epsilon)
+O(\eta)
\, . 
\label{split_etapole_plusdistr_2}
\end{equation}
The first term in Eq.~\eqref{split_etapole_plusdistr_2}, the strict $\eta$ pole term, is now independent of which of the three rapidity regulators  
\eqref{def_etaplus_reg} or \eqref{def_etaminus_reg} or \eqref{def_pure_rap_reg} is chosen. By contrast, the second term, with the derivative in $\eta$, carries the rapidity regularization scheme dependence. Then, by comparing the relation \eqref{split_etapole_plusdistr_2} with the definitions \eqref{def_etaplus_reg}, \eqref{def_etaminus_reg} and  \eqref{def_pure_rap_reg} of the rapidity regulators, it is clear that the result obtained with the pure rapidity regulator \eqref{def_pure_rap_reg} for a quantity at NLO is the mean between the results obtained with the $\eta^+$ regulator and with the $\eta^-$ regulator.  
Hence, in the present study, we will explicitly calculate the NLO corrections to $F_L$ both in the $\eta^+$ and $\eta^-$ cases, and deduce from them the corresponding results in the case of the pure rapidity regulator. 

Due to the fact that the $\eta\rightarrow 0$ limit should always be taken first, at finite $\epsilon$, before the $\epsilon\rightarrow 0$ limit, it is not safe to expand the first term in Eq.~\eqref{split_etapole_plusdistr_2} around $\epsilon=0$, due to the $1/\eta$ factor. In the following, we will always keep the full dependence on $\epsilon$ of such contribution proportional to $1/\eta$. Due to this order of limits, any term proportional to a positive power of $\eta$ is neglected regardless of its dependence on $\epsilon$, even if it is singular at $\epsilon=0$. In expressions which are expanded both in $\eta$ and $\epsilon$, we will remind this fact by the notation $+O(\eta f(\epsilon))$, corresponding to order $\eta$ times any function of $\epsilon$.


\section{\texorpdfstring{NLO corrections to the $\gamma_L^*\rightarrow q\bar{q}$ LFWF}
{
NLO corrections to the gammaL qq LFWF}}
\label{sec:gamma_L_qqbar}

\begin{figure}
\setbox1\hbox to 10cm{
\includegraphics{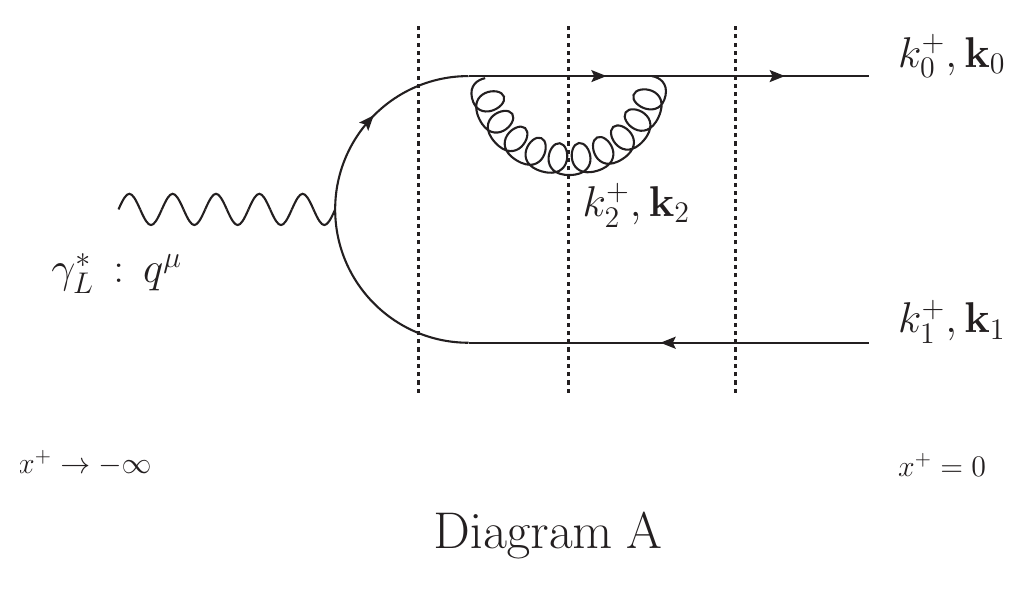}
}
\setbox2\hbox to 10cm{
\includegraphics{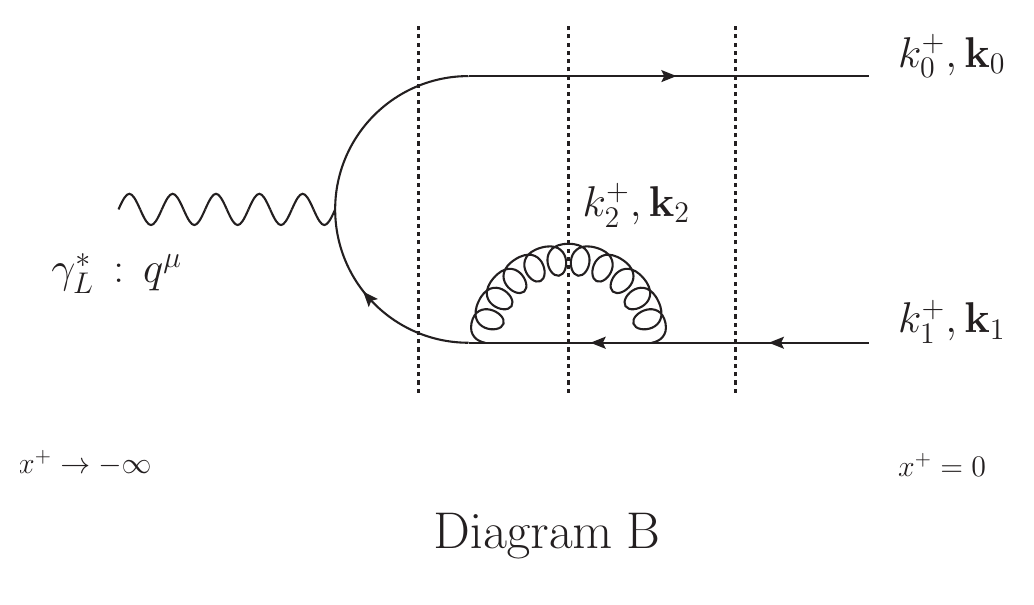}
}
\setbox3\hbox to 10cm{
\includegraphics{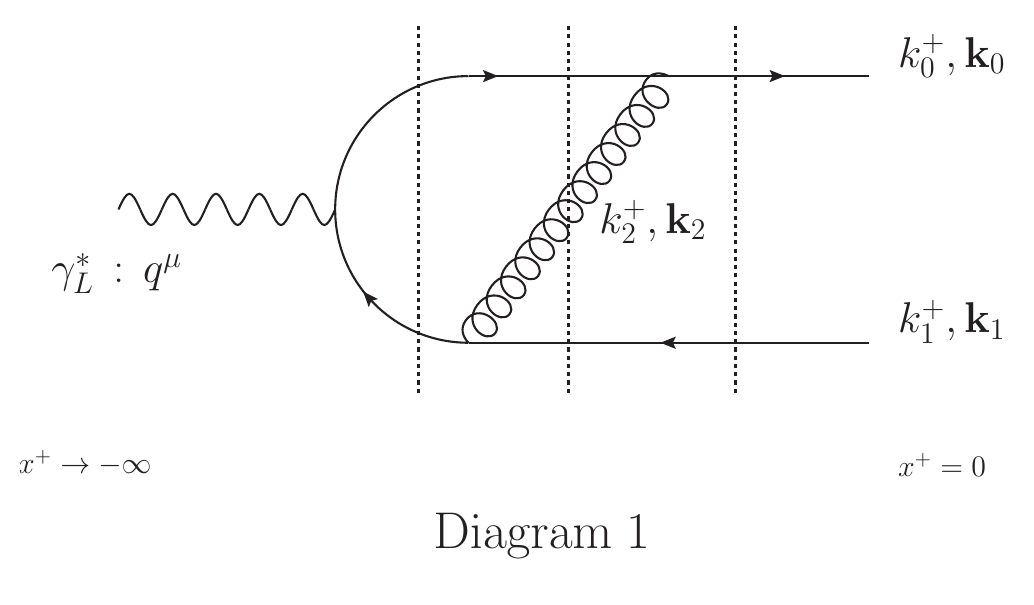}
}
\setbox4\hbox to 10cm{
\includegraphics{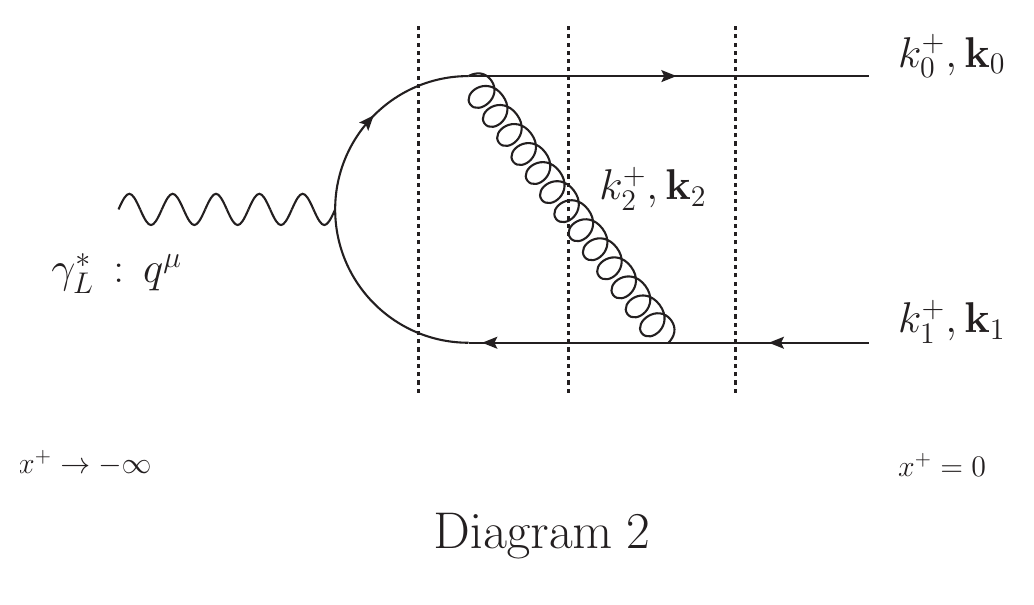}
}
\setbox5\hbox to 10cm{
\includegraphics{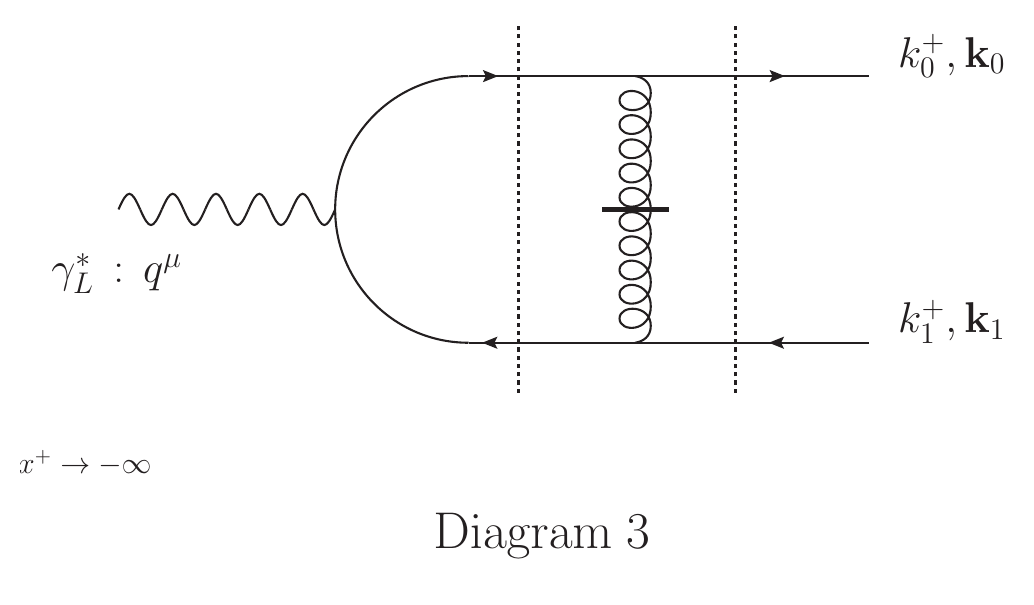}
}
\begin{center}
\resizebox*{9cm}{!}{\hspace{-7cm}\mbox{\box1 \hspace{9cm} \box2}}
\resizebox*{9cm}{!}{\hspace{-7cm}\mbox{\box3 \hspace{9cm} \box4}}
\resizebox*{9cm}{!}{\hspace{-7cm}\mbox{\hspace{9.5cm} \box5 \hspace{9.5cm}}}
\caption{\label{fig:NLO_qqbar_diagrams}Light-front perturbation theory diagrams contributing to NLO corrections to the $\gamma_L^* \to q  \bar q$ light-front wave-function. Vertical dotted lines indicate the energy denominators associated with an intermediate Fock state.}
\end{center}
\end{figure}

Next-to-leading order corrections to the $\gamma_L^* \to q + \bar q$ LFWF can be computed by considering the LFPT diagrams shown in Fig.~\ref{fig:NLO_qqbar_diagrams}.
As these have already been computed in Ref.~\cite{Beuf:2016wdz} in the cut-off scheme, we will largely follow the structure of the computation there and use the same notation and naming of the diagrams.
In the momentum space, each NLO diagram from Fig.~\ref{fig:NLO_qqbar_diagrams} is found to factorize into the leading-order LFWF and a loop factor. 
Hence, the $\gamma_L^* \to q + \bar q$ LFWF has an expression of the form:
\begin{align}
\Psi_{\gamma^*_L\rightarrow q_0+\bar{q}_1}
=&\,
\qty[1+\frac{\as \cf}{2\pi}\,
{\mathcal{V}}^{L} 
]\:
\Psi^{\lo}_{\gamma^*_L\rightarrow q_0+\bar{q}_1}
+\order{e\, \as^2}
\, ,
\label{NLO_qqbar_LFWF_L}
\end{align}
where
\begin{equation}
    \mathcal{V}^L
    =
    \mathcal{V}_A
    +
    \mathcal{V}_B
    +
    \mathcal{V}_1^L
    +
    \mathcal{V}_2^L
    +
    \mathcal{V}_3^L
\end{equation}
in which the different terms correspond to the different diagrams in Fig.~\ref{fig:NLO_qqbar_diagrams}.

\subsection{Self-energy diagrams}

From Eq.~(20) in Ref.~\cite{Beuf:2016wdz}, the self-energy diagram $A$ on the quark line gives the loop factor contribution
\begin{align}
{\cal V}_{A} 
=&\, 
-\int_{0}^{k_0^+}\!\!\frac{\dd{k_2^+}}{k_{0}^+}\;
\qty{2 \qty(\frac{k_0^+\!-\!k_2^+}{k_2^+}) +\frac{(D_s\!-\!2) k_2^+}{2k_0^+} }\;
\mathcal{A}_0(\Delta_1),
\label{VA_1}
\end{align}
where 
\begin{equation}
\Delta_1 =     
\frac{ k_2^+ q^+ ( k_0^+ - k_2^+)}{(k_0^+)^2 k_1^+} \qty[\Pt^2 + \overline{Q}^2],
\end{equation}
and
\begin{equation}
\mathcal{A}_0(\Delta) 
\equiv
4\pi\, \mu^{2\epsilon}
\int \frac{d^{2-2\epsilon}\Kt}{(2\pi)^{2-2\epsilon}}\,
\frac{1}{\left[\Kt^2+\Delta\right]}
=  
\Gamma\qty(\epsilon)
\qty[
\frac{\Delta}{ 4\pi \mu^2}
]^{-\epsilon}.
\end{equation}
As a reminder,  $k_0^+$ and $k_1^+$ are the lightcone momenta of the quark and antiquark in the considered Fock state, and $q^+$ the one of the incoming photon, so that $k_0^++k_1^+ =q^+$. $\Pt\equiv \kt_0 -k_0^+ \qt/q^+= -\kt_1 +k_1^+ \qt/q^+$ is the relative transverse momentum of the quark with respect to the photon.  We also use the notation 
\begin{align}
\overline{Q}^2 \equiv\, \frac{k_0^+k_1^+}{(q^+)^2}\, Q^2\, .
\end{align}
We can simplify the integral by writing the plus momenta as 
\begin{align}
\label{eq:xi_variable}
    k_0^+ &= z q^+,
    &
    k_1^+ &=(1-z) q^+,
    &
    k_2^+ &= z \xi q^+,
\end{align}
(so that $\overline{Q}^2=z(1-z)Q^2$) and changing the integration variable to $\xi$:
\begin{equation}
{\cal V}_{A} 
=\, 
-\int_{0}^{1}\!\!\dd{\xi}
\left\{\frac{2(1\!-\!\xi)}{\xi} +(1\!-\!\deltas \epsilon) \xi \right\}\;
\Gamma\qty(\epsilon)\;
    \qty[\frac{\xi(1\!-\!\xi)\qty(\Pt^2\!+\!\overline{Q}^2)}{4\pi\,\mu^2 (1\!-\!z)}]^{-\epsilon}    .
\end{equation}
Because $\Delta_1$ is proportional to $k_2^+$, or equivalently to $\xi$, dimensional regularization is enough to regularize not only the transverse integration, but also the integration over $k_2^+$ or $\xi$ with its potential divergence at $0$, since it produces a factor $\xi^{-\epsilon}$ as a result of the transverse integration. 
For this reason, we do not need to introduce a rapidity regulator for this integral, and carrying it out using dimensional regularization yields:
\begin{equation}
\begin{split}
{\cal V}_{A} 
=&\, 
2\,  \frac{\Gamma\!\left(1\!+\!\epsilon\right)}{\epsilon^2}\, \left[\frac{\Pt^2\!+\!\overline{Q}^2}{4\pi\,\mu^2 (1\!-\!z)}\right]^{-\epsilon}\,
\frac{\Gamma\left(1\!-\!\epsilon\right)^2}{\Gamma\left(1\!-\!2\epsilon\right)}\,
\frac{1}{(1\!-\!2\epsilon)}\,  \left[1\!-\!\epsilon  \!-\!\frac{\epsilon(1\!-\!\deltas \epsilon)}{4} \right]
\\
=&\, 
2\,  \frac{S_{\epsilon} }{\epsilon^2}\, \left[\frac{\Pt^2\!+\!\overline{Q}^2}{\mu^2 (1\!-\!z)}\right]^{-\epsilon}\,
+ \frac{3}{2}\,  \frac{S_{\epsilon} }{\epsilon} \left[\frac{\Pt^2\!+\!\overline{Q}^2}{\mu^2 (1\!-\!z)}\right]^{-\epsilon}\,
-\frac{\pi^2}{6}+\frac{\deltas}{2}+3+\order{\epsilon}
\label{VA_2}
\, ,
\end{split}
\end{equation}
where we have used the expansion
\begin{align}
\Gamma\!\left(1\!+\!\alpha\right)
=&\, e^{\alpha\, \Psi(1)}
\left[1+ \frac{\alpha^2\, \pi^2}{12} + \order{\alpha^3}
\right]
\end{align}
and the definition
\begin{align}
S_{\epsilon} \equiv &\, \left[4\pi\, e^{\Psi(1)}\right]^{\epsilon} =  \left[4\pi\, e^{-\gamma_{E}}\right]^{\epsilon}
\end{align}
for the factor collecting the universal constants following the $\msbar$ scheme.

The diagram $B$, with self-energy on the antiquark line, is the  same up to the exchange of $z$ and $1\!-\!z$:
\begin{align}
\mathcal{V}_{B} 
=&\, 
\mathcal{V}_{A} \qty(z  \leftrightarrow 1\!-\!z)
\label{VA_VB}
\, .
\end{align}

\subsection{Vertex correction diagrams}
\label{sec:vertex_corrections}

\subsubsection{Rapidity-safe contributions}

The contribution from the vertex correction diagram $1$, with the gluon emitted from the antiquark and then absorbed by the quark, is given in Eq.~(113) from Ref.~\cite{Beuf:2016wdz} as
\begin{equation}
\begin{split}
{\cal V}_{1}^{L}
=\,
 \int_{0}^{k_0^+}\!\! \frac{\dd{k_2^+}}{q^+}\;
\Bigg\{
&
-\frac{2 k_0^+k_1^+}{(k_2^+)^2}
       +\frac{(3k_1^+\!-\!k_0^+)}{k_2^+}
       +\frac{(k_0^+\!-\!k_1^+)}{k_0^+}
      +\frac{(D_s\!-\!2)}{2}\, \left(\frac{k_0^+\!-\!k_2^+}{k_0^+}\right)
\Bigg\}\; \mathcal{A}_0(\Delta_2)
\\
&
+\int_{0}^{k_0^+}\!\! d{k_2^+}\;
\Bigg\{\frac{2}{k_2^+} +\frac{(k_0^+\!-\!k_1^+)}{k_0^+k_1^+}
        -\frac{(D_s\!-\!2)k_2^+}{2k_0^+k_1^+}
\Bigg\}\; \mathcal{I}_+
\label{V1L_decomp}
 \, ,
\end{split}
\end{equation}
with the notation
\begin{equation}
\mathcal{I}_{+}
 \equiv \,
 \frac{1}{2}\,
\bigg\{\mathcal{A}_0(\Delta_1)-\mathcal{A}_0(\Delta_2)
+\left(\frac{k_0^+\!-\!k_2^+}{k_0^+}\right)\,
\left[\left(\frac{k_1^+\!+\!k_2^+}{k_1^+}\right)\, \Pt^2
  + \left(\frac{k_0^+\!-\!k_2^+}{k_0^+}\right)\, \overline{Q}^2 \right]
\mathcal{B}_0\bigg\}
\label{I_plus_def}
 \, .
\end{equation}
Here we have denoted
\begin{equation}
\label{eq:B0}
\begin{split}
    \mathcal{B}_0 =& 
    4\pi \mu^{2\epsilon}
\int \frac{\dd[2-2\epsilon]{\Kt}}{(2\pi)^{2-2\epsilon}}
\frac{1}{\qty[\Kt^2 + \Delta_1]  \qty[(\Kt+\Lt)^2 + \Delta_2]}
\\
    =&
    \left[4\pi\,\mu^2\right]^{\epsilon}\;
\Gamma\left(1\!+\!\epsilon\right)\;
    \int_0^1 \dd{x} \qty[ x(1-x) \Lt^2 + (1-x) \Delta_1 + x \Delta_2 ]^{-1-\epsilon}
\\    
=& \left[4\pi\,\mu^2\right]^{\epsilon}\;
\Gamma\left(1\!+\!\epsilon\right)\;
(1\!-\!\xi)^{-1-\epsilon}\;
\int_{0}^{1} \dd{x}
\left[ x (1\!-\!\xi) + \frac{ \xi}{(1\!-\!z)} \right]^{-1-\epsilon}\; 
\Big[(1\!-\!x)\Pt^2 + \overline{Q}^2\Big]^{-1-\epsilon}
\end{split}
\end{equation}
with the definitions \eqref{eq:xi_variable}, 
\begin{equation}
    \Delta_2 = \frac{(k_0^+ - k_2^+)(k_1^+ + k_2^+)}{k_0^+ k_1^+} \overline{Q}^2
\end{equation}
and 
\begin{equation}
    \Lt = - \frac{k_0^+ - k_2^+}{k_0^+} \Pt.
\end{equation}
It is useful to combine this contribution with a part of the instantaneous gluon exchange diagram $3$.
We can divide Diagram $3$ into two contributions based on the momentum flow between the quark and the antiquark, which we will denote by $3a$ and $3b$.
The contribution $3a$, corresponding to the light-cone momentum flowing from the antiquark into the quark, is given  in Eq.~(126) from Ref.~\cite{Beuf:2016wdz} as
\begin{align}
{\cal V}_{3a}^{L}
=&\,
 \int_{0}^{k_0^+} \frac{d k_2^+}{q^+} \;\; \bigg[
   \frac{2k_0^+k_1^+}{(k_2^+)^2}
   +\frac{2(k_0^+\!-\!k_1^+)}{k_2^+}
   -2
  \bigg]\;\;
 \mathcal{A}_0\left(\Delta_2\right)
\label{V3aL_decomp}
 \, .
\end{align}
We will combine this contribution with the one from Diagram 1, and
similarly the contribution $3b$ will be combined with Diagram 2.
We then need to calculate explicitly only Diagram 1 and the contribution $3a$,  as Diagram 2 and the contribution $3b$ can be determined by symmetry:
\begin{equation}
{\cal V}_{2}^{L}+
{\cal V}_{3b}^{L}=
\qty[
{\cal V}_{1}^{L}+
{\cal V}_{3a}^{L}
]
\qty(z \leftrightarrow 1-z).
\end{equation}

Now, it is more convenient to split the the vertex corrections contributions ${\cal V}_{1}^{L}$ and ${\cal V}_{3a}^{L}$ according to the basis of scalar master integrals $\mathcal{A}_0(\Delta_1)$, $\mathcal{A}_0(\Delta_2)$ and $\mathcal{B}_0$.
For the $\mathcal{A}_0(\Delta_1)$ contribution, present in  ${\cal V}_{1}^{L}$ via $\mathcal{I}_+$, dimensional regularization is enough to regulate both the transverse and the $k_2^+$ integral for the same reason as in the self-energy diagram $A$, and one has
\begin{equation}
    \begin{split}
{\cal V}_{1}^{L}\bigg|_{\mathcal{A}_0(\Delta_1)}
=&\,
\int_{0}^{k_0^+}\!\! \dd{k_2^+}
\Bigg\{\frac{1}{k_2^+} +\frac{(k_0^+\!-\!k_1^+)}{2k_0^+k_1^+}
        -\frac{(1\!-\!\deltas \epsilon)k_2^+}{2k_0^+k_1^+}
\Bigg\}\;\mathcal{A}_0(\Delta_1)
\\
=&\,
\int_{0}^{1}\!\! \dd{\xi}
\Bigg\{\frac{1}{\xi} +\frac{(2z\!-\!1)}{2(1\!-\!z)}
        -\frac{(1\!-\!\deltas \epsilon)z \xi}{2(1\!-\!z)}
\Bigg\}\;
\Gamma\left(\epsilon\right)\;
    \left[\frac{\xi(1\!-\!\xi)(\Pt^2\!+\!\overline{Q}^2)}{4\pi\,\mu^2 (1\!-\!z)}\right]^{-\epsilon}
   \\
=&\,
- \frac{\Gamma\!\left(1\!+\!\epsilon\right)}{\epsilon^2}\, \left[\frac{\Pt^2\!+\!\overline{Q}^2}{4\pi\,\mu^2 (1\!-\!z)}\right]^{-\epsilon}\,
\frac{\left(\Gamma\left(1\!-\!\epsilon\right)\right)^2}{\Gamma\left(1\!-\!2\epsilon\right)}\,
\left\{
1-\frac{\epsilon\left[3z\!-\!2 \!+\!\deltas \epsilon z\right]}{4(1\!-\!z)(1\!-\!2\epsilon)}
\right\}
   \\
=&\,
-  \frac{S_{\epsilon} }{\epsilon^2}\, \left[\frac{\Pt^2\!+\!\overline{Q}^2}{\mu^2 (1\!-\!z)}\right]^{-\epsilon}\,
+ \frac{(3z\!-\!2)}{4(1\!-\!z)}\,  \frac{S_{\epsilon} }{\epsilon} \left[\frac{\Pt^2\!+\!\overline{Q}^2}{\mu^2 (1\!-\!z)}\right]^{-\epsilon}\,
\\
&
+\frac{\pi^2}{12}+\frac{\deltas\, z}{4(1\!-\!z)}+\frac{(3z\!-\!2)}{2(1\!-\!z)}
+\order{\epsilon}
\label{V1L_A0_Delta1}
 \, ,
    \end{split}
\end{equation}
using the change of variables from Eq.~\eqref{eq:xi_variable}.

Concerning the contributions proportional $\mathcal{A}_0(\Delta_2)$ in ${\cal V}_{1}^{L}$ and ${\cal V}_{3a}^{L}$, since $\Delta_2$ is not proportional to $k_2^+$ and has a finite limit for $k_2^+\rightarrow 0$, dimensional regularization is not enough to regulate the terms in $\dd{k_2^+}/k_2^+$ and  $\dd{k_2^+}/(k_2^+)^2$ at $k_2^+=0$ after performing the transverse integral. 
However, collecting all the  $\mathcal{A}_0(\Delta_2)$ contributions from Eqs.~\eqref{V1L_decomp} (including the ones present in $\mathcal{I}_{+}$) and \eqref{V3aL_decomp} one finds
\begin{equation}
{\cal V}_{3a}^{L}+{\cal V}_{1}^{L}\bigg|_{\mathcal{A}_0(\Delta_2)}
=\,
\int_{0}^{k_0^+}\!\! \frac{\dd{k_2^+}}{q^+}\;
\Bigg\{-\frac{(k_0^+\!-\!k_1^+)^2}{2k_0^+k_1^+}
-1-\deltas \epsilon
+\frac{(1\!-\!\deltas \epsilon)k_2^+(k_0^+\!-\!k_1^+)}{2k_0^+k_1^+}
\Bigg\}\;\mathcal{A}_0(\Delta_2)
\label{V1L_V3aL_A0_Delta2_1}
 \, .
\end{equation}
Interestingly, all the terms in $\dd{k_2^+}/k_2^+$ and  $\dd{k_2^+}/(k_2^+)^2$ have cancelled between ${\cal V}_{1}^{L}$ and ${\cal V}_{3a}^{L}$, so that now the $k_2^+$ integral is finite and dimensional regularization for the transverse integration is enough. 
Writing Eq.~\eqref{V1L_V3aL_A0_Delta2_1} in terms of the variables given in Eq.~\eqref{eq:xi_variable}, we get
\begin{equation}
    \begin{split}
&{\cal V}_{3a}^{L}+{\cal V}_{1}^{L}\bigg|_{\mathcal{A}_0(\Delta_2)}
\\
=&\,
z\int_{0}^{1}\!\! \dd{\xi}\;
\Bigg\{-\frac{(2z\!-\!1)^2}{2z(1\!-\!z)}
-1-\deltas \epsilon
+\frac{(1\!-\!\deltas \epsilon) (2z\!-\!1)\xi}{2(1\!-\!z)}
\Bigg\}\;\Gamma\left(\epsilon\right)\;
    \left[\frac{(1\!-\!\xi)\overline{Q}^2}{4\pi\,\mu^2 } \left(1\!+\!\frac{z\xi}{(1\!-\!z)}\right)\right]^{-\epsilon}
        \\
=&\,
\left[-\frac{1}{4(1\!-\!z)} +\frac{(2z\!-\!1)}{4}
\right]\frac{S_{\epsilon} }{\epsilon} \left[\frac{\overline{Q}^2}{\mu^2 }\right]^{-\epsilon}
-\frac{\log(1\!-\!z)}{4z(1\!-\!z)}
- \frac{\deltas\, z  (3\!-\!2z)}{4(1\!-\!z)}
-\frac{1}{2(1\!-\!z)}
+\frac{(2z\!-\!1)}{4}
\\
&
+\order{\epsilon}
 \, .
 \label{V1L_V3aL_A0_Delta2_2}
    \end{split}
\end{equation}

The final remaining contribution is the term propotional to $\mathcal{B}_0$ from ${\cal V}_{1}^{L}$ in Eq.~\eqref{V1L_decomp}.
In this case, dimensional regularization is not enough to regulate divergences from the $k_2^+$ integral, and actually the transverse integral in $\mathcal{B}_0$ itself is UV finite. It is convenient to isolate a divergent piece
\begin{align}
{\cal V}_{1}^{L}\bigg|_{\mathcal{B}_0\textrm{; div.}}
=&\,
\int_{0}^{k_0^+}\!\! 
\frac{\dd{k_2^+}}{k_2^+} \;
\left(\frac{k_0^+\!-\!k_2^+}{k_0^+}\right)\,
\left[\left(\frac{k_1^+\!+\!k_2^+}{k_1^+}\right)\, \Pt^2
  + \left(\frac{k_0^+\!-\!k_2^+}{k_0^+}\right)\, \overline{Q}^2 \right]
\mathcal{B}_0
\label{V1L_B0_div}
 \, ,
\end{align}
 that we will discuss below, and a fully finite piece
\begin{equation}
\begin{split}
{\cal V}_{1}^{L}\bigg|_{\mathcal{B}_0\textrm{; finite}}
=&\,
\int_{0}^{k_0^+}\!\! \dd{k_2^+}\;
\Bigg\{\frac{(k_0^+\!-\!k_1^+)}{k_0^+k_1^+}
        -\frac{(1\!-\!\deltas \epsilon)k_2^+}{k_0^+k_1^+}
\Bigg\}\; 
 \frac{1}{2}\,
\left(\frac{k_0^+\!-\!k_2^+}{k_0^+}\right)\, 
\\
&\times
\left[\left(\frac{k_1^+\!+\!k_2^+}{k_1^+}\right)\, \Pt^2
  + \left(\frac{k_0^+\!-\!k_2^+}{k_0^+}\right)\, \overline{Q}^2 \right]
\mathcal{B}_0
\label{V1L_B0_finite_1}
 \, .
\end{split}
\end{equation}
The finite piece can be calculated as
\begin{align}
{\cal V}_{1}^{L}\bigg|_{\mathcal{B}_0\textrm{; finite}}
=&\,
\frac{ (3z\!-\!2)}{4(1\!-\!z)}\, \log\left(\frac{\Pt^2\!+\!\overline{Q}^2}{\overline{Q}^2}\right)
-\frac{ (3z\!-\!1)}{4z}\, \log(1\!-\!z)
-\frac{1}{4}
+\order{\epsilon}
\label{V1L_B0_finite_2}
 \, .
\end{align}

\subsubsection{\texorpdfstring{Rapidity divergence with the $\eta^+$ regulator}
{
Rapidity divergence with the eta+ regulator}
}

We are now left with the computation of the rapidity-divergent piece of the vertex corrections using the different regulators explained in Sec.~\ref{sec:regulators}.
Let us first consider the $\eta^+$ regulator, where we include a factor
\begin{align}
(k_2^+/\nu_B^+)^{\eta} 
&\, = (\xi z q^+/\nu_B^+)^{\eta} 
\label{eta_plus_reg}
 \, ,
\end{align}
in the integrand of Eq.~\eqref{V1L_B0_div}. The divergent piece can then be written as
\begin{align}
{\cal V}_{1}^{L}\bigg|_{\mathcal{B}_0\textrm{; div.}}^{\eta+}
=&\,
\left[\frac{z q^+}{\nu_B^+}\right]^{\eta}
\int_{0}^{1} \dd{x}
\xi^{\eta-1} \;
(1\!-\!\xi)\,
\left[\left(1\!+\!\frac{z\xi}{(1\!-\!z)}\right)\, \Pt^2
  + (1\!-\!\xi)\, \overline{Q}^2 \right]
\mathcal{B}_0
\label{V1L_B0_div_eta_plus_1}
 \, ,
\end{align}
with $\mathcal{B}_0$ defined in Eq.~\eqref{eq:B0}.
The idea is to perform both the $\xi$ and $x$ integrations, and to expand around $\eta=0$ before expanding around $\epsilon=0$. 
The $\xi$ integration is clearly regulated by $\eta$ for $\xi\rightarrow 0$. Taking $\xi=0$ inside the integrand of $\mathcal{B}_0$ in Eq.~\eqref{eq:B0} produces a factor $x^{-1-\epsilon}$, corresponding to an IR divergence at $x=0$ that is regulated by dimensional regularization. 
Hence, there is a double logarithmic divergence in the regime $\xi\rightarrow 0$ and $x\rightarrow 0$, requiring both regulators. In order to disentangle these two divergences to some extent, it is convenient to make the change of variables 
\begin{align}
x\mapsto y= x(1\!-\!\xi)+\xi 
\label{cv_x_to_y}
\end{align}
and then
\begin{align}
\xi \mapsto \zeta = \frac{\xi}{y}
\label{cv_ix_to_zeta}
\, .
\end{align}
Writing Eq.~\eqref{V1L_B0_div_eta_plus_1} in terms of these new variables, we find
\begin{equation}
    \begin{split}        
&{\cal V}_{1}^{L}\bigg|_{\mathcal{B}_0\textrm{; div.}}^{\eta+}
=\,
\left[\frac{z q^+}{\nu_B^+}\right]^{\eta}\,
\Gamma\left(1\!+\!\epsilon\right)\,
\left[4\pi\,\mu^2\right]^{\epsilon}\,
\int_{0}^{1}\!\! \dd{y} y^{-1-\epsilon+\eta}
\int_{0}^{1}\!\! 
\dd{\zeta} \zeta^{\eta-1} \;
\left[1\!+\!\frac{z\zeta}{(1\!-\!z)}\right]^{-1-\epsilon}
\\
&\, \times\;
\left[\left(1\!-\!y\right)\, \Pt^2
  + (1\!-\!y\zeta)\, \overline{Q}^2 \right]^{-1-\epsilon}
\left[\left(\left(1\!-\!y\right)\, \Pt^2
  + (1\!-\!y\zeta)\, \overline{Q}^2\right)
  +y\Pt^2 \left(1\!+\!\frac{z\zeta}{(1\!-\!z)}\right)
 \right]
\label{V1L_B0_div_eta_plus_2}
 \, .
    \end{split}
\end{equation}
From this expression, it is clear that there is a $\zeta =0$ divergence which requires a rapidity regulator and a separate $y=0$ divergence for which dimensional regularization is enough.

To extract the rapidity divergence, we follow the general procedure from Eqs.~(\ref{split_etapole_plusdistr_0}-\ref{split_etapole_plusdistr}), splitting Eq.~\eqref{V1L_B0_div_eta_plus_2} as
\begin{equation}
\label{eq:V1L_div_decomposition}
{\cal V}_{1}^{L}\bigg|_{\mathcal{B}_0\textrm{; div.}}^{\eta+}
=
{\cal V}_{1}^{L}\bigg|_{\mathcal{B}_0\textrm{;}\eta\textrm{ pole}}^{\eta+}
+
{\cal V}_{1}^{L}\bigg|_{\mathcal{B}_0\textrm{;+ distr.}}^{\eta+}
+O(\eta)
\end{equation}
where in the first term we set $\zeta =0$ everywhere except for the $\zeta^{\eta -1}$ factor, and the second term contains a plus-distribution for the $\zeta$ denominator.
The first term is then given by (see Appendix~\ref{app:details} for details)
\begin{equation}
\begin{split}
{\cal V}_{1}^{L}\bigg|_{\mathcal{B}_0\textrm{;}\eta\textrm{ pole}}^{\eta+}
=&\,
\left[\frac{z q^+}{\nu_B^+}\right]^{\eta}\,
\Gamma\left(1\!+\!\epsilon\right)\,
\left[4\pi\,\mu^2\right]^{\epsilon}\,
\int_{0}^{1}\!\! \dd{y}y^{-1-\epsilon+\eta}
\int_{0}^{1}\!\! \dd{\zeta}\ \zeta^{\eta-1} \;
\\
&\times
\left[\left(1\!-\!y\right)\, \Pt^2
  +  \overline{Q}^2 \right]^{-1-\epsilon}
\left[ \Pt^2  +  \overline{Q}^2 \right]
 \\
=&\, 
\left[\frac{1}{\eta}
+\log\left(\frac{z q^+}{\nu_B^+}\right)
\right]\,
\qty[\frac{4\pi \mu^2}{\Pt^2 + \overline Q^2}]^\epsilon
\qty[\frac{\Pt^2}{\Pt^2 + \overline Q^2}]^{\epsilon}
\Gamma(1+\epsilon)
\text{B}\qty(\frac{\Pt^2}{\Pt^2 + \overline Q^2};-\epsilon,-\epsilon)
\\
&\, 
-  \frac{S_{\epsilon} }{\epsilon^2}\,\left[\frac{\Pt^2\!+\!  \overline{Q}^2}{\mu^2}\right]^{-\epsilon}
-\textrm{Li}_2\left(\frac{\Pt^2}{\Pt^2\!+\!  \overline{Q}^2}\right) -\frac{\pi^2}{12}+\order{\epsilon}+\order{\eta f(\epsilon)}
\label{V1L_B0_div_eta_plus_eta_pole}
 \, ,
\end{split}
\end{equation}
where $\text{B}$ is the incomplete Beta function.
As a reminder, the $\eta\rightarrow 0$ limit should be taken at finite $\epsilon$, before taking the $\epsilon\rightarrow 0$ limit.
For that reason, it is not safe to expand around $\epsilon=0$ the coefficient of $1/\eta$ in Eq.~\eqref{V1L_B0_div_eta_plus_eta_pole}, and instead we keep the full $\epsilon$ dependence in the first line. By contrast, any contribution proportional to a positive power of $\eta $ is dropped, regardless of its dependence on $\epsilon$. This is indicated in Eq.~\eqref{V1L_B0_div_eta_plus_eta_pole} by the term $+\order{\eta f(\epsilon)}$. 

The incomplete Beta function could be expanded around $\epsilon=0$ as
\begin{equation}
\label{eq:inc_beta}
   x^\epsilon \text{B}\qty(x;-\epsilon,-\epsilon) 
    \equiv  
\int_0^1 \dd{y}
\frac{1}{\qty[ y (1  -xy) ]^{1+\epsilon}}
=
    \frac{-1}{\epsilon}
    - \log(1-x)
    + \order{\epsilon}.
\end{equation}
However, as explained above, it is not safe to do so due to the ordering of limits for the regulators. 

The second term in Eq.~\eqref{eq:V1L_div_decomposition}, defined with a plus-distribution and at $\eta=0$, can be written as:
\begin{equation}
    \begin{split}
{\cal V}_{1}^{L}\bigg|_{\mathcal{B}_0\textrm{;+ distr.}}^{\eta+}
=&\, 
\Gamma\left(1\!+\!\epsilon\right)\,
\left[4\pi\,\mu^2\right]^{\epsilon}\,
\int_{0}^{1}\!\! \frac{\dd{\zeta}}{(\zeta)_+}\; 
\int_{0}^{1}\!\! \dd{y} y^{-1-\epsilon} \left[1\!+\!\frac{z\zeta}{(1\!-\!z)}\right]^{-1-\epsilon}
\\
&\times
\left[\left(1\!-\!y\right)\, \Pt^2+ (1\!-\!y\zeta)\, \overline{Q}^2 \right]^{-1-\epsilon}
\\
&\, \times\;
\left\{\left[\left(1\!-\!y\right)\, \Pt^2+ (1\!-\!y\zeta)\, \overline{Q}^2 \right] + y\Pt^2 \left(1\!+\!\frac{z\zeta}{(1\!-\!z)}\right)
\right\}
\\
=&\, 
-\log(1\!-\!z) \frac{S_{\epsilon}}{\epsilon}\,
\left[\frac{\Pt^2  + \overline{Q}^2}{\mu^2}\right]^{-\epsilon}\,
-\frac{1}{2}\Big[\log(1\!-\!z)\Big]^{2}
\\
&
- \textrm{Li}_2\left(-\frac{z}{(1\!-\!z)}\right) 
+ \textrm{Li}_2\left(\frac{\Pt^2}{\Pt^2\!+\!  \overline{Q}^2}\right) 
+\order{\epsilon}
\label{V1L_B0_div_eta_plus_plus_prescr}
 \, ,
    \end{split}
\end{equation}
with the plus-distribution defined in Eq.~\eqref{def_plus_distr}.
The details of this computation are shown in Appendix~\ref{app:details}.

\subsubsection{\texorpdfstring{Rapidity divergence with the $\eta^-$ regulator}
{
Rapidity divergence with the eta- regulator}
}
An alternative way of regulating the rapidity divergence is using the $\eta^-$ regulator, where instead of powers of $(k_2^+)^\eta$ we introduce powers of $(k_2^-)^{-\eta}$, see Eq.~\eqref{def_etaminus_reg}. In LFPT, all light-front energies $k^-$ are integrated over from the start. Hence, 
 one should interpret $k_2^-$ as its on-shell value in the rapidity regularization factor from Eq.~\eqref{def_etaminus_reg}, which becomes
\begin{align}
\left[\frac{2k_2^+ \nu_B^-}{\kt_2^2}\right]^{\eta}
=&\,
 \left[\frac{2k_2^+ \nu_B^-}{\left(\Kt+\frac{k_2^+}{k_0^+}\kt_0\right)^2}\right]^{\eta} 
 \sim 
 \left[\frac{2\xi z q^+ \nu_B^-}{\Kt^2}\right]^{\eta} 
 \label{k_min_rap_reg_def}
\end{align}
in the integrand of \eqref{V1L_B0_div}, this time inside the $\mathcal{B}_0$ integral. Note that the last expression in Eq.~\eqref{k_min_rap_reg_def} for this factor can be used since it is relevant only to regulate the potential divergence at $k_2^+=0$. 
Grouping the denominators in $\mathcal{B}_0$ by introducing a Feynman parameter $x$
and then changing the integrations variables according to Eqs.~\eqref{cv_x_to_y} and \eqref{cv_ix_to_zeta}, one finds
\begin{equation}
    \begin{split}
{\cal V}_{1}^{L}\bigg|_{\mathcal{B}_0\textrm{; div.}}^{\eta-}
=&\,
4\pi\, \mu^{2\epsilon}\left[2z q^+\nu_B^-\right]^{\eta}\,
\int_{0}^{1}\!\! \dd{y}\; y^{\eta}
\int_{0}^{1}\!\! \dd{\zeta}\; \zeta^{\eta-1} \;
\int \frac{\dd[2-2\epsilon]{\Kt'}}{(2\pi)^{2-2\epsilon}}\;
\nonumber\\
&\, \times\;
\frac{\left\{\left[\left(1\!-\!y\right)\, \Pt^2+ (1\!-\!y\zeta)\, \overline{Q}^2 \right] + y\Pt^2 \left(1\!+\!\frac{z\zeta}{(1\!-\!z)}\right)
\right\}}{\left[{\Kt'}^2 
+ y \left(1\!+\!\frac{z\zeta}{(1\!-\!z)}\right) \left(\left(1\!-\!y\right)\, \Pt^2+ (1\!-\!y\zeta)\, \overline{Q}^2 \right) \right]^2
\left[\left(\Kt'+y(1\!-\!\zeta)\Pt\right)^2\right]^{\eta} }
\label{V1L_B0_div_eta_minus_1}
 \, 
    \end{split}
\end{equation}
where we have also shifted the transverse loop momentum by
\begin{equation}
    \Kt = \Kt' + y (1-\zeta) \Pt.
\end{equation}
At this stage, there is a $\zeta=0$ divergence regulated by $\eta$, and another divergence at $y=0$ where the transverse integral becomes divergent for $\Kt'=0$ which is regulated by dimensional regularization.
We can again isolate the rapidity pole contribution as in Eq.~\eqref{eq:V1L_div_decomposition}, following the general procedure from Eqs.~(\ref{split_etapole_plusdistr_0}-\ref{split_etapole_plusdistr}). In such a way, one finds the rapidity pole contribution (see Appendix~\ref{app:details}):
\begin{equation}
    \begin{split}
{\cal V}_{1}^{L}\bigg|_{\mathcal{B}_0\textrm{;}\eta\textrm{ pole}}^{\eta-}
=&\,
4\pi\, \mu^{2\epsilon}\left[2z q^+\nu_B^-\right]^{\eta}\,
\int_{0}^{1}\!\! \dd{y} y^{\eta}
\int_{0}^{1}\!\! \dd{\zeta} \zeta^{\eta-1} \;
\\
&\times
\int \frac{\dd[2-2\epsilon]{\Kt'}}{(2\pi)^{2-2\epsilon}}\;
\frac{\left[ \Pt^2+  \overline{Q}^2 \right] 
}{\left[{\Kt'}^2 
+ y \left(\left(1\!-\!y\right)\, \Pt^2+  \overline{Q}^2 \right) \right]^2
\left[\left(\Kt'+y\Pt\right)^2\right]^{\eta} }
 \\
=&\, 
\left[\frac{1}{\eta} 
+\log\left(\frac{2zq^+ \nu_B^-}{\Pt^2+\overline{Q}^2}\right)
\right]
\Gamma(1+\epsilon)
\qty[\frac{4\pi \mu^2}{\Pt^2 + \overline Q^2}]^\epsilon
\qty[\frac{\Pt^2}{\Pt^2 + \overline Q^2}]^{\epsilon}
\text{B}\qty(\frac{\Pt^2}{\Pt^2 + \overline Q^2};-\epsilon,-\epsilon)
\\
& 
-\frac{\pi^2}{3}+\order{\epsilon}+\order{\eta f(\epsilon)}
\label{V1L_B0_div_eta_minus_eta_pole}
 \, .
    \end{split}
\end{equation}
and the leftover piece from Eq.~\eqref{V1L_B0_div_eta_minus_1} written in terms of the plus-distribution for $\zeta$ as
\begin{equation}
    \begin{split}
{\cal V}_{1}^{L}\bigg|_{\mathcal{B}_0\textrm{;+ distr.}}^{\eta-}
=&\,
4\pi\, \mu^{2\epsilon}\,
\int_{0}^{1}\!\! \dd{y} 
\int_{0}^{1}\!\! \frac{\dd{\zeta}}{(\zeta)_+} 
\int \frac{\dd[2-2\epsilon]{\Kt'}}{(2\pi)^{2-2\epsilon}}\;
\\
&\times
\frac{\left\{\left[\left(1\!-\!y\right)\, \Pt^2+ (1\!-\!y\zeta)\, \overline{Q}^2 \right] + y\Pt^2 \left(1\!+\!\frac{z\zeta}{(1\!-\!z)}\right)
\right\}}{\left[{\Kt'}^2 
+ y \left(1\!+\!\frac{z\zeta}{(1\!-\!z)}\right) \left(\left(1\!-\!y\right)\, \Pt^2+ (1\!-\!y\zeta)\, \overline{Q}^2 \right) \right]^2}
\\
=&\, 
-\log(1\!-\!z) \frac{S_{\epsilon}}{\epsilon}\,
\left[\frac{\Pt^2  + \overline{Q}^2}{\mu^2}\right]^{-\epsilon}\,
-\frac{1}{2}\Big[\log(1\!-\!z)\Big]^{2}
\\
&
- \textrm{Li}_2\left(-\frac{z}{(1\!-\!z)}\right) 
+ \textrm{Li}_2\left(\frac{\Pt^2}{\Pt^2\!+\!  \overline{Q}^2}\right) 
+\order{\epsilon}
\label{V1L_B0_div_eta_minus_plus_prescr}
 \, ,
    \end{split}
\end{equation}
where the detailed computation is again done in Appendix~\ref{app:details}.
This contribution is actually the same as the analogous one with $\eta^+$ regulator, Eq.~\eqref{V1L_B0_div_eta_plus_plus_prescr}, which is expected as these terms do not need a rapidity regulator.

\subsection{Collecting the pieces}

Let us first collect all the rapidity-safe contributions to the NLO factor for the $\gamma^*_L \to q \bar q$ LFWF. 
Adding the contributions from Eqs.~\eqref{VA_2}, \eqref{V1L_A0_Delta1}, \eqref{V1L_V3aL_A0_Delta2_2} and \eqref{V1L_B0_finite_2}, as well as their symmetric contributions by the exchange of $z$ and $(1\!-\!z)$, one obtains
\begin{equation}
    \begin{split}
{\cal V}^{L}\bigg|_{\textrm{rap. safe}}
=&\,
2\,  \frac{S_{\epsilon} }{\epsilon^2}\, \left[\frac{\Pt^2\!+\!\overline{Q}^2}{\mu^2}\right]^{-\epsilon}\,
+\log\left(z(1\!-\!z)\right)\frac{S_{\epsilon} }{\epsilon} \left[\frac{\Pt^2\!+\!\overline{Q}^2}{\mu^2}\right]^{-\epsilon}\,
\\
&
-\frac{3}{2}\,\left[-\frac{S_{\epsilon} }{\epsilon} \left[\frac{\overline{Q}^2}{\mu^2}\right]^{-\epsilon}\,
+2\, \log\left(\frac{\Pt^2\!+\!\overline{Q}^2}{\overline{Q}^2}\right)
\right]
\\
&\, 
+\frac{1}{2}\Big[\log(1\!-\!z)\Big]^{2}+\frac{1}{2} \Big[\log(z)\Big]^{2}
-\frac{\pi^2}{6}+\frac{(5\!+\!\deltas)}{2}
+\order{\epsilon}
\label{VL_rap_safe_terms}
 \, .
    \end{split}
\end{equation}

From the rapidity-sensitive contributions, using either version of the rapidity regulator, one obtains scheme-dependent terms including the $1/\eta$ pole and the same terms involving a  plus-distribution, see Eqs.~\eqref{V1L_B0_div_eta_plus_plus_prescr} and \eqref{V1L_B0_div_eta_minus_plus_prescr}. Adding to the common result from Eqs.~\eqref{V1L_B0_div_eta_plus_plus_prescr} and \eqref{V1L_B0_div_eta_minus_plus_prescr} its symmetric by exchange of 
$z$ and $(1\!-\!z)$, one finds
\begin{equation}
    \begin{split}
{\cal V}^{L}\bigg|_{\textrm{+ distr.}}
=&\,
{\cal V}_{1}^{L}\bigg|_{\mathcal{B}_0\textrm{;+ distr.}}^{\eta\pm}
+ \big(z  \leftrightarrow(1\!-\!z)\big)
\\
=&\,
-\log\left(z\qty(1\!-\!z)\right)\frac{S_{\epsilon} }{\epsilon} \left[\frac{\Pt^2\!+\!\overline{Q}^2}{\mu^2}\right]^{-\epsilon}\,
-\frac{1}{2}\Big[\log(1\!-\!z)\Big]^{2}
-\frac{1}{2} \Big[\log(z)\Big]^{2}
\\
&\, 
+\frac{1}{2} \qty[\log\left(\frac{z}{1\!-\!z}\right)]^{2}
+\frac{\pi^2}{6}
+2 \textrm{Li}_2\left(\frac{\Pt^2}{\Pt^2\!+\!  \overline{Q}^2}\right) 
+\order{\epsilon}
\label{VL_+_prescr_terms}
 \, .
    \end{split}
\end{equation}
Here, we have used the identity
\begin{align}
\label{eq:dilog_identity}
\textrm{Li}_2\left(-X\right) 
+\textrm{Li}_2\left(-\frac{1}{X}\right)
=&\,
-\frac{1}{2}\, \big[\log\left(X\right)\big]^2
-\frac{\pi^2}{6}
 \end{align}
for the dilogarithm function.

\subsubsection{\texorpdfstring{$\eta^+$ regulator}
{
eta+ regulator}
}

For the $\eta^+$ regulator, the contribution from Diagram 1 including the $\eta$ pole is given in Eq.~\eqref{V1L_B0_div_eta_plus_eta_pole}. 
Adding that contribution to its 
symmetric by exchange of 
$z$ and $(1\!-\!z)$ leads to
\begin{equation}
    \begin{split}
{\cal V}^{L}\bigg|_{\eta \textrm{  pole}}^{\eta+}
=&\,
{\cal V}_{1}^{L}\bigg|_{\mathcal{B}_0\textrm{; }\eta\textrm{ pole}}^{\eta+}
+ \big(z  \leftrightarrow(1\!-\!z)\big)
\\
=&\, 
2
\left[\frac{1}{\eta}
+\log\left(\frac{\sqrt{z(1-z)} q^+}{\nu_B^+}\right)
\right]
\Gamma(1+\epsilon)
\qty[\frac{4\pi \mu^2}{\Pt^2 + \overline Q^2}]^\epsilon
\qty[\frac{\Pt^2}{\Pt^2 + \overline Q^2}]^{\epsilon}
\text{B}\qty(\frac{\Pt^2}{\Pt^2 + \overline Q^2};-\epsilon,-\epsilon)
\\
&\, 
- 2 \frac{S_{\epsilon} }{\epsilon^2}\,\left[\frac{\Pt^2\!+\!  \overline{Q}^2}{\mu^2}\right]^{-\epsilon}
-2\textrm{Li}_2\left(\frac{\Pt^2}{\Pt^2\!+\!  \overline{Q}^2}\right) -\frac{\pi^2}{6}+\order{\epsilon}+\order{\eta f(\epsilon)}
\label{VL_eta_pole_plus}
 \, ,
    \end{split}
\end{equation}
The full result for the loop factor in momentum space using the $\eta^+$ regulator is then obtained by adding the contributions
\eqref{VL_rap_safe_terms}, \eqref{VL_+_prescr_terms}
and
\eqref{VL_eta_pole_plus} together, as
\begin{equation}
    \begin{split}
{\cal V}^{L}\bigg|^{\eta+}
=&\,
2
\left[\frac{1}{\eta}
+\log\left(\frac{\sqrt{z(1-z)} q^+}{\nu_B^+}\right)
\right]
\Gamma(1+\epsilon)
\qty[\frac{4\pi \mu^2}{\Pt^2 + \overline Q^2}]^\epsilon
\qty[\frac{\Pt^2}{\Pt^2 + \overline Q^2}]^{\epsilon}
\text{B}\qty(\frac{\Pt^2}{\Pt^2 + \overline Q^2};-\epsilon,-\epsilon)
 \\
& -\frac{3}{2}
\left[-\frac{S_{\epsilon} }{\epsilon} \left[\frac{\overline{Q}^2}{\mu^2}\right]^{-\epsilon}\,
+2\, \log\left(\frac{\Pt^2\!+\!\overline{Q}^2}{\overline{Q}^2}\right)
\right]
\\
&\,
+\frac{1}{2}\qty[\log\left(\frac{z}{1\!-\!z}\right)]^{2}
-\frac{\pi^2}{6}+\frac{(5\!+\!\deltas)}{2}
+\order{\epsilon}+\order{\eta f(\epsilon)}
\label{VL_eta_plus_mom}
 \, .
    \end{split}
\end{equation}
Note that the $\epsilon$ double pole cancels in this case.

\subsubsection{\texorpdfstring{$\eta^-$ regulator}
{eta- regulator}
}

For the $\eta^-$ regulator, the contribution from Diagram 1 including the pole in $\eta$ is given in Eq.~\eqref{V1L_B0_div_eta_minus_eta_pole}. 
Adding that contribution to its 
symmetric by exchange of 
$z$ and $(1\!-\!z)$ leads to
\begin{equation}
    \begin{split}
{\cal V}^{L}\bigg|_{\eta \textrm{  pole}}^{\eta-}
=&\,
{\cal V}_{1}^{L}\bigg|_{\mathcal{B}_0\textrm{; }\eta\textrm{ pole}}^{\eta-}
+ \big(z  \leftrightarrow(1\!-\!z)\big)
\\
=&\, 
2 \left[\frac{1}{\eta} 
+\log\left(\frac{2\sqrt{z(1\!-\!z)}q^+ \nu_B^-}{\Pt^2\!+\!\overline{Q}^2}\right)
\right]
\Gamma(1+\epsilon)
\qty[\frac{4\pi \mu^2}{\Pt^2 \!+\! \overline Q^2}]^\epsilon
\qty[\frac{\Pt^2}{\Pt^2 \!+\! \overline Q^2}]^{\epsilon}
\\
& 
\times\, 
\text{B}\qty(\frac{\Pt^2}{\Pt^2 \!+\! \overline Q^2};-\epsilon,-\epsilon)
-\frac{2\pi^2}{3}+\order{\epsilon}+\order{\eta f(\epsilon)}
\\
=&\, 
2 \left[\frac{1}{\eta} 
+\log\left(\frac{2\sqrt{z(1\!-\!z)}q^+ \nu_B^-}{\overline{Q}^2}\right)
\right]
\Gamma(1+\epsilon)
\qty[\frac{4\pi \mu^2}{\Pt^2 \!+\! \overline Q^2}]^\epsilon
\qty[\frac{\Pt^2}{\Pt^2 \!+\! \overline Q^2}]^{\epsilon}
\\
& 
\times\, 
\text{B}\qty(\frac{\Pt^2}{\Pt^2 \!+\! \overline Q^2};-\epsilon,-\epsilon)
-\frac{2\pi^2}{3}
\\
& 
+2\log\left(\frac{\Pt^2 \!+\! \overline Q^2}{\Pt^2}\right)
\left[\frac{S_{\epsilon} }{\epsilon} \left(\frac{\overline{Q}^2}{\mu^2}\right)^{-\epsilon}\,
-2\log\left(\frac{\Pt^2 \!+\! \overline Q^2}{\Pt^2}\right)\right]
+\order{\epsilon}+\order{\eta f(\epsilon)}
\label{VL_eta_pole_minus}
 \, .
    \end{split}
\end{equation}
The full result for the loop factor in momentum space using the $\eta^-$ regulator is then obtained by adding the contributions
\eqref{VL_rap_safe_terms}, \eqref{VL_+_prescr_terms}
and
\eqref{VL_eta_pole_minus} together, as
 \begin{equation}
     \begin{split}
{\cal V}^{L}\bigg|^{\eta-}
=&\, 
2 \left[\frac{1}{\eta} 
+\log\left(\frac{2\sqrt{z(1-z)}q^+ \nu_B^-}{\overline{Q}^2}\right)
\right]
\Gamma(1+\epsilon)
\qty[\frac{4\pi \mu^2}{\Pt^2 + \overline Q^2}]^\epsilon
\qty[\frac{\Pt^2}{\Pt^2 + \overline Q^2}]^{\epsilon}
\\
& 
\times\, 
\text{B}\qty(\frac{\Pt^2}{\Pt^2 + \overline Q^2};-\epsilon,-\epsilon)
-\frac{3}{2}
\left[-\frac{S_{\epsilon} }{\epsilon} \left[\frac{\overline{Q}^2}{\mu^2}\right]^{-\epsilon}\,
+2 \log\left(\frac{\Pt^2\!+\!\overline{Q}^2}{\overline{Q}^2}\right)\right]
\\
&
+2\,  \frac{S_{\epsilon} }{\epsilon^2} \left[\frac{\overline{Q}^2}{\mu^2}\right]^{-\epsilon}\,
+2\, \textrm{Li}_2\left(\frac{\Pt^2}{\Pt^2\!+\!\overline{Q}^2}\right)
-3 \left[ \log\left(\frac{\Pt^2\!+\!\overline{Q}^2}{\overline{Q}^2}\right)\right]^2
\\
&
+\frac{1}{2} \left[ \log\left(\frac{z}{1\!-\!z}\right)\right]^2
-\frac{2\pi^2}{3}+\frac{(5+\deltas)}{2}
  +\order{\epsilon}
  +\order{\eta f(\epsilon)}
\label{VL_eta_minus_mom}
 \, .
     \end{split}
 \end{equation}
In that case, the double pole term in $\epsilon$ survives, as well as extra terms with dilogarithmic  or double-logarithmic dependence on $\Pt^2$.


\subsection{Fourier transform to mixed space}

So far, we have derived the $\gamma_L^*\rightarrow q\bar{q}$ LFWF in full momentum space at NLO. 
The dipole picture of DIS is formulated in terms of intermediate Fock states in mixed space where the transverse momenta are Fourier transformed into position space. 
Hence, we need to Fourier transform the LFWF to the mixed space:
\begin{align}
\widetilde{\Psi}_{\gamma^*_L\rightarrow q_0+\bar{q}_1}
\equiv &\,
\int\frac{\dd[2-2\epsilon]{\kt_0}}{(2\pi)^{2-2\epsilon}}
\int\frac{\dd[2-2\epsilon]{\kt_1}}{(2\pi)^{2-2\epsilon}}\,
e^{i\kt_0\vdot\xt_0 +i \kt_1\vdot\xt_1}\,
{\Psi}_{\gamma^*_L\rightarrow q_0+\bar{q}_1}
\nonumber\\
=&\,
\int\frac{\dd[2-2\epsilon]{\kt_0}}{(2\pi)^{2-2\epsilon}}
\int\frac{\dd[2-2\epsilon]{\kt_1}}{(2\pi)^{2-2\epsilon}}\,
e^{i\kt_0\vdot\xt_0 +i \kt_1\vdot\xt_1}\,
\left[1+\frac{\as \cf}{2\pi}\,
{\cal V}^{L} 
\right]\:
{\Psi}^{\lo}_{\gamma^*_L\rightarrow q_0+\bar{q}_1}
+\order{\as^2}
\, .
\label{FT_of_qqbar_LFWF_L}
\end{align}
At leading order, the momentum-space LFWF is given by~\cite{Beuf:2016wdz}:
\begin{align}
{\Psi}^{\lo}_{\gamma^*_L\rightarrow q_0+\bar{q}_1}
=&\,
2\pi
\delta(k_0^+ \!+\! k_1^+ \!-\! q^+)
(2\pi)^{2-2\epsilon}
\delta^{(2-2\epsilon)}
(\kt_0 + \kt_1 - \qt)\, \delta_{\alpha_0 \alpha_1}
\nonumber\\
&\,
\times\,
(-1)
\mu^{\epsilon}
e e_f
\qty( \frac{2 k_0^+ k_1^+}{(q^+)^2} )
\frac{1}{\qty[\Pt^2 + \overline Q^2 - i\epsilon]}
\frac{Q}{q^+}
\overline u(0) \gamma^+ v(1),
\end{align}
where $\Pt = \frac{k_1^+}{q^+} \kt_0 - \frac{k_0^+}{q^+} \kt_1$, and $\alpha_0$ and $\alpha_1$ are the color indices of the quark and antiquark.
The combination $\overline u(0) \gamma^+ v(1)$ involves only the so-called good components of the spinors $u(0)$ and $v(1)$ associated with the quark and antiquark. Hence, this combination depends on their light-cone momenta $k^+_0$ and $k^+_1$ as well as their light-front helicities, but not on their transverse momenta (see for example Appendix A.1 of Ref.~\cite{Beuf:2016wdz}). We are thus free to compute the Fourier transform \eqref{FT_of_qqbar_LFWF_L} over $\kt_i$, using Eq.~\eqref{eq:FT_LO}, and obtain:
\begin{equation}
\begin{split}
\widetilde{\Psi}^\lo_{\gamma^*_L\rightarrow q_0+\bar{q}_1}
=&
2\pi
\delta(k_0^+ \!+\! k_1^+ \!-\! q^+)
\delta_{\alpha_0 \alpha_1}
\overline u(0) \gamma^+ v(1)
\frac{e e_f}{2\pi}
\qty(
- \frac{2 k_0^+ k_1^+}{(q^+)^2}
)
Q
e^{i \frac{\qt}{q^+}
\vdot (k_0^+ \xt_0 + k_1^+ \xt_1)
}
\\
& \times
\qty(
\frac{\overline Q}{2\pi \abs{\xt_{01}} \mu})^{-\epsilon}
\Kb_{-\epsilon} \qty( \abs{\xt_{01}} \overline Q ).
\end{split}
\label{FT_of_qqbar_LFWF_L_LO}
\end{equation}
In general, the reduced LFWF for $\gamma^*_L\rightarrow q_0+\bar{q}_1$ appearing in Eq.~\eqref{eq:sigma_NLO} is defined as 
\begin{equation}
\begin{split}
\widetilde{\Psi}_{\gamma^*_L\rightarrow q_0+\bar{q}_1}
=&
(2q^+)2\pi
\delta(k_0^+ \!+\! k_1^+ \!-\! q^+)
\delta_{\alpha_0 \alpha_1}\,
e^{i \frac{\qt}{q^+}
\vdot (k_0^+ \xt_0 + k_1^+ \xt_1)
}\, 
\widetilde{\psi}_{\gamma^*_L\rightarrow q_0+\bar{q}_1}\, ,
\end{split}
\label{def_reduced_LFWF}
\end{equation}
so that at LO, one finds
\begin{equation}
\begin{split}
\widetilde{\psi}^\lo_{\gamma^*_L\rightarrow q_0+\bar{q}_1}
=&
\frac{1}{2q^+}\,
\overline u(0) \gamma^+ v(1)
\frac{e e_f}{2\pi}
\qty(
- \frac{2 k_0^+ k_1^+}{(q^+)^2}
)
Q
\qty(
\frac{\overline Q}{2\pi \abs{\xt_{01}} \mu})^{-\epsilon}
\Kb_{-\epsilon} \qty( \abs{\xt_{01}} \overline Q ).
\end{split}
\label{FT_of_qqbar_LFWF_L_LO_reduced}
\end{equation}

\subsubsection{\texorpdfstring{$\eta^+$ regulator}
{eta+ regulator}
}

Using Eqs.~\eqref{eq:FT_LO}, \eqref{eq:FT_log}, and \eqref{eq:FT_beta} from Appendix~\ref{app:FT},  one can calculate the Fourier transform \eqref{FT_of_qqbar_LFWF_L} for the $\eta^+$ regulator with the loop factor \eqref{VL_eta_plus_mom}.
 The result can still be written in a factorized way as
\begin{align}
\widetilde{\Psi}_{\gamma^*_L\rightarrow q_0+\bar{q}_1}
=&\,
\left[1+\frac{\as \cf}{2\pi}\,
\widetilde{\cal V}^{L} 
\right]\:
\widetilde{\Psi}^{\lo}_{\gamma^*_L\rightarrow q_0+\bar{q}_1}
+\order{e\, \as^2}
\, ,
\label{NLO_qqbar_LFWF_L_mixed}
\end{align}
or equivalently at the level of reduced LFWF as
\begin{align}
\widetilde{\psi}_{\gamma^*_L\rightarrow q_0+\bar{q}_1}
=&\,
\left[1+\frac{\as \cf}{2\pi}\,
\widetilde{\cal V}^{L} 
\right]\:
\widetilde{\psi}^{\lo}_{\gamma^*_L\rightarrow q_0+\bar{q}_1}
+\order{e\, \as^2}
\, ,
\label{NLO_qqbar_LFWF_L_mixed_reduced}
\end{align}
with the mixed-space loop factor
\begin{equation}
\begin{split}
\widetilde{\cal V}^{L}\bigg|^{\eta+}
=&\,
-2\left[\frac{1}{\eta}
+ \log\left(
\frac{\sqrt{z(1-z)} q^+}{\nu_B^+}\right)
\right]
\frac{\Gamma(1\!-\!\epsilon)}{\epsilon}
\qty[ \pi \mu^2 \xt_{01}^2]^\epsilon
 +\frac{3}{2}
\frac{S_{\epsilon} }{\epsilon} \left[\frac{\xt_{01}^2 \mu^2}{c_0^2}\right]^{\epsilon}\,
\\
&\,
+\frac{1}{2}\left[\log\left(\frac{z}{1\!-\!z}\right)\right]^{2}
-\frac{\pi^2}{6}+\frac{(5\!+\!\deltas)}{2}
+\order{\epsilon}+\order{\eta  f(\epsilon)}
\label{VL_eta_plus_pos}
 \, ,
\end{split}
\end{equation}
where $c_0 = 2 e^{\Psi(1)}=2 e^{-\gamma_E}$. 
We can now compare the final expressions \eqref{VL_eta_plus_mom} and \eqref{VL_eta_plus_pos}, obtained with the $\eta^+$ regulator, to those 
with the cut-off regulator in Ref.~\cite{Beuf:2016wdz}:
\begin{align}
\begin{split}
    \mathcal{V}^L\bigg|^{\text{cut-off}} =& 
    \qty[
    \log(\frac{\kmin^+}{k_0^+})
    +
    \log(\frac{\kmin^+}{k_1^+})
    + \frac{3}{2}
    ]
    \qty[
    \frac{\Gamma(1\!+\!\epsilon)}{\epsilon}
    \qty(\frac{\overline Q^2}{4\pi \mu^2})^{-\epsilon}
    -2 \log(\frac{\Pt^2 +\overline Q^2}{\overline Q^2})
    ] 
    \\
    &+ \frac{1}{2} \qty[\log(\frac{k_0^+}{k_1^+})]^2
    - \frac{\pi^2}{6}
    +\frac{(5\!+\!\deltas)}{2}
+\order{\frac{\kmin^+}{q^+}}+\order{\epsilon},
\end{split}
\\
\begin{split}
  \widetilde{  \mathcal{V}}^L\bigg|^{\text{cut-off}} =& 
    \qty[
    \log(\frac{\kmin^+}{k_0^+})
    +
    \log(\frac{\kmin^+}{k_1^+})
    + \frac{3}{2}
    ]
    \qty[
    \frac{\Gamma(1\!+\!\epsilon)}{\epsilon}
    \qty(4\pi)^{\epsilon}
    + \log(\frac{\xt_{01}^2 \mu^2}{c_0^2})
    ] 
    \\
    &+ \frac{1}{2} \qty[\log(\frac{k_0^+}{k_1^+})]^2
    - \frac{\pi^2}{6}
    +\frac{(5\!+\!\deltas)}{2}
+\order{\frac{\kmin^+}{q^+}}+\order{\epsilon}.
\end{split}
\label{VL_cutoff_pos}
\end{align}
Once these expressions have been expanded in terms of $\epsilon$, we note that the terms of order $1/\epsilon$ and $\epsilon^0$ obtained either with the $\eta^+$ regulator or with the cut-off are identical up to the identification
\begin{align}
\frac{1}{\eta}+\log\left(\frac{q^+}{ \nu_B^+}\right)  \mapsto &\, \log\left(\frac{q^+}{k^+_{\textrm{min}}}\right)
\, .
 \end{align}
A more detailed comparison would require a more rigorous attention to the order of the limit $\epsilon\rightarrow 0$ and $k^+_{\textrm{min}}\rightarrow 0$ in the calculation with the cut-off.

Note also that the dependence on the photon virtuality $Q^2$ completely drops out in mixed space for both the $\eta^+$ and the cut-off regulators in Eqs.~\eqref{VL_eta_plus_pos} and~\eqref{VL_cutoff_pos}.

\subsubsection{\texorpdfstring{$\eta^-$ regulator}
{eta- regulator}}

Using the relations Eqs.~\eqref{eq:FT_LO}, \eqref{eq:FT_log}, \eqref{eq:FT_beta}, \eqref{FT_dilog_and_log2}, one can calculate the Fourier transform
\eqref{FT_of_qqbar_LFWF_L} for the $\eta^-$ regulator with the loop factor \eqref{VL_eta_minus_mom}. 
The result can still be written in the factorized form \eqref{NLO_qqbar_LFWF_L_mixed}, but now with the loop factor
\begin{equation}
    \begin{split}
\widetilde{\cal V}^{L}\bigg|^{\eta-}
=&\,
-2\left[\frac{1}{\eta}
+\log\left(
\frac{2\sqrt{z(1-z)} q^+ \nu_B^- \xt_{01}^2}{c_0^2}\right)
\right]
\frac{\Gamma(1\!-\!\epsilon)}{\epsilon}
\qty[ \pi \mu^2 \xt_{01}^2]^\epsilon
+2\,  \frac{S_{\epsilon} }{\epsilon^2}\,  \left[\frac{\xt_{01}^2 \mu^2}{c_0^2}\right]^{\epsilon}\,
\\
&\, 
+\frac{3}{2}
\frac{S_{\epsilon} }{\epsilon} \left[\frac{\xt_{01}^2 \mu^2}{c_0^2}\right]^{\epsilon}\,
+\frac{1}{2} \left[ \log\left(\frac{z}{1\!-\!z}\right)\right]^2
-\frac{\pi^2}{3}+\frac{(5+\deltas)}{2}
  +\order{\epsilon}+\order{\eta  f(\epsilon)}
\label{VL_eta_minus_pos}
 \, .        
    \end{split}
\end{equation}
The dependence on $Q^2$ still drops out in the mixed-space expression \eqref{VL_eta_minus_pos}, but there is now a leftover double pole in $\epsilon$.


\subsection{Pure rapidity regulator}

As an alternative to the $\eta^+$ and $\eta^-$ regulators (see Eqs.~\eqref{eta_plus_reg} and~\eqref{k_min_rap_reg_def}), one can instead use the pure rapidity regulator, defined in Eq.~\eqref{def_pure_rap_reg}.
As discussed at the end of Sec.~\ref{sec:regulators}, the results obtained at NLO with this rapidity regulator are automatically the mean of the ones obtained with the $\eta^+$ and with the $\eta^-$ regulators. 
In particular, the loop factor with the pure rapidity regulator is
\begin{equation}
    \begin{split}
&{\cal V}^{L}\bigg|^{\textrm{pure rap. reg.}}
=\, 
\frac{1}{2}\, 
{\cal V}^{L}\bigg|^{\eta+}
+\frac{1}{2}\,
{\cal V}^{L}\bigg|^{\eta-}
\\
=&\, 
\frac{2}{\eta}
\left[\frac{2 z(1-z) (q^+)^2 \nu_B^-}{\qty[\Pt^2+\overline{Q}^2] \nu_B^+}\right]^{\eta/2}
\qty[\frac{4\pi \mu^2}{\Pt^2 + \overline Q^2}]^\epsilon
\qty[\frac{\Pt^2}{\Pt^2 + \overline Q^2}]^{\epsilon}
\Gamma(1+\epsilon)
\text{B}\qty(\frac{\Pt^2}{\Pt^2 + \overline Q^2};-\epsilon,-\epsilon)
\\
&-\frac{3}{2}
\left[-\frac{S_{\epsilon} }{\epsilon} \left[\frac{\overline{Q}^2}{\mu^2}\right]^{-\epsilon}\,
+2 \log\left(\frac{\Pt^2\!+\!\overline{Q}^2}{\overline{Q}^2}\right)\right]
+ \frac{S_{\epsilon} }{\epsilon^2} \left[\frac{\overline{Q}^2}{\mu^2}\right]^{-\epsilon}\,
+ \textrm{Li}_2\left(\frac{\Pt^2}{\Pt^2\!+\!\overline{Q}^2}\right)
\\
&\, 
-\frac{3}{2} \left[ \log\left(\frac{\Pt^2\!+\!\overline{Q}^2}{\overline{Q}^2}\right)\right]^2
+\frac{1}{2} \left[ \log\left(\frac{z}{1\!-\!z}\right)\right]^2
-\frac{5\pi^2}{12}+\frac{(5+\deltas)}{2}
  +\order{\epsilon}+\order{\eta}
\label{VL_pure_rap_reg_mom}
    \end{split}
\end{equation}
in momentum space, and
\begin{equation}
    \begin{split}
&\widetilde{\cal V}^{L}\bigg|^{\textrm{pure rap. reg.}}
=\, 
\frac{1}{2}\, 
\widetilde{\cal V}^{L}\bigg|^{\eta+}
+\frac{1}{2}\,
\widetilde{\cal V}^{L}\bigg|^{\eta-}
\\
=&\, 
-\left[
\frac{2}{\eta}
+
\log\left(
\frac{2 z(1-z) (q^+)^2 \nu_B^- \xt_{01}^2}{c_0^2 \nu_B^+}
\right)
\right]
\frac{\Gamma(1\!-\!\epsilon)}{\epsilon}
\qty[ \pi \mu^2 \xt_{01}^2]^\epsilon
+  \frac{S_{\epsilon} }{\epsilon^2}\,  \left[\frac{\xt_{01}^2 \mu^2}{c_0^2}\right]^{\epsilon}\,
\\
&\, 
+\frac{3}{2}
\frac{S_{\epsilon} }{\epsilon} \left[\frac{\xt_{01}^2 \mu^2}{c_0^2}\right]^{\epsilon}\,
+\frac{1}{2} \left[ \log\left(\frac{z}{1\!-\!z}\right)\right]^2
-\frac{\pi^2}{4}
+\frac{(5+\deltas)}{2}
  +\order{\epsilon}+\order{\eta}
\label{VL_pure_rap_reg_pos}
 \,
    \end{split}
\end{equation}
in mixed space.


\subsection{\texorpdfstring{Contribution to $F_L$ from the $q\bar{q}$ Fock states at NLO}
{Contribution to FL from the qq Fock states at NLO}}

Thanks to the factorized form \eqref{NLO_qqbar_LFWF_L_mixed_reduced} of the loop corrections to the 
$\gamma_L^*\rightarrow q\bar{q}$ LFWF in mixed space, the integrand of the $q\bar{q}$ contribution to the $\gamma_L^*$--target cross section, Eq.~\eqref{eq:sigma_NLO}, factorizes as well:
\begin{equation}
    \begin{split}
\sigma^{\gamma^*}_{L} \Big|_{q\bar{q}}
=&\, 4 N_c \aem
\sum_{f}  e_f^2
\int_{0}^{1} \!\!
\dd{z}
\int\frac{\dd[2-2\epsilon]{\xt_0}}{2\pi}
\int\frac{\dd[2-2\epsilon]{\xt_1}}{2\pi}
\;
4 Q^2\, z^2 (1\!-\!z)^2\, \\
&\, \times\,
\left(\frac{(2\pi)^2\mu^2\xt_{01}^2}{\overline{Q}^2}\right)^{\epsilon}
\left[\Kb_{-\epsilon}\qty(\overline{Q}|\xt_{01}|)\right]^2
{\textrm{Re}}\left[1-\dipole\right]\,
\left[1+\frac{\as \cf}{\pi}\,
\widetilde{\cal V}^{L} 
\right]\:
+\order{\aem \as^2}
\, .
\label{sigma_L_qqbar_1}
    \end{split}
\end{equation}
Here, the first term in the last bracket corresponds to the dipole factorization formula at LO, and the other  is the NLO contribution due to a loop in the LFWF before interaction with the target, either in the amplitude or in the complex conjugate amplitude. 
The general form of Eq.~\eqref{sigma_L_qqbar_1}
is valid for any of the considered rapidity regulators. Only the expression of the loop factor $\widetilde{\cal V}^{L}$ depends on the choice of rapidity regulator: Eq.~\eqref{VL_cutoff_pos} in the cut-off case, Eq.~\eqref{VL_eta_plus_pos} in the $\eta^+$ case, 
Eq.~\eqref{VL_eta_minus_pos} in the $\eta^-$ case,
and \eqref{VL_pure_rap_reg_pos} in the pure rapidity regulator case.


\section{\texorpdfstring{Contributions to $F_L$ at NLO from the $ q\bar{q}g$ Fock states}
{Contributions to FL at NLO from the qqg Fock states
}
}
\label{sec:FL_qqbarg}

Let us now consider the contribution from the $q \bar q g$ Fock state to the structure function $F_L$, given by the second line in Eq.~\eqref{eq:sigma_NLO}.
We will follow the calculation from Ref.~\cite{Beuf:2017bpd}, using the same notations and naming of diagrams.
This allows us to avoid repeating large parts of the computation, as we can start from the intermediate results where the rapidity regulator has not been taken into account yet.

\begin{figure}
\setbox1\hbox to 10cm{
\includegraphics{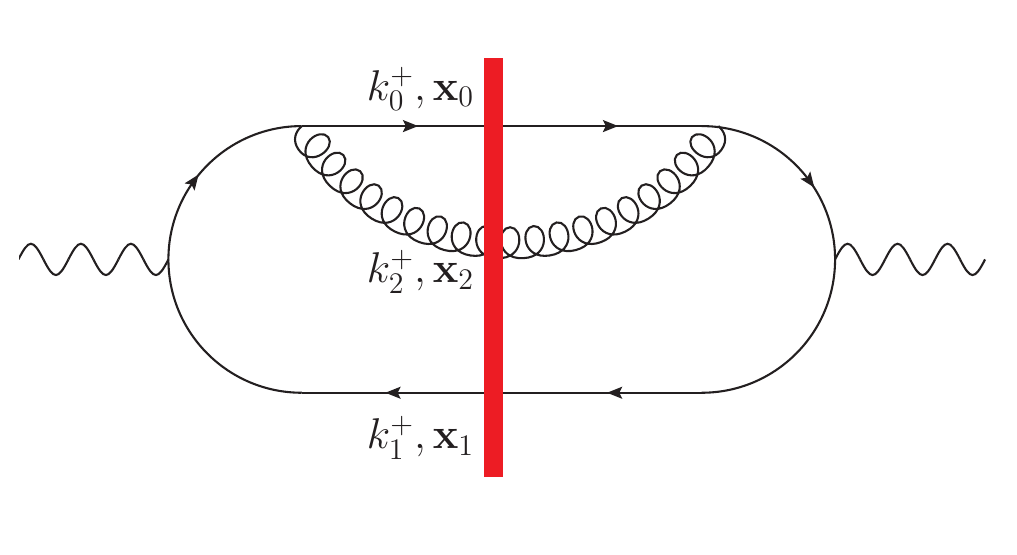}
}
\setbox2\hbox to 10cm{
\includegraphics{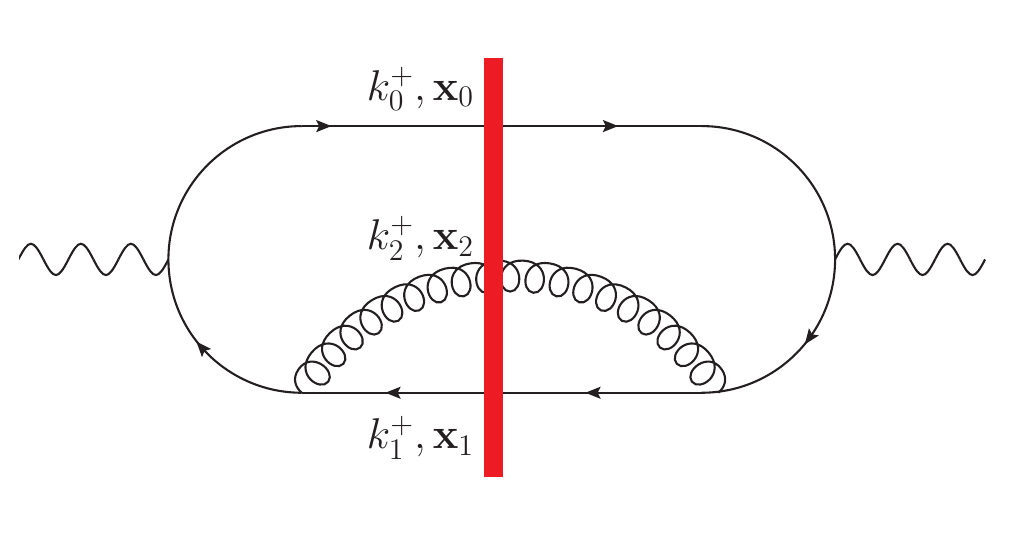}
}
\setbox3\hbox to 10cm{
\includegraphics{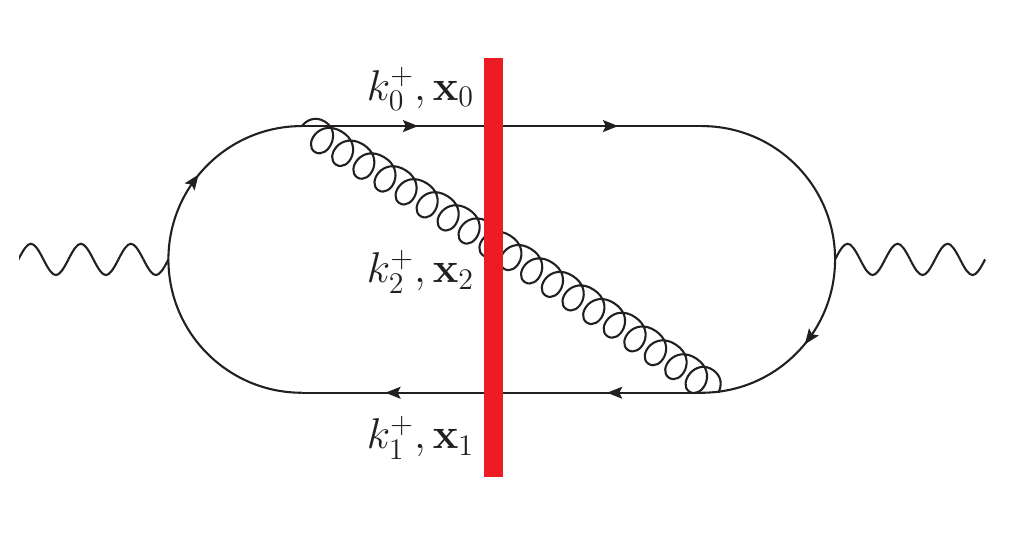}
}
\setbox4\hbox to 10cm{
\includegraphics{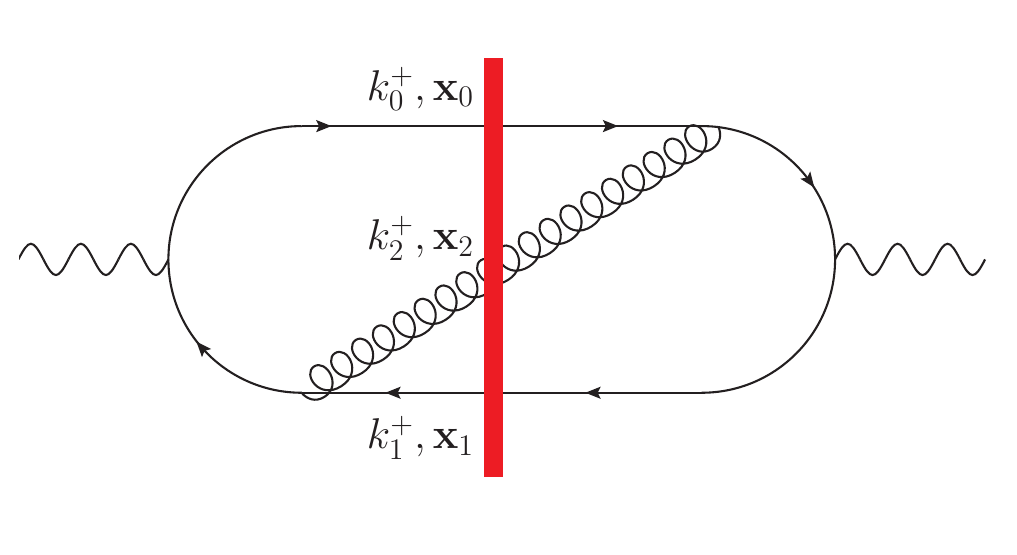}
}
\begin{center}
\resizebox*{9cm}{!}{\hspace{-7cm}\mbox{\box1 \hspace{9cm} \box2}}
\resizebox*{9cm}{!}{\hspace{-7cm}\mbox{\box3 \hspace{9cm} \box4}}
\caption{\label{fig:NLO_qqbarg_diagrams} Diagrams contributing to the $F_L$ structure function at NLO with a $q\bar q g$ intermediate Fock state scattering on the target.}
\end{center}
\end{figure}

At the LFWF level, there are two diagrams to produce a $q\bar{q}g$ Fock state from a longitudinal photon before interaction with the target, with the gluon emitted either from the quark or from the antiquark.
Thus, there are four diagrams at the level of the cross section: squared emission from the quark, squared emission from the antiquark, and  two interference diagrams, represented in Fig.~\ref{fig:NLO_qqbarg_diagrams}.  

The raw results from the calculation of these four diagrams are given in Eq.~(73) of Ref.~\cite{Beuf:2017bpd}, which we recall here for convenience:
\begin{equation}
    \begin{split}
\sigma^{\gamma^*}_{L} \Big|_{q\bar{q}g}
=&\, 4 N_c \aem \left(\frac{\as \cf}{\pi}\right)
\sum_{f}  e_f^2\,
\int_{0}^{\infty} \dd{k^+_0} \int_{0}^{\infty}  \dd{k^+_1}
\int_{0}^{\infty}  
\frac{\dd{k^+_2}}{k^+_2}\; \frac{\delta(k_0^+\!+\!k_1^+\!+\!k_2^+\!-\!q^+)}{q^+}\; \\
& \times
\int \frac{\dd[D-2]{\xt_0}}{2\pi} \int \frac{\dd[D-2]{\xt_1}}{2\pi} \int \frac{\dd[D-2]{\xt_2}}{2\pi}\:
{\textrm{Re}}\left[1-\tripole\right]\;
\frac{4 Q^2}{(q^+)^4}\; \frac{(2\pi)^4}{2}\;
\\
\, \times
\Bigg\{
& (k_{1}^+)^2\, \Big[4k_{0}^+(k_{0}^+\!+\!k_{2}^+) + (D_s\!-\!2)(k_{2}^+)^2\Big]\,
\Big|{\cal I}^{m}\!\left(a\right)\Big|^2
\\
+&
(k_{0}^+)^2\, \Big[4k_{1}^+(k_{1}^+\!+\!k_{2}^+) + (D_s\!-\!2)(k_{2}^+)^2\Big]\,
\Big|{\cal I}^{m}\!\left(b\right)\Big|^2
\\
-& 2k_{0}^+ k_{1}^+ \Big[2k_{1}^+(k_{0}^+\!+\!k_{2}^+)
+2k_{0}^+(k_{1}^+\!+\!k_{2}^+)-(D_s\!-\!4)(k_{2}^+)^2\Big]\,
{\textrm{Re}}\Big({\cal I}^{m}\!\left(a\right)^*\, {\cal I}^{m}\!\left(b\right) \Big)
\Bigg\}
\, ,
\label{sigma_L_qqbarg_1}
    \end{split}
\end{equation}
where the first line in the bracket comes from the top left diagram on Fig.~\ref{fig:NLO_qqbarg_diagrams}, the second line in the bracket comes from the top right diagram on Fig.~\ref{fig:NLO_qqbarg_diagrams}, and the third line in the bracket comes from the two bottom diagrams  on Fig.~\ref{fig:NLO_qqbarg_diagrams}. 
In Eq.~\eqref{sigma_L_qqbarg_1}, the Fourier integrals ${\cal I}^{m}\!\left(a\right)$ associated with gluon emission from the quark and ${\cal I}^{m}\!\left(b\right)$ associated with gluon emission from the antiquark are defined as
\begin{align}
{\cal I}^{m}\!\left(a\right) \equiv &\,
{\cal I}^{m}\!\left(\xt_{0+2;1},\xt_{20};\overline{Q}_{(a)}^2,{\cal C}_{(a)}\right),
\\
{\cal I}^{m}\!\left(b\right) \equiv &\,
{\cal I}^{m}\!\left(\xt_{0;1+2},\xt_{21};\overline{Q}_{(b)}^2,{\cal C}_{(b)}\right)
\, ,
\end{align}
where
\begin{align}
\overline{Q}_{(a)}^2 
\equiv &\, 
\frac{k_1^+ (q^+\!-\!k_1^+)}{(q^+)^2}\; Q^2,
&
\overline{Q}_{(b)}^2
\equiv &\,
\frac{k_0^+ (q^+\!-\!k_0^+)}{(q^+)^2}\; Q^2,
\\
{\cal C}_{(a)}
\equiv &\,
\frac{q^+\, k_0^+\,k_2^+}{k_1^+ (k_0^+\!+\!k_2^+)^2},
&
{\cal C}_{(b)}
\equiv &\,
\frac{q^+\, k_1^+\,k_2^+}{k_0^+ (k_1^+\!+\!k_2^+)^2},
\\
\xt_{n+m;p}
= &\,
 -\xt_{p;n+m}
\equiv \left(\frac{k_n^+\xt_n + k_m^+\xt_m }{k_n^+\!+\!k_m^+}\right) -\xt_p
\label{def_pre-split_parent_dipole}
\, ,
\end{align}
as well as
\begin{align}
{\cal I}^{m}\!\left(\rt,\rt';\overline{Q}^2,{\cal C}\right)
\equiv &\,
(\mu^2)^{2-\frac{D}{2}}\,
\int \frac{\dd[D-2]{\Pt}}{(2\pi)^{D-2}}\; \int \frac{\dd[D-2]{\Kt}}{(2\pi)^{D-2}}\;
\frac{\Kt^m\:   e^{i\Kt\vdot \rt'}\:
e^{i\Pt\vdot \rt}\,}{ \qty[\Pt^2+\overline{Q}^2] \qty{\Kt^2 + {\cal C} \qty[\Pt^2+\overline{Q}^2]}}
\label{def_int_Im}
\, .
\end{align}

In the first term in Eq.~\eqref{sigma_L_qqbarg_1}, there is a potential UV divergence in the regime $\xt_2 \rightarrow \xt_0$, corresponding to the quark and the gluon at the same transverse position when they cross the target. It is associated with the regime of large $\Kt$ in the Fourier integrals \eqref{def_int_Im}.   
There is a similar potential UV divergence also for the second term in Eq.~\eqref{sigma_L_qqbarg_1} for $\xt_2 \rightarrow \xt_1$, corresponding to the antiquark and the gluon at the same transverse position when they cross the target. 
By contrast, the interference contribution is UV safe. 
Moreover, each of the three terms is collinear safe, corresponding to the regime of gluon emitted at large transverse distance $|\xt_{20}|\sim |\xt_{21}| \gg |\xt_{01}|$, which is associated with the regime of small $\Kt$ in the Fourier integral \eqref{def_int_Im}. 

Since the phase-space integration in Eq.~\eqref{sigma_L_qqbarg_1} contains $\dd{k_2^+}/k_2^+$, any contribution in the bracket in Eq.~\eqref{sigma_L_qqbarg_1} with a finite limit at $k^+_2=0$ will lead to a logarithmic rapidity divergence upon integration over $k^+_2$ and requires the introduction of a rapidity regulator. 
When extracting the $1/\eta$ rapidity pole, 
one has to be careful in taking the limit $k_2^+ \to 0$ as this can potentially reintroduce collinear divergences.
This can be seen from the Fourier integral~\eqref{def_int_Im}, as both ${\cal C}_{(a)}$ and ${\cal C}_{(b)}$ vanish for $k_2^+ = 0$.
Squaring this term and integrating over $\xt_2$ introduces a delta function setting the integration momenta $\Kt$ to be the same in $ {\cal I}^{m}$ and $ ({\cal I}^{m})^*$, such that we are eventually left with  integrals of the form $\int \dd[D-2]{\Kt}/\Kt^2$ for small $\Kt$.
This integral contains a collinear divergence that does not appear in the original integral~\eqref{sigma_L_qqbarg_1}.

In order to avoid these collinear divergences, it is convenient to combine parts of the interference diagrams together with the squared diagrams as
\begin{equation}
    \begin{split}
\sigma^{\gamma^*}_{L} \Big|_{q\bar{q}g}
=&\, 4 N_c \aem \left(\frac{\as \cf}{\pi}\right)
\sum_{f}  e_f^2
\int_{0}^{\infty} \dd{k^+_0} \int_{0}^{\infty}  \dd{k^+_1}
\int_{0}^{\infty}  
\frac{\dd{k^+_2}}{k^+_2}\; \frac{\delta(k_0^+\!+\!k_1^+\!+\!k_2^+\!-\!q^+)}{q^+}\; \\
& \times
\int \frac{\dd[D-2]{\xt_0}}{2\pi} \int \frac{\dd[D-2]{\xt_1}}{2\pi} \int \frac{\dd[D-2]{\xt_2}}{2\pi}\:
{\textrm{Re}}\left[1-\tripole\right]\;
\frac{4 Q^2}{(q^+)^4}\; \frac{(2\pi)^4}{2}\;
\\
\, \times
\Bigg\{
&(k_{1}^+)^2\, \Big[4k_{0}^+(k_{0}^+\!+\!k_{2}^+) + 2(1\!-\!\deltas \epsilon)(k_{2}^+)^2\Big]\,
\bigg[
\Big|{\cal I}^{m}\!\left(a\right)\Big|^2
-{\textrm{Re}}\Big({\cal I}^{m}\!\left(a\right)^*\, {\cal I}^{m}\!\left(b\right) \Big)
\bigg]
\\
+&
(k_{0}^+)^2\, \Big[4k_{1}^+(k_{1}^+\!+\!k_{2}^+) + 2(1\!-\!\deltas \epsilon)(k_{2}^+)^2\Big]\,
\bigg[
\Big|{\cal I}^{m}\!\left(b\right)\Big|^2
-{\textrm{Re}}\Big({\cal I}^{m}\!\left(a\right)^*\, {\cal I}^{m}\!\left(b\right) \Big)
\bigg]
\\ 
+&2(k_{2}^+)^2  \Big[(k_{1}^+)^2
+(k_{0}^+)^2
-\deltas \epsilon
(k_{0}^+\!+\!k_{1}^+)^2
\Big]\,
{\textrm{Re}}\Big({\cal I}^{m}\!\left(a\right)^*\, {\cal I}^{m}\!\left(b\right) \Big)
\Bigg\}
\, .
\label{sigma_L_qqbarg_2}
    \end{split}
\end{equation}
Because of these subtraction terms, the contributions in the first line and in the second line of the bracket in Eq.~\eqref{sigma_L_qqbarg_2} stay collinear safe even when isolating the rapidity divergences. 
Additionally, the contributions in the third line of the bracket are rapidity safe thanks to the $(k_2^+)^2$ prefactor, and 
since they are also UV and collinear safe, we can drop both the rapidity regularization and dimensional regularization for these terms. 
In particular, the last term proportional to $\epsilon$ goes to zero.
This allows us to write the cross section as
\begin{align}
\sigma^{\gamma^*}_{L} \Big|_{q\bar{q}g}
=&\,\sigma^{\gamma^*}_{L} \Big|_{q\bar{q}g,\, |a|^2 \textrm{ like}}
+
\sigma^{\gamma^*}_{L} \Big|_{q\bar{q}g,\, |b|^2 \textrm{ like}}
+\order{\epsilon}
\, ,
\label{sigma_L_qqbarg_3}
\end{align}
with 
\begin{equation}
    \begin{split}
\sigma^{\gamma^*}_{L} \Big|_{q\bar{q}g,\, |a|^2 \textrm{ like}}
=&\, 4 N_c \aem \left(\frac{\as \cf}{\pi}\right)
\sum_{f}  e_f^2
\int_{0}^{\infty} \dd{k^+_0} \int_{0}^{\infty}  \dd{k^+_1}
\int_{0}^{\infty}  
\frac{\dd{k^+_2}}{k^+_2}\; \frac{\delta(k_0^+\!+\!k_1^+\!+\!k_2^+\!-\!q^+)}{q^+}\; \\
& \times
\int \frac{\dd[D-2]{\xt_0}}{2\pi} \int \frac{\dd[D-2]{\xt_1}}{2\pi} \int \frac{\dd[D-2]{\xt_2}}{2\pi}\: 
{\textrm{Re}}\left[1-\tripole\right]\;
\frac{4 Q^2}{(q^+)^4}\; \frac{(2\pi)^4}{2}\;
\\
 \times
\Bigg\{
&
(k_{1}^+)^2\, \Big[4k_{0}^+(k_{0}^+\!+\!k_{2}^+) + 2(1\!-\!\deltas \epsilon)(k_{2}^+)^2\Big]\,
\bigg[
\Big|{\cal I}^{m}\!\left(a\right)\Big|^2
-{\textrm{Re}}\Big({\cal I}^{m}\!\left(a\right)^*\, {\cal I}^{m}\!\left(b\right) \Big)
\bigg]
\\
+& 2(k_{2}^+)^2  (k_{1}^+)^2
\,
{\textrm{Re}}\Big({\cal I}^{m}\!\left(a\right)^*\, {\cal I}^{m}\!\left(b\right) \Big)
\Bigg\}
\, ,
\label{sigma_L_qqbarg_a_like}
    \end{split}
\end{equation}
and
\begin{equation}
    \begin{split}
\sigma^{\gamma^*}_{L} \Big|_{q\bar{q}g,\, |b|^2 \textrm{ like}}
=&\, 4 N_c \aem \left(\frac{\as \cf}{\pi}\right)
\sum_{f}  e_f^2
\int_{0}^{\infty} \dd{k^+_0} \int_{0}^{\infty}  \dd{k^+_1}
\int_{0}^{\infty}  
\frac{\dd{k^+_2}}{k^+_2}\; \frac{\delta(k_0^+\!+\!k_1^+\!+\!k_2^+\!-\!q^+)}{q^+}\; \\
& \times
\int \frac{\dd[D-2]{\xt_0}}{2\pi} \int \frac{\dd[D-2]{\xt_1}}{2\pi} \int \frac{\dd[D-2]{\xt_2}}{2\pi}\: 
{\textrm{Re}}\left[1-\tripole\right]\;
\frac{4 Q^2}{(q^+)^4}\; \frac{(2\pi)^4}{2}\;
\\
 \times
\Bigg\{
&
(k_{0}^+)^2\, \Big[4k_{1}^+(k_{1}^+\!+\!k_{2}^+) + 2(1\!-\!\deltas \epsilon)(k_{2}^+)^2\Big]\,
\bigg[
\Big|{\cal I}^{m}\!\left(b\right)\Big|^2
-{\textrm{Re}}\Big({\cal I}^{m}\!\left(a\right)^*\, {\cal I}^{m}\!\left(b\right) \Big)
\bigg]
\\
+& 2(k_{2}^+)^2  (k_{0}^+)^2
\,
{\textrm{Re}}\Big({\cal I}^{m}\!\left(a\right)^*\, {\cal I}^{m}\!\left(b\right) \Big)
\Bigg\}
\, .
\label{sigma_L_qqbarg_b_like}
    \end{split}
\end{equation}
The two terms in Eq.~\eqref{sigma_L_qqbarg_3} are symmetric under the  exchange of $(k^+_0,\xt_0)$ and $(k^+_1,\xt_1)$, corresponding to the interchange of the quark and the antiquark. 
Because of this,
they should be equal at the integrated level. For simplicity, we will present a detailed derivation only for the contribution given in Eq.~\eqref{sigma_L_qqbarg_a_like}, whereas the derivation for the contribution given in Eq.~\eqref{sigma_L_qqbarg_b_like} can be performed in a similar way.
In the case of Eq.~\eqref{sigma_L_qqbarg_a_like}, it is convenient to perform the change of variables\footnote{\label{fn_cv_b_like}In the expression \eqref{sigma_L_qqbarg_b_like}, one instead makes the change of variables 
$k_0^+\mapsto z= k_0^+/q^+$ and $k_2^+\mapsto \xi= k_2^+/(k_1^+\!+\!k_2^+)$, so that $k_0^+=z q^+$, $k_1^+=(1\!-\!\xi)(1\!-\!z) q^+$ and $k_2^+=\xi (1\!-\!z) q^+$.
}
\begin{align}
k_1^+\mapsto &\, z = \frac{(q^+\!-\!k_1^+)}{q^+},
&
k_2^+\mapsto &\, \xi = \frac{k_2^+}{(k_0^+\!+\!k_2^+)}
\, .\label{a_like_cv}
\end{align}
This change of variables corresponds to
\begin{equation}
   \int_{0}^{\infty} \dd{k^+_0} \int_{0}^{\infty}  \dd{k^+_1}
\int_{0}^{\infty}  
\frac{\dd{k^+_2}}{k^+_2}\; \frac{\delta(k_0^+\!+\!k_1^+\!+\!k_2^+\!-\!q^+)}{q^+}\, (\cdots)
= \int_0^1 \dd{z} \int_0^1 \frac{\dd{\xi}}{\xi}\, (\cdots)
\end{equation}
with
\begin{align}
k_0^+= &\, (1\!-\!\xi) z q^+, 
&
k_1^+ = &\, (1\!-\!z){q^+},
&
k_2^+= &\, \xi z q^+ 
\, .\label{a_like_cv_2}
\end{align}
In terms of these new variables, we get:
\begin{equation}
    \begin{split}
&\sigma^{\gamma^*}_{L} \Big|_{q\bar{q}g,\, |a|^2 \textrm{ like}}
=\, 4 N_c \aem \left(\frac{\as \cf}{\pi}\right)
\sum_{f}  e_f^2\;
(2\pi)
\\
&\times
\int \dd[2-2\epsilon]{\xt_0} \int \dd[2-2\epsilon]{\xt_1} \int \dd[2-2\epsilon]{\xt_2}\:
\int_{0}^{1}
\dd{z}\,  4 Q^2\, z^2(1\!-\!z)^2
\int_0^1 \dd{\xi}
{\textrm{Re}}\left[1-\tripole\right]\;
\\
&\, \times
\Bigg\{
\left[\frac{2}{\xi}-2+
(1\!-\!\deltas \epsilon)\xi\right]\,
\bigg[
\Big|{\cal I}^{m}\!\left(a\right)\Big|^2
-{\textrm{Re}}\Big({\cal I}^{m}\!\left(a\right)^*\, {\cal I}^{m}\!\left(b\right) \Big)
\bigg]
+\xi\: 
{\textrm{Re}}\Big({\cal I}^{m}\!\left(a\right)^*\, {\cal I}^{m}\!\left(b\right) \Big)
\Bigg\}
\, .
\label{sigma_L_qqbarg_a_like_2}
    \end{split}
\end{equation}
In these variables, we have
%
\begin{equation}
{\cal I}^{m}\!\left(a\right)
= \,
\mu^{2\epsilon}\,
\int \frac{\dd[2-2\epsilon]{\Pt}}{(2\pi)^{2-2\epsilon}}\; \int \frac{\dd[2-2\epsilon]{\Kt}}{(2\pi)^{2-2\epsilon}}\;
\frac{\Kt^m\:   e^{i\Kt\vdot \xt_{20}}\:
e^{i\Pt\vdot (\xt_{01} +\xi \xt_{20})}}{ \big[\Pt^2+z(1\!-\!z){Q}^2\big] \left\{\Kt^2 + \frac{\xi(1\!-\!\xi)}{(1\!-\!z)} \big[\Pt^2+z(1\!-\!z){Q}^2\big]\right\}},
\label{def_int_Im_a_alike}
\end{equation}

\begin{equation}
    \begin{split}
        {\cal I}^{m}\!\left(b\right)
= &\,
\mu^{2\epsilon}\,
\int \frac{\dd[2-2\epsilon]{\Pt}}{(2\pi)^{2-2\epsilon}}\; \int \frac{\dd[2-2\epsilon]{\Kt}}{(2\pi)^{2-2\epsilon}}\;
\frac{\Kt^m\:   e^{i\Kt\vdot \xt_{21}}\:
}{\big[\Pt^2+z(1\!-\!\xi)\big(1\!-\!z(1\!-\!\xi)\big){Q}^2\big]}\:
e^{i\Pt\vdot \frac{
[(1\!-\!z)\xt_{01} -z\xi \xt_{20}]
}{(1\!-\!z\!+\!z\xi)}}
\\
&\, \times\,
\frac{1}{\left\{\Kt^2 + \frac{\xi(1\!-\!z)}{(1\!-\!\xi)[1\!-\!z\!+\!z\xi]^2} 
\big[\Pt^2+z(1\!-\!\xi)\big(1\!-\!z(1\!-\!\xi)\big){Q}^2\big]
\right\}}
\label{def_int_Im_b_alike}
\, .
    \end{split}
\end{equation}
As a remark, for the $|b|^2$ like term in Eq.~\eqref{sigma_L_qqbarg_b_like},
a different change of variables is used, defined in the footnote \ref{fn_cv_b_like}, so that 
different expressions in $z$ and $\xi$ variables than Eq.~\eqref{def_int_Im_a_alike} and \eqref{def_int_Im_b_alike} are obtained for the integrals ${\cal I}^{m}\!\left(a\right)$ and ${\cal I}^{m}\!\left(b\right)$.

In Eq.~\eqref{sigma_L_qqbarg_a_like_2}, only the first term in the integrand, proportional to $2/\xi$, has a rapidity divergence corresponding to $\xi\rightarrow 0$ and requires a rapidity regularization. By contrast, the rest is rapidity safe, and reads
\begin{equation}
    \begin{split}
&\sigma^{\gamma^*}_{L} \Big|_{q\bar{q}g,\, |a|^2 \textrm{ like}}^{\textrm{rap. safe}}
=\, 4 N_c \aem \left(\frac{\as \cf}{\pi}\right)
\sum_{f}  e_f^2\;
(2\pi)
\int_{0}^{1}
\dd{z}\,  4 Q^2\, z^2(1\!-\!z)^2\\
&\times
\int \dd[2-2\epsilon]{\xt_0} \int \dd[2-2\epsilon]{\xt_1} \int \dd[2-2\epsilon]{\xt_2}\:
{\textrm{Re}}\left[1-\tripole\right]\;
\\
&\, \times
\int_0^1 \dd{\xi}\Bigg\{
\big[-2+
(1\!-\!\deltas \epsilon)\xi\big]\,
\bigg[
\Big|{\cal I}^{m}\!\left(a\right)\Big|^2
-{\textrm{Re}}\Big({\cal I}^{m}\!\left(a\right)^*\, {\cal I}^{m}\!\left(b\right) \Big)
\bigg]
+\xi\: 
{\textrm{Re}}\Big({\cal I}^{m}\!\left(a\right)^*\, {\cal I}^{m}\!\left(b\right) \Big)
\Bigg\}
\, .
\label{sigma_L_qqbarg_a_like_rap_safe}
    \end{split}
\end{equation}
This contribution still contains a UV divergence in the regime $\xt_{2}\rightarrow \xt_{0}$, which can be isolated in the same way as in Ref.~\cite{Beuf:2017bpd}.


\subsection{\texorpdfstring{Rapidity-sensitive terms: $\eta^+$ regulator}
{Rapidity-sensitive terms: eta+ regulator
}
}

Let us now consider the rapidity sensitive contribution in Eq.~\eqref{sigma_L_qqbarg_a_like_2} not included in Eq.~\eqref{sigma_L_qqbarg_a_like_rap_safe}, with its integrand proportional to $2/\xi$. 
Taking into account the relation~\eqref{a_like_cv_2}, using the $\eta^+$ regulator corresponds to including the factor
\begin{align}
\left[\frac{k_2^+}{\nu_B^+}\right]^{\eta}
=&\,
\xi^{\eta}\, \left[\frac{z q^+}{\nu_B^+}\right]^{\eta} 
\end{align}
in the integrand:
\begin{equation}
    \begin{split}
\sigma^{\gamma^*}_{L} \Big|_{q\bar{q}g,\, |a|^2 \textrm{ like}}^{\textrm{rap. sensitive};\, \eta+}
=&\, 4 N_c \aem \left(\frac{\as \cf}{\pi}\right)
\sum_{f}  e_f^2\;
(2\pi)
\int_{0}^{1}
\dd{z}\,  4 Q^2\, z^2(1\!-\!z)^2
\\
&\times 
\int \dd[2-2\epsilon]{\xt_0} \int \dd[2-2\epsilon]{\xt_1} \int \dd[2-2\epsilon]{\xt_2}\:
{\textrm{Re}}\left[1-\tripole\right]\;
\\
&\, \times
\left[\frac{z q^+}{\nu_B^+}\right]^{\eta} 
\int_0^1 \dd{\xi}\, 
2\,\xi^{\eta-1}\,
\bigg[
\Big|{\cal I}^{m}\!\left(a\right)\Big|^2
-{\textrm{Re}}\Big({\cal I}^{m}\!\left(a\right)^*\, {\cal I}^{m}\!\left(b\right) \Big)
\bigg]
\, .
\label{sigma_L_qqbarg_a_like_rap_unsafe_plus}
    \end{split}
\end{equation}
As usual, in order to perform the expansion at small $\eta$, it is convenient to subtract and add back the value of the bracket at $\xi=0$ to isolate 
the $\eta$ pole.
This corresponds to dividing the integral into a singular term and a plus-distribution term as
\begin{align}
\sigma^{\gamma^*}_{L} \Big|_{q\bar{q}g,\, |a|^2 \textrm{ like}}^{\textrm{rap. sensitive};\, \eta+}
=&\, 
\sigma^{\gamma^*}_{L} \Big|_{q\bar{q}g,\, |a|^2 \textrm{ like}}^{\eta \textrm{ pole};\, \eta+}
+
\sigma^{\gamma^*}_{L} \Big|_{q\rightarrow g}^{\textrm{+ distr.}}
+\order{\eta}
\, ,
\label{sigma_L_qqbarg_a_like_rap_unsafe_plus_2}
\end{align}
with
\begin{equation}
\begin{split}
\sigma^{\gamma^*}_{L} \Big|_{q\bar{q}g,\, |a|^2 \textrm{ like}}^{\eta \textrm{ pole};\, \eta+}
=&\, 4 N_c \aem \left(\frac{\as \cf}{\pi}\right)
\sum_{f}  e_f^2\;
(2\pi)
\int_{0}^{1}
\dd{z}\,  4 Q^2\, z^2(1\!-\!z)^2
\\
&\times
\int \dd[2-2\epsilon]{\xt_0} \int \dd[2-2\epsilon]{\xt_1} \int \dd[2-2\epsilon]{\xt_2}\:
{\textrm{Re}}\left[1-\tripole\right]\;
\\
&\, \times\, 
2\,\left[\frac{z q^+}{\nu_B^+}\right]^{\eta}\:\bigg[
\Big|{\cal I}^{m}\!\left(a\right)\Big|^2
-{\textrm{Re}}\Big({\cal I}^{m}\!\left(a\right)^*\, {\cal I}^{m}\!\left(b\right) \Big)
\bigg]\bigg|_{\xi=0} 
\int_0^1 \dd{\xi}\, 
\xi^{\eta-1}\,
\label{sigma_L_qqbarg_a_like_eta_pole_plus}
\end{split}
\end{equation}
and
\begin{equation}
    \begin{split}
\sigma^{\gamma^*}_{L} \Big|_{q\rightarrow g}^{\textrm{+ distr.}}
=&\, 4 N_c \aem \left(\frac{\as \cf}{\pi}\right)
\sum_{f}  e_f^2\;
(2\pi)
\int_{0}^{1}
\dd{z}\,  4 Q^2\, z^2(1\!-\!z)^2
\\
&\times
\int \dd[2-2\epsilon]{\xt_0} \int \dd[2-2\epsilon]{\xt_1} \int \dd[2-2\epsilon]{\xt_2}\:
{\textrm{Re}}\left[1-\tripole\right]\;
\\
&\, \times\,
2 
\int_0^1 \frac{\dd{\xi}}{(\xi)_+}\, 
\,
\bigg[
\Big|{\cal I}^{m}\!\left(a\right)\Big|^2
-{\textrm{Re}}\Big({\cal I}^{m}\!\left(a\right)^*\, {\cal I}^{m}\!\left(b\right) \Big)
\bigg]
\, .
\label{sigma_L_qqbarg_a_like_plus_prescr}
    \end{split}
\end{equation}

In order to go further, one should then consider  the integrals ${\cal I}^{m}\!\left(a\right)$ and ${\cal I}^{m}\!\left(b\right)$, given in Eqs.~\eqref{def_int_Im_a_alike} and \eqref{def_int_Im_b_alike}, at $\xi=0$. Interestingly, in that case, these integrals factorize as
\begin{align}
{\cal I}^{m}\!\left(a\right)
\Big|_{\xi=0}
= &\,
\mu^{2\epsilon}\,
\int \frac{\dd[2-2\epsilon]{\Pt}}{(2\pi)^{2-2\epsilon}}\; 
\frac{
e^{i\Pt\vdot \xt_{01} }}{ \big[\Pt^2+z(1\!-\!z){Q}^2\big] }
\int \frac{\dd[2-2\epsilon]{\Kt}}{(2\pi)^{2-2\epsilon}}\;
\frac{\Kt^m
}{ \Kt^2 }\:   e^{i\Kt\vdot \xt_{20}}\:
\label{def_int_Im_a_alike_0_xi}
\\
{\cal I}^{m}\!\left(b\right)
\Big|_{\xi=0}
= &\,
\mu^{2\epsilon}\,
\int \frac{\dd[2-2\epsilon]{\Pt}}{(2\pi)^{2-2\epsilon}}\; 
\frac{
e^{i\Pt\vdot \xt_{01} }}{ \big[\Pt^2+z(1\!-\!z){Q}^2\big] }
\int \frac{\dd[2-2\epsilon]{\Kt}}{(2\pi)^{2-2\epsilon}}\;
\frac{\Kt^m
}{ \Kt^2 }\:   e^{i\Kt\vdot \xt_{21}}\:
\label{def_int_Im_b_alike_0_xi}
\, .
\end{align}
Using the integrals~\eqref{eq:FT_LO} and \eqref{eq:FT_power},
one gets
\begin{align}
{\cal I}^{m}\!\left(a\right)
\Big|_{\xi=0}
= &\,
\frac{i}{(2\pi)^2}\,
\Gamma(1\!-\!\epsilon)\,
\left(\frac{2\pi^2\mu^2|\xt_{01}|}{\overline{Q}}\right)^{\epsilon}
\Kb_{-\epsilon}\left(\overline{Q}|\xt_{01}|\right)\, 
\frac{ \xt_{20}^m}{\left(\xt_{20}^2\right)^{1-\epsilon}},
\label{def_int_Im_a_alike_0_xi_2}
\\
{\cal I}^{m}\!\left(b\right)
\Big|_{\xi=0}
= &\,
\frac{i}{(2\pi)^2}\,
\Gamma(1\!-\!\epsilon)\,
\left(\frac{2\pi^2\mu^2|\xt_{01}|}{\overline{Q}}\right)^{\epsilon}
\Kb_{-\epsilon}\left(\overline{Q}|\xt_{01}|\right)\, 
\frac{ \xt_{21}^m}{\left(\xt_{21}^2\right)^{1-\epsilon}}
\label{def_int_Im_b_alike_0_xi_2}
\, ,
\end{align}
and thus
\begin{equation}
    \begin{split}        
&\bigg[
\Big|{\cal I}^{m}\!\left(a\right)\Big|^2
-{\textrm{Re}}\Big({\cal I}^{m}\!\left(a\right)^*\, {\cal I}^{m}\!\left(b\right) \Big)
\bigg]\bigg|_{\xi=0} \\
=&\,
\frac{\Gamma(1\!-\!\epsilon)^2}{(2\pi)^4}\,
\left(\frac{4\pi^4\mu^4\xt_{01}^2}{\overline{Q}^2}\right)^{\epsilon}
\left[\Kb_{-\epsilon}\left(\overline{Q}|\xt_{01}|\right)\right]^2
\left[\frac{\xt_{20}}{(\xt_{20}^2)^{1-\epsilon}}\vdot\left(\frac{\xt_{20}}{(\xt_{20}^2)^{1-\epsilon}}\!-\!\frac{\xt_{21}}{(\xt_{21}^2)^{1-\epsilon}}\right)\right]
\, .
    \end{split}
\end{equation}
Thanks to this result, the $\eta$ pole contribution \eqref{sigma_L_qqbarg_a_like_eta_pole_plus} can be simplified into
\begin{equation}
    \begin{split}
\sigma^{\gamma^*}_{L} \Big|_{q\bar{q}g,\, |a|^2 \textrm{ like}}^{\eta \textrm{ pole};\, \eta+}
= &
4 N_c \aem
\sum_{f}  e_f^2\;
\int_{0}^{1}
\dd{z}  
\int \frac{\dd[2-2\epsilon]{\xt_0}}{2\pi} \int \frac{\dd[2-2\epsilon]{\xt_1}}{2\pi}\,
\\
&\times
4 Q^2\, z^2(1\!-\!z)^2\,
\left(\frac{(2\pi)^2\mu^2\xt_{01}^2}{\overline{Q}^2}\right)^{\epsilon}
\left[\Kb_{-\epsilon}\left(\overline{Q}|\xt_{01}|\right)\right]^2\,
\\
& \times 
\left(\frac{\as \cf}{\pi}\right)
\frac{2}{\eta}
\left[\frac{z q^+}{\nu_B^+}\right]^{\eta}\,
\Gamma(1\!-\!\epsilon)^2\, \pi^{2\epsilon}
\mu^{2\epsilon}
\\
&\times
\int \frac{\dd[2-2\epsilon]{\xt_2}}{2\pi}
\left[\frac{\xt_{20}}{(\xt_{20}^2)^{1-\epsilon}}\vdot\left(\frac{\xt_{20}}{(\xt_{20}^2)^{1-\epsilon}}\!-\!\frac{\xt_{21}}{(\xt_{21}^2)^{1-\epsilon}}\right)\right]
{\textrm{Re}}\left[1\!-\!\tripole\right]
\label{sigma_L_qqbarg_a_like_eta_pole_plus_2}
\, .
    \end{split}
\end{equation}

In the integral over $\xt_2$, the first term in the bracket still leads to a UV divergence at $\xt_2\rightarrow\xt_0$ for $\epsilon=0$, whereas the potential collinear divergences for $|\xt_2|\rightarrow+\infty$ cancel between the two terms in the bracket. By splitting the tripole operator as
\begin{align}
{\textrm{Re}}\left[1\!-\!\tripole\right]
=&\,
{\textrm{Re}}\left[1\!-\!\dipole\right]
+
{\textrm{Re}}\left[\dipole\!-\! \tripole\right]
\, ,
\label{tripole_split}
\end{align}
the UV-sensitive contribution can be isolated into a piece proportional to the dipole operator that resembles the $q\bar{q}$ contribution considered in Sec.~\ref{sec:gamma_L_qqbar}, see Eq.~\eqref{sigma_L_qqbar_1}. Indeed, in the second term in Eq.~\eqref{tripole_split}, the  operator $\dipole\!-\! \tripole$ itself vanishes for $\xt_2\rightarrow\xt_0$ (and for $\xt_2\rightarrow\xt_1$), which removes the potential UV divergence. Note that this operator $\dipole\!-\! \tripole$ is the same one that appears on the right hand side of the BK equation at LO, or more precisely, in the right hand side of the first equation of the Balitsky hierarchy of equations.   
In this way, the expression \eqref{sigma_L_qqbarg_a_like_eta_pole_plus_2} is further split as 
\begin{align}
\sigma^{\gamma^*}_{L} \Big|_{q\bar{q}g,\, |a|^2 \textrm{ like}}^{\eta \textrm{ pole};\, \eta+}
=&\, 
\sigma^{\gamma^*}_{L} \Big|_{q\bar{q}g,\, |a|^2 \textrm{ like}}^{\textrm{dipole-like } \eta \textrm{ pole};\, \eta+}
+
\sigma^{\gamma^*}_{L} \Big|_{q\bar{q}g,\, |a|^2 \textrm{ like}}^{\textrm{BK-like } \eta \textrm{ pole};\, \eta+}
\label{sigma_L_qqbarg_a_like_eta_pole_plus_3}
\, ,
\end{align}
where
\begin{equation}
    \begin{split}
&\sigma^{\gamma^*}_{L} \Big|_{q\bar{q}g,\, |a|^2 \textrm{ like}}^{\textrm{dipole-like } \eta \textrm{ pole};\, \eta+}
=\, 4 N_c \aem
\sum_{f}  e_f^2\;
\int_{0}^{1}
\dd{z}  
\int \frac{\dd[2-2\epsilon]{\xt_0}}{2\pi} \int \frac{\dd[2-2\epsilon]{\xt_1}}{2\pi}\,
\\
&\times
4 Q^2\, z^2(1\!-\!z)^2\,
\left(\frac{(2\pi)^2\mu^2\xt_{01}^2}{\overline{Q}^2}\right)^{\epsilon}
\left[\Kb_{-\epsilon}\left(\overline{Q}|\xt_{01}|\right)\right]^2\,
 \times 
\left(\frac{\as \cf}{\pi}\right)\,
\widetilde{\cal V}^{L}_{|a|^2,\textrm{ pole; }\eta+}\;
{\textrm{Re}}\left[1\!-\! \dipole\right],
\label{sigma_L_qqbarg_a_like_dipole_eta_pole_plus}
    \end{split}
\end{equation}
and
\begin{equation}
    \begin{split}
\sigma^{\gamma^*}_{L} \Big|_{q\bar{q}g,\, |a|^2 \textrm{ like}}^{\textrm{BK-like } \eta \textrm{ pole};\, \eta+}
=&\, 4 N_c \aem
\sum_{f}  e_f^2\;
\int_{0}^{1}
\dd{z}  
\int \frac{\dd[2-2\epsilon]{\xt_0}}{2\pi} \int \frac{\dd[2-2\epsilon]{\xt_1}}{2\pi}\,
\\
&\times
4 Q^2\, z^2(1\!-\!z)^2\,
\left(\frac{(2\pi)^2\mu^2\xt_{01}^2}{\overline{Q}^2}\right)^{\epsilon}
\left[\Kb_{-\epsilon}\left(\overline{Q}|\xt_{01}|\right)\right]^2\,
\\
& \times 
\frac{2\as \cf}{\pi}
\frac{1}{\eta}
\left[\frac{z q^+}{\nu_B^+}\right]^{\eta}\,
\Gamma(1\!-\!\epsilon)^2\, \pi^{2\epsilon}
\mu^{2\epsilon}
\\
&\, \times\,
\int \frac{\dd[2-2\epsilon]{\xt_2}}{2\pi}
\left[\frac{\xt_{20}}{(\xt_{20}^2)^{1-\epsilon}}\vdot\left(\frac{\xt_{20}}{(\xt_{20}^2)^{1-\epsilon}}\!-\!\frac{\xt_{21}}{(\xt_{21}^2)^{1-\epsilon}}\right)\right]
{\textrm{Re}}\left[\dipole\!-\!\tripole\right]
\label{sigma_L_qqbarg_a_like_BK_eta_pole_plus_2}
\, .
    \end{split}
\end{equation}
In Eq.~\eqref{sigma_L_qqbarg_a_like_dipole_eta_pole_plus}, in analogy to the $q\bar{q}$ contributions from Eq.~\eqref{sigma_L_qqbar_1}, we have introduced the loop factor 
\begin{equation}
    \begin{split}
\widetilde{\cal V}^{L}_{|a|^2,\textrm{ pole; }\eta+}
=&\,
\frac{2}{\eta}
\left[\frac{z q^+}{\nu_B^+}\right]^{\eta}\,
\Gamma(1\!-\!\epsilon)^2\, \pi^{2\epsilon}
\mu^{2\epsilon}
\int \frac{\dd[2-2\epsilon]{\xt_2}}{2\pi}
\left[\frac{\xt_{20}}{(\xt_{20}^2)^{1-\epsilon}}\vdot\left(\frac{\xt_{20}}{(\xt_{20}^2)^{1-\epsilon}}\!-\!\frac{\xt_{21}}{(\xt_{21}^2)^{1-\epsilon}}\right)\right]
\\
=&\,
\frac{1}{\eta}
\left[\frac{z q^+}{\nu_B^+}\right]^{\eta}\;
\frac{\Gamma(1\!-\!\epsilon)}{\epsilon}\,
\left[\pi \mu^2 \xt_{01}^2\right]^{\epsilon}
\\
=&\,
\left[\frac{1}{\eta}
+
\log\left(\frac{z q^+}{\nu_B^+}\right)\;
\right]
\frac{\Gamma(1\!-\!\epsilon)}{\epsilon}\,
\left[\pi \mu^2 \xt_{01}^2\right]^{\epsilon}
+O(\eta)
\, .
\label{V_L_a_like_dipole_eta_pole_plus}
    \end{split}
\end{equation}
This factor contains both a rapidity pole at $\eta=0$ and a UV pole at $\epsilon=0$.
By contrast, the rapidity pole is the only leftover divergence in the BK-like contribution \eqref{sigma_L_qqbarg_a_like_BK_eta_pole_plus_2}.

So far, we have studied the $\eta$ pole contribution \eqref{sigma_L_qqbarg_a_like_eta_pole_plus}. Let us now consider the plus-distribution contribution \eqref{sigma_L_qqbarg_a_like_plus_prescr}.
Writing the subtraction term associated with the plus-distribution explicitly, one has
\begin{equation}
    \begin{split}        
&\sigma^{\gamma^*}_{L} \Big|_{q\rightarrow g}^{\textrm{+ distr.}}
=\, 4 N_c \aem \left(\frac{\as \cf}{\pi}\right)
\sum_{f}  e_f^2\;
(2\pi)
\int_{0}^{1}
dz\,  4 Q^2\, z^2(1\!-\!z)^2
\\
&\times
\int \dd[2-2\epsilon]{\xt_0} \int \dd[2-2\epsilon]{\xt_1} \int \dd[2-2\epsilon]{\xt_2}\:
{\textrm{Re}}\left[1-\tripole\right]\;
\\
& \times
2 
\int_0^1 \frac{\dd{\xi}}{\xi}
\Bigg\{
\bigg[
\Big|{\cal I}^{m}\!\left(a\right)\Big|^2
-{\textrm{Re}}\Big({\cal I}^{m}\!\left(a\right)^*\, {\cal I}^{m}\!\left(b\right) \Big)
\bigg]
-
\bigg[
\Big|{\cal I}^{m}\!\left(a\right)\Big|^2
-{\textrm{Re}}\Big({\cal I}^{m}\!\left(a\right)^*\, {\cal I}^{m}\!\left(b\right) \Big)
\bigg]\bigg|_{\xi=0}
\Bigg\}
 .
\label{sigma_L_qqbarg_a_like_plus_prescr_2}
    \end{split}
\end{equation}
The plus-distribution is by definition regulating the potential divergence at $\xi=0$. As mentioned earlier, the term $|{\cal I}^{m}\!\left(a\right)|^2$ brings a potential UV divergence for $\xt_2\rightarrow\xt_0$. However, the value of the integral ${\cal I}^{m}\!\left(a\right)$ at $\xi=0$ reproduces
the leading behavior of ${\cal I}^{m}\!\left(a\right)$ in the UV regime $\xt_2\rightarrow\xt_0$.
Hence, the plus distribution used to subtract the rapidity divergence in Eq.~\eqref{sigma_L_qqbarg_a_like_plus_prescr_2} is also subtracting the UV divergence, so that the expression \eqref{sigma_L_qqbarg_a_like_plus_prescr_2} is finite at $\epsilon=0$. In particular, one can use the expression of the integrals ${\cal I}^{m}\!\left(a\right)$ and ${\cal I}^{m}\!\left(b\right)$ at $\epsilon=0$ in order to evaluate the contribution 
\eqref{sigma_L_qqbarg_a_like_plus_prescr_2}.
From Eqs.~\eqref{def_int_Im_a_alike} and \eqref{def_int_Im_b_alike}, these can be evaluated as (see for example Eq.~(A7) in Ref.~\cite{Beuf:2017bpd}):
\begin{align}
{\cal I}^{m}\!\left(a\right)
= &\,
\frac{i}{(2\pi)^2}\,
\frac{ \xt_{20}^m}{\xt_{20}^2}\,
\Kb_{0}\left(\overline{Q}\sqrt{
(1\!-\!\xi)\xt_{01}^2
+\frac{z\xi(1\!-\!\xi)}{(1\!-\!z)}\xt_{20}^2
+\xi \xt_{21}^2
}
\right)\, 
+\order{\epsilon},
\label{def_int_Im_a_alike_0_xi_D4}
\\
{\cal I}^{m}\!\left(b\right)
= &\,
\frac{i}{(2\pi)^2}\,
\frac{ \xt_{21}^m}{\xt_{21}^2}\,
\Kb_{0}\left(\overline{Q}\sqrt{
(1\!-\!\xi)\xt_{01}^2
+\frac{z\xi(1\!-\!\xi)}{(1\!-\!z)}\xt_{20}^2
+\xi \xt_{21}^2
}
\right)\, 
+\order{\epsilon}
\label{def_int_Im_b_alike_0_xi_D4}
\, .
\end{align}
Hence, the expression \eqref{sigma_L_qqbarg_a_like_plus_prescr_2} becomes
\begin{equation}
    \begin{split}
&\,\sigma^{\gamma^*}_{L} \Big|_{q\rightarrow g}^{\textrm{+ distr.}}
= 4 N_c \aem 
\sum_{f}  e_f^2\;
\int_{0}^{1}
\dd{z}\,  4 Q^2\, z^2(1\!-\!z)^2
\int \frac{\dd[2]{\xt_0}}{2\pi} \int \frac{\dd[2]{\xt_1}}{2\pi}\; 
\\
&\times
\frac{2\as \cf}{\pi}\!\!\int \frac{\dd[2]{\xt_2}}{2\pi}\:
\left[\frac{\xt_{20}}{\xt_{20}^2}\vdot\left(\frac{\xt_{20}}{\xt_{20}^2}\!-\!\frac{\xt_{21}}{\xt_{21}^2}\right)\right]
\;
{\textrm{Re}}\left[1-\tripole\right]
\\
&\, \times\,
\int_0^1 \frac{\dd{\xi}}{\xi}
\Bigg\{
\left[
\Kb_{0}\left(\overline{Q}\sqrt{
(1\!-\!\xi)\xt_{01}^2
+\frac{z\xi(1\!-\!\xi)}{(1\!-\!z)}\xt_{20}^2
+\xi \xt_{21}^2
}
\right)
\right]^2
-
\Big[
\Kb_{0}\left(\overline{Q}|\xt_{01}|
\right)
\Big]^2
\Bigg\}
+\order{\epsilon}
\, .
\label{sigma_L_qqbarg_a_like_plus_prescr_3}
    \end{split}
\end{equation}
This can be written in a more compact way as
\begin{equation}
    \begin{split}
&\sigma^{\gamma^*}_{L} \Big|_{q\rightarrow g}^{\textrm{+ distr.}}
=\,
4 N_c \aem 
\sum_{f}  e_f^2\;
\int_{0}^{1}
\dd{z}\,  4 Q^2\, z^2(1\!-\!z)^2
\int \frac{\dd[2]{\xt_0}}{2\pi} \int \frac{\dd[2]{\xt_1}}{2\pi}\;
\\
&\times
\frac{2\as \cf}{\pi}\!\!\int \frac{\dd[2]{\xt_2}}{2\pi}\:
\left[\frac{\xt_{20}}{\xt_{20}^2}\vdot\left(\frac{\xt_{20}}{\xt_{20}^2}\!-\!\frac{\xt_{21}}{\xt_{21}^2}\right)\right]
\;
{\textrm{Re}}\left[1-\tripole\right]
\int_0^1 \frac{\dd{\xi}}{(\xi)_+}\, 
\,
\Big[
\Kb_{0}\left({Q}X^{(a)}_{012}
\right)
\Big]^2
+\order{\epsilon}
\, ,
\label{sigma_L_qqbarg_a_like_plus_prescr_4}
    \end{split}
\end{equation}
with the notation
\begin{equation}
    \begin{split}
{X_{012}^{(a)}}^2
\equiv &\,
z(1\!-\!z)(1\!-\!\xi)\xt_{01}^2
+z^2\xi(1\!-\!\xi)\xt_{20}^2
+z(1\!-\!z)\xi \xt_{21}^2
\, ,
\label{def_X012_q2g}
    \end{split}
\end{equation}
which corresponds to the quantity 
\begin{equation}
    \begin{split}
X_{012}^2
\equiv &\,
\frac{1}{(q^+)^2}\Big[k^+_0 k^+_1\xt_{01}^2
+k^+_2 k^+_0\xt_{20}^2
+k^+_2 k^+_1\xt_{21}^2\Big]
\label{def_X012_generic}
    \end{split}
\end{equation}
after applying the change of variables~\eqref{a_like_cv}.

A few comments are in order concerning Eq.~\eqref{sigma_L_qqbarg_a_like_plus_prescr_3}. As already mentioned, the bracket in the last line vanishes both for $\xi=0$ and for $\xt_{20}=0$, making this whole contribution finite. Let us first note that this cancellation extends to the whole regime $\xt_{20}^2\ll \xt_{01}^2 \sim \xt_{21}^2$ independently of the value of $\xi$. Then, in the regimes $\xt_{21}^2\ll \xt_{01}^2 \sim \xt_{20}^2$ and  $\xt_{20}^2\sim \xt_{21}^2\sim \xt_{01}^2 $, the cancellation occurs as soon as $\xi\ll 1$. In these regimes, the integration over $\xi$ is thus not expected to give a large result.
By contrast, in the regime $\xt_{20}^2\sim \xt_{21}^2\gg 1/Q^2 \gtrsim \xt_{01}^2 $, the first term in the bracket in Eq.~\eqref{sigma_L_qqbarg_a_like_plus_prescr_3} is exponentially suppressed compared to the second term over a wide range in $\xi$, for $1>\xi \gg 1/Q^2\xt_{20}^2$. Hence, in the regime $\xt_{20}^2\sim \xt_{21}^2\gg 1/Q^2 \gtrsim \xt_{01}^2 $, the third line of Eq.~\eqref{sigma_L_qqbarg_a_like_plus_prescr_3} can be approximated as 
\begin{align}
\int_{\frac{c_0^2}{Q^2 \xt_{20}^2}}^1 \frac{\dd{\xi}}{\xi}
\Bigg\{
-
\Big[
\Kb_{0}\left(\overline{Q}|\xt_{01}|
\right)
\Big]^2
\Bigg\}
= &\,
- \log\left(\frac{Q^2 \xt_{20}^2}{c_0^2}\right)\,
\Big[
\Kb_{0}\left(\overline{Q}|\xt_{01}|
\right)
\Big]^2
\, ,
\end{align}
which includes a large transverse log and the same Bessel function factor as in the LO term \eqref{sigma_L_qqbar_1}. The contribution from the regime $\xt_{20}^2\sim \xt_{21}^2\gg 1/Q^2 \gtrsim \xt_{01}^2 $ to Eq.~\eqref{sigma_L_qqbarg_a_like_plus_prescr_3} can then be written in a form analogous to the LO term, up to the replacement of the dipole operator ${\textrm{Re}}\left[1-\dipole\right]$ by the factor 
\begin{align}
&
- 
\frac{2\as \cf}{\pi}
\!\!\int_{\xt_{20}^2\gg 1/Q^2} \frac{\dd[2]{\xt_2}}{2\pi}\:
{\textrm{Re}}\left[1-\tripole\right]\:
\frac{\xt_{01}\!\cdot\!\xt_{21}}{\xt_{20}^2\xt_{21}^2}\,
\;
\log\left(\frac{Q^2 \xt_{20}^2}{c_0^2}\right)\,
\, 
\label{large_coll_log_plus_dist_q2g}
\end{align}
in the integrand.

Similarly, from the $\bar q \to g$ analogue of Eq.~\eqref{sigma_L_qqbarg_a_like_plus_prescr_3}, extracted from Eq.~\eqref{sigma_L_qqbarg_b_like}, one would obtain 
from the regime $\xt_{20}^2\sim \xt_{21}^2\gg 1/Q^2 \gtrsim \xt_{01}^2 $ the contribution 
\begin{align}
&
+ 
\frac{2\as \cf}{\pi}
\!\!\int_{\xt_{21}^2\gg 1/Q^2} \frac{\dd[2]{\xt_2}}{2\pi}\:
{\textrm{Re}}\left[1-\tripole\right]\:
\frac{\xt_{01}\!\cdot\!\xt_{20}}{\xt_{20}^2\xt_{21}^2}\,
\;
\log\left(\frac{Q^2 \xt_{21}^2}{c_0^2}\right)\,
\, 
\label{large_coll_log_plus_dist_qbar2g}
\end{align}
symmetric to \eqref{large_coll_log_plus_dist_q2g}.

The domain $\xt_{20}^2\sim \xt_{21}^2\gg 1/Q^2 \gtrsim \xt_{01}^2 $ corresponds to the collinear regime to the target. Hence, the terms \eqref{large_coll_log_plus_dist_q2g} and \eqref{large_coll_log_plus_dist_qbar2g} are qualitatively the large logarithms expected to occur at higher orders when the high-energy evolution equation is formulated in terms of the  variable $k^+$, as discussed in the introduction and in Refs.~\cite{Salam:1998tj,Beuf:2014uia} for example.

\subsection{Rapidity-safe terms}

Let us now consider the rapidity-safe contribution from Eq.~\eqref{sigma_L_qqbarg_a_like_rap_safe}. It still contains a potential UV divergence in the regime $\xt_2\rightarrow \xt_0$ from the term in $|{\cal I}^{m}\!\left(a\right)|^2$. In order to isolate it, we subtract and add back a term that is equivalent in this UV regime. In particular, in this subtraction term we make the replacement
\begin{align}
{\textrm{Re}}\left[1-\tripole\right] 
\mapsto &\,
{\textrm{Re}}\left[1-\dipole\right] 
\end{align}
since the tripole operator reduces to the dipole operator in this UV regime by color coherence. Moreover, we also make the replacement 
\begin{align}
\bigg[
\Big|{\cal I}^{m}\!\left(a\right)\Big|^2
-{\textrm{Re}}\Big({\cal I}^{m}\!\left(a\right)^*\, {\cal I}^{m}\!\left(b\right) \Big)
\bigg]
\mapsto &\,
\bigg[
\Big|{\cal I}^{m}\!\left(a\right)\Big|^2
-{\textrm{Re}}\Big({\cal I}^{m}\!\left(a\right)^*\, {\cal I}^{m}\!\left(b\right) \Big)
\bigg]\bigg|_{\xi=0} 
\end{align}
since the $\xi=0$ value of the ${\cal I}^{m}\!\left(a\right)$ integral corresponds to its UV approximation as discussed earlier. 
We also leave the term involving the ${\cal I}^{m}\!\left(b\right)$ from the interference diagrams in order to prevent the appearance of collinear divergences at $|\xt_2|\rightarrow +\infty$.

In this way, we build from the expression \eqref{sigma_L_qqbarg_a_like_rap_safe} the term containing its UV contribution as 
\begin{equation}
    \begin{split}
&\sigma^{\gamma^*}_{L} \Big|_{q\bar{q}g,\, |a|^2 \textrm{ like}}^{\textrm{dipole-like; rap. safe}}
\\
=&\, 4 N_c \aem \left(\frac{\as \cf}{\pi}\right)
\sum_{f}  e_f^2\;
(2\pi)
\int_{0}^{1}
\dd{z}\,  4 Q^2\, z^2(1\!-\!z)^2
\int \dd[2-2\epsilon]{\xt_0} \int \dd[2-2\epsilon]{\xt_1} \int \dd[2-2\epsilon]{ \xt_2}\:
\\
&\, \times\,
{\textrm{Re}}\left[1-\dipole\right]\;
\bigg[
\Big|{\cal I}^{m}\!\left(a\right)\Big|^2
-{\textrm{Re}}\Big({\cal I}^{m}\!\left(a\right)^*\, {\cal I}^{m}\!\left(b\right) \Big)
\bigg]\bigg|_{\xi=0}
\int_0^1 \dd{\xi}
\big[-2+
(1\!-\!\deltas \epsilon)\xi\big]\,
\\
=& 4 N_c \aem
\sum_{f}  e_f^2\;
\int_{0}^{1}
\dd{z}
\int \frac{\dd[2-2\epsilon]{\xt_0}}{2\pi} \int \frac{\dd[2-2\epsilon]{\xt_1}}{2\pi} 
4 Q^2\, z^2(1\!-\!z)^2 
\left(\frac{(2\pi)^2\mu^2\xt_{01}^2}{\overline{Q}^2}\right)^{\epsilon}
\left[\Kb_{-\epsilon}\left(\overline{Q}|\xt_{01}|\right)\right]^2\,
\\
&\, \times\,
\left(\frac{\as \cf}{\pi}\right)\,
\widetilde{\cal V}^{L}_{|a|^2,\textrm{rap. safe}}\;
{\textrm{Re}}\left[1\!-\!{\cal S}_{01}\right]
\, ,
\label{sigma_L_qqbarg_a_like_rap_safe_dipole}
    \end{split}
\end{equation}
with
\begin{equation}
    \begin{split}
\widetilde{\cal V}^{L}_{|a|^2,\textrm{rap. safe}}
=&\,
\Gamma(1\!-\!\epsilon)^2\, \pi^{2\epsilon}
\mu^{2\epsilon}
\int_0^1 \dd{\xi}
\big[-2+
(1\!-\!\deltas \epsilon)\xi\big]\,
\\
&\times
\int \frac{\dd[2-2\epsilon]{\xt_2}}{2\pi}
\left[\frac{\xt_{20}}{(\xt_{20}^2)^{1-\epsilon}}\vdot\left(\frac{\xt_{20}}{(\xt_{20}^2)^{1-\epsilon}}\!-\!\frac{\xt_{21}}{(\xt_{21}^2)^{1-\epsilon}}\right)\right]
\\
=&\,
\left[-\frac{3}{2}
-\frac{\deltas \epsilon}{2}\right]\,
\frac{\Gamma(1\!-\!\epsilon)}{2\epsilon}\,
\left[\pi \mu^2 \xt_{01}^2\right]^{\epsilon}
\\
=&\,
-\frac{3}{4}\,
\frac{S_{\epsilon} }{\epsilon} \left(\frac{\xt_{01}^2 \mu^2}{c_0^2}\right)^{\epsilon}\,
-\frac{\deltas }{4}
+\order{\epsilon}
\, .
\label{V_L_a_like_dipole_rap_safe}
    \end{split}
\end{equation}

Once we extract the dipole-like UV contribution~\eqref{sigma_L_qqbarg_a_like_rap_safe_dipole} from Eq.~\eqref{sigma_L_qqbarg_a_like_rap_safe}, the finite leftover piece is given by
\begin{equation}
    \begin{split}
&\sigma^{\gamma^*}_{L} \Big|_{q\rightarrow g}^{\textrm{reg.}}
=\, 
\sigma^{\gamma^*}_{L} \Big|_{q\bar{q}g,\, |a|^2 \textrm{ like}}^{\textrm{rap. safe}}
-
\sigma^{\gamma^*}_{L} \Big|_{q\bar{q}g,\, |a|^2 \textrm{ like}}^{\textrm{dipole-like; rap. safe}}
\\
=&\, 
4 N_c \aem \left(\frac{\as \cf}{\pi}\right)
\sum_{f}  e_f^2\;
(2\pi)
\int_{0}^{1}
\dd{z}\,  4 Q^2\, z^2(1\!-\!z)^2
\int \dd[2-2\epsilon]{\xt_0} \int \dd[2-2\epsilon]{\xt_1} \int \dd[2-2\epsilon]{\xt_2}
\int_0^1 \dd{\xi}\:
\\
& \times
\Bigg\{
\big[-2+
(1\!-\!\deltas \epsilon)\xi\big]\,
\Bigg[
\bigg[
\Big|{\cal I}^{m}\!\left(a\right)\Big|^2
-{\textrm{Re}}\Big({\cal I}^{m}\!\left(a\right)^*\, {\cal I}^{m}\!\left(b\right) \Big)
\bigg]
{\textrm{Re}}\left[1\!-\!\tripole\right]\;
\\
&-
\bigg[
\Big|{\cal I}^{m}\!\left(a\right)\Big|^2
-{\textrm{Re}}\Big({\cal I}^{m}\!\left(a\right)^*\, {\cal I}^{m}\!\left(b\right) \Big)
\bigg]\bigg|_{\xi=0}
{\textrm{Re}}\left[1\!-\!\dipole\right]
\Bigg]
\\
&+\xi\: 
{\textrm{Re}}\Big({\cal I}^{m}\!\left(a\right)^*\, {\cal I}^{m}\!\left(b\right) \Big)
{\textrm{Re}}\left[1\!-\!\tripole\right]
\Bigg\}
\, .
\label{sigma_L_qqbarg_a_like_rap_safe_finite_leftover}
    \end{split}
\end{equation}
In Eq.~\eqref{sigma_L_qqbarg_a_like_rap_safe_finite_leftover}, it is now possible to take $\epsilon=0$, since all UV and collinear divergences cancel at the integrand level. Using the relations \eqref{def_int_Im_a_alike_0_xi_D4} and \eqref{def_int_Im_b_alike_0_xi_D4}, one gets
\begin{equation}
    \begin{split}
\sigma^{\gamma^*}_{L} \Big|_{q\rightarrow g}^{\textrm{reg.}}
=&\, 
4 N_c \aem \left(\frac{\as \cf}{\pi}\right)
\sum_{f}  e_f^2\;
\int_{0}^{1}
\dd{z}\,  4 Q^2\, z^2(1\!-\!z)^2
\int \frac{\dd[2]{ \xt_0}}{2\pi} 
\int \frac{\dd[2]{ \xt_1}}{2\pi}
\int \frac{\dd[2]{\xt_2}}{2\pi}
\int_0^1 \dd{\xi}
\\
\, 
\times
\Bigg\{
\Biggl\{
&
\qty[
\Kb_{0}\qty(\overline{Q}\sqrt{
(1\!-\!\xi)\xt_{01}^2
+\frac{z\xi(1\!-\!\xi)}{(1\!-\!z)}\xt_{20}^2
+\xi \xt_{21}^2
}
)
]^2
{\textrm{Re}}\left[1\!-\!\tripole\right]
\\
&-
\qty[
\Kb_{0}\qty(\overline{Q}|\xt_{01}|
)
]^2
{\textrm{Re}}\qty[1\!-\!{\cal S}_{01}]
\Biggr\}
\big(-2+
\xi\big)\,
\left[\frac{\xt_{20}}{\xt_{20}^2}\vdot\left(\frac{\xt_{20}}{\xt_{20}^2}\!-\!\frac{\xt_{21}}{\xt_{21}^2}\right)\right]
\\
\,
+\xi\, 
&
\frac{\xt_{20}\!\vdot\!\xt_{21}}{\xt_{20}^2\,\xt_{21}^2}\,
\left[
\Kb_{0}\left(\overline{Q}\sqrt{
(1\!-\!\xi)\xt_{01}^2
+\frac{z\xi(1\!-\!\xi)}{(1\!-\!z)}\xt_{20}^2
+\xi \xt_{21}^2
}
\right)
\right]^2
{\textrm{Re}}\left[1\!-\!\tripole\right]
\Bigg\}
+\order{\epsilon}
\, .
\label{sigma_L_qqbarg_a_like_rap_safe_finite_leftover_2}
    \end{split}
\end{equation}
This contribution can be written in a more compact way using the notation \eqref{def_X012_q2g} as
\begin{equation}
    \begin{split}
&\sigma^{\gamma^*}_{L} \Big|_{q\rightarrow g}^{\textrm{reg.}}
=\, 
4 N_c \aem \left(\frac{\as \cf}{\pi}\right)
\sum_{f}  e_f^2\;
\int_{0}^{1}
\dd{z}\,  4 Q^2\, z^2(1\!-\!z)^2
\int \frac{\dd[2]{ \xt_0}}{2\pi} 
\int \frac{\dd[2]{ \xt_1}}{2\pi}
\int \frac{\dd[2]{ \xt_2}}{2\pi}
\int_0^1 \dd{\xi}\:
\\
&\, 
\times
\Bigg\{
\big(-2+
\xi\big)\,
\left[\frac{\xt_{20}}{\xt_{20}^2}\vdot\left(\frac{\xt_{20}}{\xt_{20}^2}\!-\!\frac{\xt_{21}}{\xt_{21}^2}\right)\right]
\Bigg\{
\Big[
\Kb_{0}\left({Q} X^{(a)}_{012}
\right)
\Big]^2
{\textrm{Re}}\left[1\!-\!\tripole\right]\;
-
\Big(\xt_2\rightarrow \xt_0\Big)
\Bigg\}
\\
&\,
+\xi\, 
\frac{\xt_{20}\!\vdot\!\xt_{21}}{\xt_{20}^2\,\xt_{21}^2}\,
\Big[
\Kb_{0}\left({Q} X^{(a)}_{012}
\right)
\Big]^2
{\textrm{Re}}\left[1\!-\!\tripole\right]
\Bigg\}
+\order{\epsilon}
\, .
\label{sigma_L_qqbarg_a_like_rap_safe_finite_leftover_3}
    \end{split}
\end{equation}

\subsection{\texorpdfstring{Rapidity sensitive terms: $\eta^-$ regulator}
{Rapidity sensitive terms: eta- regulator
}
}

Let us now discuss the treatment of the rapidity sensitive $q\bar{q}g$ contributions using the $\eta^-$ regulator. 
An additional complication compared to the $\eta^+$ regulator is that in this case the rapidity regulator depends on the transverse momentum of the gluon, see Eq.~\eqref{k_min_rap_reg_def}. 
However, the gluon has a different transverse momentum before and after crossing the shockwave target, and therefore it is not immediately clear which momentum to choose in the rapidity regulator.
Up to an irrelevant shift proportional to $k_2^+$, these two transverse momenta of the gluon corresponds to the momentum $\Kt$ in each of the two factors ${\cal I}^{m}\!\left(a\right)$ or ${\cal I}^{m}\!\left(b\right)$ present in each $q\bar{q}g$ contribution from Eq.~\eqref{sigma_L_qqbarg_1}.  
We suggest applying the square root of this rapidity regulator factor with the gluon momentum from before the shockwave and another square root with the gluon momentum from after the shockwave, as
\begin{align}
\left[\frac{2k_2^+ \nu_B^-}{\kt_{2,\textrm{ bef.}}^2}\right]^{\frac{\eta}{2}} \, 
\left[\frac{2k_2^+ \nu_B^-}{\kt_{2,\textrm{ aft.}}^2}\right]^{\frac{\eta}{2}}
\sim &\,
 \frac{\left[2k_2^+ \nu_B^-\right]^{\eta}}{
\left[\Kt^2_{\textrm{ bef.}}\right]^{\frac{\eta}{2}}
 \left[\Kt^2_{\textrm{ aft.}}\right]^{\frac{\eta}{2}}
 }
 \, .
 \label{k_min_rap_reg_def_qqbarg}
\end{align}
Our implementation of the $\eta^-$ regulator is thus the following. First, we modify the ${\cal I}^{m}\!\left(a\right)$ or ${\cal I}^{m}\!\left(b\right)$ integrals into
\begin{align}
\begin{split}
{\cal I}_{\eta}^{m}\!\left(a\right)
=& \,
\mu^{2\epsilon}\,
\int \frac{\dd[2-2\epsilon]{\Pt}}{(2\pi)^{2-2\epsilon}}\; \int \frac{\dd[2-2\epsilon]{\Kt}}{(2\pi)^{2-2\epsilon}}\;
\frac{\Kt^m\:   e^{i\Kt\vdot \xt_{20}}\:
}{ \big[\Pt^2+z(1\!-\!z){Q}^2\big] }
e^{i\Pt\vdot (\xt_{01} +\xi \xt_{20})}
\\
&\times
\frac{1}{\left\{\Kt^2 + \frac{\xi(1\!-\!\xi)}{(1\!-\!z)} \big[\Pt^2+z(1\!-\!z){Q}^2\big]\right\}
\big[\Kt^2\big]^{\frac{\eta}{2}}},
\label{def_int_Im_a_alike_eta}
\end{split}
\\
    \begin{split}
{\cal I}_{\eta}^{m}\!\left(b\right)
= &\,
\mu^{2\epsilon}\,
\int \frac{\dd[2-2\epsilon]{\Pt}}{(2\pi)^{2-2\epsilon}}\; \int \frac{\dd[2-2\epsilon]{\Kt}}{(2\pi)^{2-2\epsilon}}\;
\frac{\Kt^m\:   e^{i\Kt\vdot \xt_{21}}\:
}{\big[\Pt^2+z(1\!-\!\xi)\big(1\!-\!z(1\!-\!\xi)\big){Q}^2\big]}\:
e^{i\Pt\vdot \frac{
[(1\!-\!z)\xt_{01} -z\xi \xt_{20}]
}{(1\!-\!z\!+\!z\xi)}}
\\
&\, \times\,
\frac{1}{\left\{\Kt^2 + \frac{\xi(1\!-\!z)}{(1\!-\!\xi)[1\!-\!z\!+\!z\xi]^2} 
\big[\Pt^2+z(1\!-\!\xi)\big(1\!-\!z(1\!-\!\xi)\big){Q}^2\big]
\right\}
\big[\Kt^2\big]^{\frac{\eta}{2}}}
\label{def_int_Im_b_alike_eta}
\, .
    \end{split}
\end{align}
Second, we include a factor 
\begin{align}
\left[2k_2^+ \nu_B^-\right]^{\eta}
= &\,
 \left[2 zq^+ \nu_B^-\right]^{\eta} \xi^{\eta}
 \label{k_min_rap_reg_def_qqbarg_bis}
\end{align}
in the integration measure in Eq.~\eqref{sigma_L_qqbarg_a_like_2}. 
In this way, the rapidity sensitive contribution from Eq.~\eqref{sigma_L_qqbarg_a_like_2} can be written as
\begin{equation}
    \begin{split}
\sigma^{\gamma^*}_{L} \Big|_{q\bar{q}g,\, |a|^2 \textrm{ like}}^{\textrm{rap. sensitive};\, \eta-}
=&\, 4 N_c \aem \left(\frac{\as \cf}{\pi}\right)
\sum_{f}  e_f^2\;
(2\pi)
\int_{0}^{1}
\dd{z}\,  4 Q^2\, z^2(1\!-\!z)^2
\\
&\times
\int \dd[2-2\epsilon]{\xt_0} \int \dd[2-2\epsilon]{\xt_1} \int \dd[2-2\epsilon]{\xt_2}\:
{\textrm{Re}}\left[1-\tripole\right]\;
\\
&\, \times
\left[{2z q^+\nu_B^-}\right]^{\eta} 
\int_0^1 \dd{\xi}\, 
2\,\xi^{\eta-1}\,
\bigg[
\Big|{\cal I}_{\eta}^{m}\!\left(a\right)\Big|^2
-{\textrm{Re}}\Big({\cal I}_{\eta}^{m}\!\left(a\right)^*\, {\cal I}_{\eta}^{m}\!\left(b\right) \Big)
\bigg]
\, .
\label{sigma_L_qqbarg_a_like_rap_unsafe_plus_3}
    \end{split}
\end{equation}
Again, one can split this expression into a contribution with a pole at $\eta$ and a leftover piece involving a plus distribution, as
\begin{align}
\sigma^{\gamma^*}_{L} \Big|_{q\bar{q}g,\, |a|^2 \textrm{ like}}^{\textrm{rap. sensitive};\, \eta-}
=&\, 
\sigma^{\gamma^*}_{L} \Big|_{q\bar{q}g,\, |a|^2 \textrm{ like}}^{\eta \textrm{ pole};\, \eta-}
+
\sigma^{\gamma^*}_{L} \Big|_{q\rightarrow g}^{\textrm{+ distr.}}
+\order{\eta}
\, .
\label{sigma_L_qqbarg_a_like_rap_unsafe_minus_2}
\end{align}
The plus-distribution contribution, taken at $\eta=0$, is identical to the one obtained for $\eta^+$ regulator, given in Eq.~\eqref{sigma_L_qqbarg_a_like_plus_prescr_2} or \eqref{sigma_L_qqbarg_a_like_plus_prescr_4}.
The part containing the rapidity pole is then
\begin{equation}
    \begin{split}
\sigma^{\gamma^*}_{L} \Big|_{q\bar{q}g,\, |a|^2 \textrm{ like}}^{\eta \textrm{ pole};\, \eta-}
=&\, 4 N_c \aem \left(\frac{\as \cf}{\pi}\right)
\sum_{f}  e_f^2\;
(2\pi)
\int_{0}^{1}
\dd{z}\,  4 Q^2\, z^2(1\!-\!z)^2
\\
&\times
\int \dd[2-2\epsilon]{\xt_0} \int \dd[2-2\epsilon]{\xt_1} \int \dd[2-2\epsilon]{ \xt_2}\:
{\textrm{Re}}\left[1-\tripole\right]\;
\\
&\, \times\, 
2\,\left[{2z q^+\nu_B^-}\right]^{\eta} \:\bigg[
\Big|{\cal I}_{\eta}^{m}\!\left(a\right)\Big|^2
-{\textrm{Re}}\Big({\cal I}_{\eta}^{m}\!\left(a\right)^*\, {\cal I}_{\eta}^{m}\!\left(b\right) \Big)
\bigg]\bigg|_{\xi=0} 
\int_0^1 \dd{\xi}\, 
\xi^{\eta-1}
\, .
\label{sigma_L_qqbarg_a_like_eta_pole_minus}
    \end{split}
\end{equation}

Using Eq.~\eqref{eq:FT_power}
as well as \eqref{eq:FT_LO}, we can evaluate the Fourier integrals \eqref{def_int_Im_a_alike_eta} and \eqref{def_int_Im_b_alike_eta} at $\xi=0$ to be
\begin{align}
{\cal I}_{\eta}^{m}\!\left(a\right)
\Big|_{\xi=0}
= &\,
\frac{i}{(2\pi)^2}\,
\frac{\Gamma\left(1\!-\!\epsilon\!-\!\frac{\eta}{2}\right)}{2^{\eta}\, \Gamma\left(1\!+\!\frac{\eta}{2}\right)}\,
\left(\frac{2\pi^2\mu^2|\xt_{01}|}{\overline{Q}}\right)^{\epsilon}
\Kb_{-\epsilon}\left(\overline{Q}|\xt_{01}|\right)\, 
\frac{ \xt_{20}^m}{\left(\xt_{20}^2\right)^{1-\epsilon-\frac{\eta}{2}}},
\label{def_int_Im_a_alike_0_xi_2_eta}
\\
{\cal I}_{\eta}^{m}\!\left(b\right)
\Big|_{\xi=0}
= &\,
\frac{i}{(2\pi)^2}\,
\frac{\Gamma\left(1\!-\!\epsilon\!-\!\frac{\eta}{2}\right)}{2^{\eta}\, \Gamma\left(1\!+\!\frac{\eta}{2}\right)}\,
\left(\frac{2\pi^2\mu^2|\xt_{01}|}{\overline{Q}}\right)^{\epsilon}
\Kb_{-\epsilon}\left(\overline{Q}|\xt_{01}|\right)\, 
\frac{ \xt_{21}^m}{\left(\xt_{21}^2\right)^{1-\epsilon-\frac{\eta}{2}}}
\label{def_int_Im_b_alike_0_xi_2_eta}
\, .
\end{align}
With these results, the $\eta$-pole contribution \eqref{sigma_L_qqbarg_a_like_eta_pole_minus} can be simplified into
\begin{equation}
    \begin{split}
\sigma^{\gamma^*}_{L} \Big|_{q\bar{q}g,\, |a|^2 \textrm{ like}}^{\eta \textrm{ pole};\, \eta-}
=&\, 4 N_c \aem
\sum_{f}  e_f^2\;
\int_{0}^{1}
\dd{z}  
\int \frac{\dd[2-2\epsilon]{\xt_0}}{2\pi} \int \frac{\dd[2-2\epsilon]{\xt_1}}{2\pi}\,
4 Q^2\, z^2(1\!-\!z)^2\,
\\
&\times
\left(\frac{(2\pi)^2\mu^2\xt_{01}^2}{\overline{Q}^2}\right)^{\epsilon}
\left[\Kb_{-\epsilon}\left(\overline{Q}|\xt_{01}|\right)\right]^2\,
\\
& \times 
\left(\frac{\as \cf}{\pi}\right)
\frac{2}{\eta}
\left[\frac{2z q^+\nu_B^-}{4}\right]^{\eta}\,
\frac{\Gamma\left(1\!-\!\epsilon\!-\!\frac{\eta}{2}\right)^2}{ \Gamma\left(1\!+\!\frac{\eta}{2}\right)^2}\,\, \pi^{2\epsilon}
\mu^{2\epsilon}
\int \frac{\dd[2-2\epsilon]{\xt_2}}{2\pi}
{\textrm{Re}}\left[1\!-\!\tripole\right]
\\
& \times 
\left[\frac{\xt_{20}}{\left(\xt_{20}^2\right)^{1-\epsilon-\frac{\eta}{2}}}\vdot\left(\frac{\xt_{20}}{\left(\xt_{20}^2\right)^{1-\epsilon-\frac{\eta}{2}}}\!-\!\frac{\xt_{21}}{\left(\xt_{21}^2\right)^{1-\epsilon-\frac{\eta}{2}}}\right)\right]
\label{sigma_L_qqbarg_a_like_eta_pole_minus_2}
\, .
    \end{split}
\end{equation}

As in the case of the $\eta^+$ regulator, 
we can use Eq.~\eqref{tripole_split} to isolate 
the UV-sensitive part of Eq.~\eqref{sigma_L_qqbarg_a_like_eta_pole_minus_2} into a dipole-type contribution, leading to
\begin{align}
\sigma^{\gamma^*}_{L} \Big|_{q\bar{q}g,\, |a|^2 \textrm{ like}}^{\eta \textrm{ pole};\, \eta-}
=&\, 
\sigma^{\gamma^*}_{L} \Big|_{q\bar{q}g,\, |a|^2 \textrm{ like}}^{\textrm{dipole-like } \eta \textrm{ pole};\, \eta-}
+
\sigma^{\gamma^*}_{L} \Big|_{q\bar{q}g,\, |a|^2 \textrm{ like}}^{\textrm{BK-like } \eta \textrm{ pole};\, \eta-}
\label{sigma_L_qqbarg_a_like_eta_pole_minus_3}
\, ,
\end{align}
where
\begin{equation}
    \begin{split}
&\sigma^{\gamma^*}_{L} \Big|_{q\bar{q}g,\, |a|^2 \textrm{ like}}^{\textrm{dipole-like } \eta \textrm{ pole};\, \eta-}
=\, 4 N_c \aem
\sum_{f}  e_f^2\;
\int_{0}^{1}
\dd{z}  
\int \frac{\dd[2-2\epsilon]{\xt_0}}{2\pi} \int \frac{\dd[2-2\epsilon]{\xt_1}}{2\pi}\,
\\
&\times
4 Q^2\, z^2(1\!-\!z)^2\,
\left(\frac{(2\pi)^2\mu^2\xt_{01}^2}{\overline{Q}^2}\right)^{\epsilon}
\left[\Kb_{-\epsilon}\left(\overline{Q}|\xt_{01}|\right)\right]^2\,
 \times 
\left(\frac{\as \cf}{\pi}\right)\,
\widetilde{\cal V}^{L}_{|a|^2,\textrm{ pole; }\eta-}\;
{\textrm{Re}}\left[1\!-\!{\cal S}_{01}\right]
\label{sigma_L_qqbarg_a_like_dipole_eta_pole_minus}
    \end{split}
\end{equation}
and
\begin{equation}
    \begin{split}
&\sigma^{\gamma^*}_{L} \Big|_{q\bar{q}g,\, |a|^2 \textrm{ like}}^{\textrm{BK-like } \eta \textrm{ pole};\, \eta-}
\\
=&\, 4 N_c \aem
\sum_{f}  e_f^2\;
\int_{0}^{1}
\dd{z}
\int \frac{\dd[2-2\epsilon]{\xt_0}}{2\pi} \int \frac{\dd[2-2\epsilon]{\xt_1}}{2\pi}
4 Q^2 z^2(1\!-\!z)^2
\qty(\frac{(2\pi)^2\mu^2\xt_{01}^2}{\overline{Q}^2})^{\epsilon}
\qty[\Kb_{-\epsilon}\left(\overline{Q}|\xt_{01}|\right)]^2\,
\\
& \times 
\frac{2\as \cf}{\pi}
 \bigg[\frac{1}{\eta}\left(\frac{2z q^+\nu_B^-}{c_0^2}\right)^{\eta}
 +\Psi(1)-\Psi(1\!-\!\epsilon)
\bigg]\,
\Gamma(1\!-\!\epsilon)^2\, \pi^{2\epsilon}
\mu^{2\epsilon}
\\
&\, \times\,
\int \frac{\dd[2-2\epsilon]{ \xt_2}}{2\pi}
\left[\frac{\xt_{20}}{\left(\xt_{20}^2\right)^{1-\epsilon-\frac{\eta}{2}}}\vdot\left(\frac{\xt_{20}}{\left(\xt_{20}^2\right)^{1-\epsilon-\frac{\eta}{2}}}\!-\!\frac{\xt_{21}}{\left(\xt_{21}^2\right)^{1-\epsilon-\frac{\eta}{2}}}\right)\right]
{\textrm{Re}}\left[\dipole\!-\!\tripole\right]
+\order{\eta}
\label{sigma_L_qqbarg_a_like_BK_eta_pole_minus_2}
\, .
    \end{split}
\end{equation}
By comparison to the $q\bar{q}$ contributions from Eq.~\eqref{sigma_L_qqbar_1}, the analogous loop factor is defined in Eq.~\eqref{sigma_L_qqbarg_a_like_dipole_eta_pole_minus} as
\begin{equation}
    \begin{split}
\widetilde{\cal V}^{L}_{|a|^2,\textrm{ pole; }\eta-}
=&\,
\frac{2}{\eta}
\left[\frac{2z q^+\nu_B^-}{4}\right]^{\eta}\,
\frac{\Gamma\left(1\!-\!\epsilon\!-\!\frac{\eta}{2}\right)^2}{ \Gamma\left(1\!+\!\frac{\eta}{2}\right)^2}\,\, \pi^{2\epsilon}
\mu^{2\epsilon}
\\
&\times
\int \frac{\dd[2-2\epsilon]{\xt_2}}{2\pi}
\left[\frac{\xt_{20}}{\left(\xt_{20}^2\right)^{1-\epsilon-\frac{\eta}{2}}}\vdot\left(\frac{\xt_{20}}{\left(\xt_{20}^2\right)^{1-\epsilon-\frac{\eta}{2}}}\!-\!\frac{\xt_{21}}{\left(\xt_{21}^2\right)^{1-\epsilon-\frac{\eta}{2}}}\right)\right]
\\
=&\,
-\frac{1}{\eta}
\left[\frac{2z q^+\nu_B^-\xt_{01}^2}{4}\right]^{\eta}\,
\frac{\Gamma(-\epsilon\!-\!\eta)}{\Gamma(1\!+\!\eta)}\,
\left[\pi \mu^2 \xt_{01}^2\right]^{\epsilon}
\\
=&\,
\frac{\Gamma(1\!-\!\epsilon)}{\epsilon}\, 
\left[\pi \mu^2 \xt_{01}^2\right]^{\epsilon}
\left\{
\frac{1}{\eta}
+
\log\left(\frac{2z q^+\nu_B^-\xt_{01}^2}{c_0^2}\right)
-\frac{1}{\epsilon}
+\Psi(1)-\Psi(1\!-\!\epsilon)
+\order{\eta}
\right\}
\\
=&\,
\frac{\Gamma(1\!-\!\epsilon)}{\epsilon}\, 
\left[\pi \mu^2 \xt_{01}^2\right]^{\epsilon}
\left[
\frac{1}{\eta}
+
\log\left(\frac{2z q^+\nu_B^-\xt_{01}^2}{c_0^2}\right)
\right]
-\frac{S_{\epsilon} }{\epsilon^2} \left(\frac{\xt_{01}^2 \mu^2}{c_0^2}\right)^{\epsilon}
\\
&\,
+\frac{\pi^2}{12}
+\order{\epsilon}+\order{\eta f(\epsilon)}
\, .\label{V_L_a_like_dipole_eta_pole_minus}
    \end{split}
\end{equation}


\subsection{Results for the $\bar q \to g$ contributions}

Let us now state the results for all the contributions obtained from Eq.~\eqref{sigma_L_qqbarg_b_like}, interpreted as contributions in which the gluon is emitted from the antiquark, following the same method as in the previous three subsections, after applying in Eq.~\eqref{sigma_L_qqbarg_b_like} the change of variables from the footnote \ref{fn_cv_b_like}.

First, one finds the rapidity safe and UV subtracted contribution 
\begin{equation}
    \begin{split}
&\sigma^{\gamma^*}_{L} \Big|_{\bar q\rightarrow g}^{\textrm{reg.}}
=\, 
4 N_c \aem \left(\frac{\as \cf}{\pi}\right)
\sum_{f}  e_f^2\;
\int_{0}^{1}
\dd{z}\,  4 Q^2\, z^2(1\!-\!z)^2
\int \frac{\dd[2]{ \xt_0}}{2\pi} 
\int \frac{\dd[2]{ \xt_1}}{2\pi}
\int \frac{\dd[2]{ \xt_2}}{2\pi}
\int_0^1 \dd{\xi}\:
\\
&\, 
\times
\Bigg\{
\big(-2+
\xi\big)\,
\left[\frac{\xt_{21}}{\xt_{21}^2}\vdot\left(\frac{\xt_{21}}{\xt_{21}^2}\!-\!\frac{\xt_{20}}{\xt_{20}^2}\right)\right]
\Bigg\{
\Big[
\Kb_{0}\left({Q} X^{(b)}_{012}
\right)
\Big]^2
{\textrm{Re}}\left[1\!-\!\tripole\right]\;
-
\Big(\xt_2\rightarrow \xt_1\Big)
\Bigg\}
\\
&\,
+\xi\, 
\frac{\xt_{20}\!\vdot\!\xt_{21}}{\xt_{20}^2\,\xt_{21}^2}\,
\Big[
\Kb_{0}\left({Q} X^{(b)}_{012}
\right)
\Big]^2
{\textrm{Re}}\left[1\!-\!\tripole\right]
\Bigg\}
+\order{\epsilon}
\, ,
\label{sigma_L_qqbarg_b_like_rap_safe_finite_leftover}
    \end{split}
\end{equation}
analog to Eq.~\eqref{sigma_L_qqbarg_a_like_rap_safe_finite_leftover_3}, using the notation 
\begin{equation}
    \begin{split}
{X_{012}^{(b)}}^2
\equiv &\,
z(1\!-\!z)(1\!-\!\xi)\xt_{01}^2
+(1\!-\!z)^2\xi(1\!-\!\xi)\xt_{21}^2
+z(1\!-\!z)\xi \xt_{20}^2
\, 
\label{def_X012_qbar2g}
    \end{split}
\end{equation}
corresponding to the quantity $X_{012}^2$ from Eq.~\eqref{def_X012_generic}
after applying the change of variables from the footnote~\ref{fn_cv_b_like}. 

The rapidity safe UV divergent contribution extracted from Eq.~\eqref{sigma_L_qqbarg_b_like_rap_safe_finite_leftover} is a dipole-like contribution with a NLO factor 
\begin{equation}
    \begin{split}
\widetilde{\cal V}^{L}_{|b|^2,\textrm{rap. safe}}
=&\,
-\frac{\left(3
+\deltas \epsilon\right)}{4}\,
\frac{\Gamma(1\!-\!\epsilon)}{\epsilon}\,
\left[\pi \mu^2 \xt_{01}^2\right]^{\epsilon}
\, ,
\label{V_L_b_like_dipole_rap_safe}
    \end{split}
\end{equation}
equal to the one from Eq.~\eqref{V_L_a_like_dipole_rap_safe}.

Then, out of the rapidity sensitive part of Eq.~\eqref{sigma_L_qqbarg_b_like}, one finds a scheme-independent $+$ distribution contribution
\begin{equation}
    \begin{split}
\sigma^{\gamma^*}_{L} \Big|_{\bar q\rightarrow g}^{\textrm{+ distr.}}
&=\,
4 N_c \aem 
\sum_{f}  e_f^2\;
\int_{0}^{1}
\dd{z}\,  4 Q^2\, z^2(1\!-\!z)^2
\int \frac{\dd[2]{\xt_0}}{2\pi} \int \frac{\dd[2]{\xt_1}}{2\pi}\;
\frac{2\as \cf}{\pi}\!\!\int \frac{\dd[2]{\xt_2}}{2\pi}\:
\\
&\times
\left[\frac{\xt_{21}}{\xt_{21}^2}\vdot\left(\frac{\xt_{21}}{\xt_{21}^2}\!-\!\frac{\xt_{20}}{\xt_{20}^2}\right)\right]
\;
{\textrm{Re}}\left[1-\tripole\right]
\int_0^1 \frac{\dd{\xi}}{(\xi)_+}\, 
\,
\Big[
\Kb_{0}\left({Q}X^{(b)}_{012}
\right)
\Big]^2
+\order{\epsilon}
\, ,
\label{sigma_L_qqbarg_b_like_plus_prescr_4}
    \end{split}
\end{equation}
analog to Eq.~\eqref{sigma_L_qqbarg_a_like_plus_prescr_4},

Using the $\eta^+$ rapidity regulator, one gets the UV subtracted rapidity pole contribution 
\begin{equation}
    \begin{split}
\sigma^{\gamma^*}_{L} \Big|_{q\bar{q}g,\, |b|^2 \textrm{ like}}^{\textrm{BK-like } \eta \textrm{ pole};\, \eta+}
=&\, 4 N_c \aem
\sum_{f}  e_f^2\;
\int_{0}^{1}
\dd{z}  
\int \frac{\dd[2-2\epsilon]{\xt_0}}{2\pi} \int \frac{\dd[2-2\epsilon]{\xt_1}}{2\pi}\,
\\
&\times
4 Q^2\, z^2(1\!-\!z)^2\,
\left(\frac{(2\pi)^2\mu^2\xt_{01}^2}{\overline{Q}^2}\right)^{\epsilon}
\left[\Kb_{-\epsilon}\left(\overline{Q}|\xt_{01}|\right)\right]^2\,
\\
& \times 
\frac{2\as \cf}{\pi}
\frac{1}{\eta}
\left[\frac{(1\!-\!z) q^+}{\nu_B^+}\right]^{\eta}\,
\Gamma(1\!-\!\epsilon)^2\, \pi^{2\epsilon}
\mu^{2\epsilon}
\\
&\, \times\,
\int \frac{\dd[2-2\epsilon]{\xt_2}}{2\pi}
\left[\frac{\xt_{21}}{(\xt_{21}^2)^{1-\epsilon}}\vdot\left(\frac{\xt_{21}}{(\xt_{21}^2)^{1-\epsilon}}\!-\!\frac{\xt_{20}}{(\xt_{20}^2)^{1-\epsilon}}\right)\right]
{\textrm{Re}}\left[\dipole\!-\!\tripole\right]
\label{sigma_L_qqbarg_b_like_BK_eta_pole_plus_2}
\, .
    \end{split}
\end{equation}
analog to Eq.~\eqref{sigma_L_qqbarg_a_like_BK_eta_pole_plus_2}. The UV and rapidity divergent contribution extracted from Eq.~\eqref{sigma_L_qqbarg_b_like_BK_eta_pole_plus_2} is 
\begin{equation}
    \begin{split}
\widetilde{\cal V}^{L}_{|b|^2,\textrm{ pole; }\eta+}
=&\,
\frac{1}{\eta}
\left[\frac{(1\!-\!z) q^+}{\nu_B^+}\right]^{\eta}\;
\frac{\Gamma(1\!-\!\epsilon)}{\epsilon}\,
\left[\pi \mu^2 \xt_{01}^2\right]^{\epsilon}
\\
=&\,
\left[\frac{1}{\eta}
+
\log\left(\frac{(1\!-\!z) q^+}{\nu_B^+}\right)\;
\right]
\frac{\Gamma(1\!-\!\epsilon)}{\epsilon}\,
\left[\pi \mu^2 \xt_{01}^2\right]^{\epsilon}
+O(\eta)
\, .
\label{V_L_b_like_dipole_eta_pole_plus}
    \end{split}
\end{equation}

Using instead the $\eta^-$ rapidity regulator, the UV subtracted rapidity pole contribution is obtained as
\begin{equation}
    \begin{split}
&\sigma^{\gamma^*}_{L} \Big|_{q\bar{q}g,\, |b|^2 \textrm{ like}}^{\textrm{BK-like } \eta \textrm{ pole};\, \eta-}
\\
=&\, 4 N_c \aem
\sum_{f}  e_f^2\;
\int_{0}^{1}
\dd{z}
\int \frac{\dd[2-2\epsilon]{\xt_0}}{2\pi} \int \frac{\dd[2-2\epsilon]{\xt_1}}{2\pi}
4 Q^2 z^2(1\!-\!z)^2
\qty(\frac{(2\pi)^2\mu^2\xt_{01}^2}{\overline{Q}^2})^{\epsilon}
\qty[\Kb_{-\epsilon}\left(\overline{Q}|\xt_{01}|\right)]^2\,
\\
& \times 
\frac{2\as \cf}{\pi}
 \bigg[\frac{1}{\eta}\left(\frac{2(1\!-\!z) q^+\nu_B^-}{c_0^2}\right)^{\eta}
 +\Psi(1)-\Psi(1\!-\!\epsilon)
\bigg]\,
\Gamma(1\!-\!\epsilon)^2\, \pi^{2\epsilon}
\mu^{2\epsilon}
\\
&\, \times\,
\int \frac{\dd[2-2\epsilon]{ \xt_2}}{2\pi}
\left[\frac{\xt_{21}}{\left(\xt_{21}^2\right)^{1-\epsilon-\frac{\eta}{2}}}\vdot\left(\frac{\xt_{21}}{\left(\xt_{21}^2\right)^{1-\epsilon-\frac{\eta}{2}}}\!-\!\frac{\xt_{20}}{\left(\xt_{20}^2\right)^{1-\epsilon-\frac{\eta}{2}}}\right)\right]
{\textrm{Re}}\left[\dipole\!-\!\tripole\right]
+\order{\eta}
\label{sigma_L_qqbarg_b_like_BK_eta_pole_minus_2}
\, ,
    \end{split}
\end{equation}
analog to Eq.~\eqref{sigma_L_qqbarg_a_like_BK_eta_pole_minus_2}.
And finally, the UV and rapidity divergent contribution extracted from Eq.~\eqref{sigma_L_qqbarg_b_like_BK_eta_pole_minus_2} is 
\begin{equation}
    \begin{split}
\widetilde{\cal V}^{L}_{|b|^2,\textrm{ pole; }\eta-}
=&\,
-\frac{1}{\eta}
\left[\frac{2(1\!-\!z) q^+\nu_B^-\xt_{01}^2}{4}\right]^{\eta}\,
\frac{\Gamma(-\epsilon\!-\!\eta)}{\Gamma(1\!+\!\eta)}\,
\left[\pi \mu^2 \xt_{01}^2\right]^{\epsilon}
\\
=&\,
\frac{\Gamma(1\!-\!\epsilon)}{\epsilon}\, 
\left[\pi \mu^2 \xt_{01}^2\right]^{\epsilon}
\bigg\{
\frac{1}{\eta}
+
\log\left(\frac{2(1\!-\!z) q^+\nu_B^-\xt_{01}^2}{c_0^2}\right)
\\
& \hspace{4cm}
-\frac{1}{\epsilon}
+\Psi(1)-\Psi(1\!-\!\epsilon)
+\order{\eta}
\bigg\}
\, .\label{V_L_b_like_dipole_eta_pole_minus}
    \end{split}
\end{equation}


\subsection{Total dipole-like contribution}

We have now separated the UV-divergent parts of the $q \bar q g$ contribution that have a similar form to the
$q\bar{q}$ contributions from Sec.~\ref{sec:gamma_L_qqbar}.
It is then natural to combine this with the NLO correction $\widetilde{\mathcal{V}}^L$ to the $\gamma^* \to q \bar q$ LFWF.

For the $\eta^+$ regulator, 
this total UV factor from the $q \bar q g$ contribution corresponds to the sum of the factors~\eqref{V_L_a_like_dipole_eta_pole_plus} and \eqref{V_L_a_like_dipole_rap_safe} and their symmetric counterparts~\eqref{V_L_b_like_dipole_eta_pole_plus} and \eqref{V_L_b_like_dipole_rap_safe}   extracted from the $q \bar q g$ contribution~\eqref{sigma_L_qqbarg_a_like}:
\begin{equation}
    \begin{split}
&\widetilde{\cal V}^{L}_{q\bar{q}g,\textrm{ UV; }\eta+}
=\,
\widetilde{\cal V}^{L}_{|a|^2,\textrm{ pole; }\eta+}
+
\widetilde{\cal V}^{L}_{|a|^2,\textrm{rap. safe}}
+ \widetilde{\cal V}^{L}_{|b|^2,\textrm{ pole; }\eta+}
+
\widetilde{\cal V}^{L}_{|b|^2,\textrm{rap. safe}}
\\
=&\,
2\, \frac{\Gamma(1\!-\!\epsilon)}{\epsilon}
\qty[\pi \mu^2 \xt_{01}^2]^\epsilon
\left[ \frac{1}{\eta}
+\log\left(\frac{ \sqrt{z(1\!-\!z)} q^+}{\nu_B^+}
\right)
-\frac{\left(3
+\deltas \epsilon\right)}{4}
\right]
+\order{\eta}
\\
=&\,
2\left[ \frac{1}{\eta}
+\log\left(\frac{ \sqrt{z(1\!-\!z)} q^+}{\nu_B^+}
\right)\right]
\frac{\Gamma(1\!-\!\epsilon)}{\epsilon}
\qty[\pi \mu^2 \xt_{01}^2]^\epsilon
-\frac{3}{2}
\frac{S_{\epsilon} }{\epsilon} \left(\frac{\xt_{01}^2 \mu^2}{c_0^2}\right)^{\epsilon}\,
-\frac{\deltas }{2}
+\order{\epsilon}
+\order{\eta f(\epsilon)}
.\label{V_L_dipole_from_qqbarg_eta_plus}        
    \end{split}
\end{equation}
The analogous quantity for the $\eta^-$ regulator can be obtained similarly from the expressions \eqref{V_L_a_like_dipole_eta_pole_minus} and \eqref{V_L_a_like_dipole_rap_safe} and their symmetric counterparts~\eqref{V_L_b_like_dipole_eta_pole_minus} and \eqref{V_L_b_like_dipole_rap_safe} as
\begin{equation}
    \begin{split}        
\widetilde{\cal V}^{L}_{q\bar{q}g,\textrm{ UV; }\eta-}
=&\,
\widetilde{\cal V}^{L}_{|a|^2,\textrm{ pole; }\eta-}
+
\widetilde{\cal V}^{L}_{|a|^2,\textrm{rap. safe}}
+ \widetilde{\cal V}^{L}_{|b|^2,\textrm{ pole; }\eta-}
+
\widetilde{\cal V}^{L}_{|b|^2,\textrm{rap. safe}}
\\
=&\,
2\, \frac{\Gamma(1\!-\!\epsilon)}{\epsilon}
\qty[\pi \mu^2 \xt_{01}^2]^\epsilon
\Bigg\{ \frac{1}{\eta}
+\log\left(\frac{ 2\sqrt{z(1\!-\!z)} q^+ \nu_B^- \xt_{01}^2}{c_0^2}\right)
\\
&\,
\hspace{3.5cm}
-\frac{1}{\epsilon}
+\Psi(1)-\Psi(1\!-\!\epsilon) 
-\frac{\left(3
+\deltas \epsilon\right)}{4}\Bigg\}
+\order{\eta}
\\
=&\,
2\left[ \frac{1}{\eta}
+\log\left(\frac{ 2\sqrt{z(1\!-\!z)} q^+ \nu_B^- \xt_{01}^2}{c_0^2}\right)\right]
\frac{\Gamma(1\!-\!\epsilon)}{\epsilon}
\qty[\pi \mu^2 \xt_{01}^2]^\epsilon
\\
&\,
-\frac{2 S_{\epsilon} }{\epsilon^2} \left(\frac{\xt_{01}^2 \mu^2}{c_0^2}\right)^{\epsilon}
-\frac{3}{2}
\frac{S_{\epsilon} }{\epsilon} \left(\frac{\xt_{01}^2 \mu^2}{c_0^2}\right)^{\epsilon}
+\frac{\pi^2}{6}\,
-\frac{\deltas }{2}
+\order{\epsilon}
+\order{\eta f(\epsilon)}
\label{V_L_dipole_from_qqbarg_eta_minus}
\, .
    \end{split}
\end{equation}
Finally, the result that would be obtained with the pure rapidity regulator 
is the average of
Eqs.~\eqref{V_L_dipole_from_qqbarg_eta_plus} and \eqref{V_L_dipole_from_qqbarg_eta_minus}:
\begin{equation}
    \begin{split}
\widetilde{\cal V}^{L}_{q\bar{q}g,\textrm{ UV; pure rap. reg.}}
=&\,
\frac{1}{2}\, \widetilde{\cal V}^{L}_{q\bar{q}g,\textrm{ UV; }\eta+}
+
\frac{1}{2}\,
\widetilde{\cal V}^{L}_{q\bar{q}g,\textrm{ UV; }\eta-}
\\
=&\,
\left[ \frac{2}{\eta}
+\log\left(\frac{ 2 z(1\!-\!z) (q^+)^2 \nu_B^- \xt_{01}^2}{ c_0^2 \nu_B^+}\right)\right]
\frac{\Gamma(1\!-\!\epsilon)}{\epsilon}
\qty[\pi \mu^2 \xt_{01}^2]^\epsilon
\\
&\,
-\frac{S_{\epsilon} }{\epsilon^2} \left(\frac{\xt_{01}^2 \mu^2}{c_0^2}\right)^{\epsilon}
-\frac{3}{2}
\frac{S_{\epsilon} }{\epsilon} \left(\frac{\xt_{01}^2 \mu^2}{c_0^2}\right)^{\epsilon}
+\frac{\pi^2}{12}\,
-\frac{\deltas }{2}
+\order{\epsilon}
+\order{\eta f(\epsilon)}
\label{V_L_dipole_from_qqbarg_pure_rap_reg}
\, .
    \end{split}
\end{equation}

The total dipole-like contribution at NLO is then obtained by adding the UV factor extracted from the $q\bar{q}g$ contribution, Eq.~\eqref{V_L_dipole_from_qqbarg_eta_plus} or \eqref{V_L_dipole_from_qqbarg_eta_minus} or \eqref{V_L_dipole_from_qqbarg_pure_rap_reg}, to the corresponding loop factor in the $q\bar{q}$ contribution, Eq.~\eqref{VL_eta_plus_pos} or \eqref{VL_eta_minus_pos} or \eqref{VL_pure_rap_reg_pos}, within the expression \eqref{sigma_L_qqbar_1}. 
With any of these choices for the rapidity regulator, the same result is obtained, 
\begin{equation}
    \begin{split}
\widetilde{\cal V}^{L}_{\textrm{dipole}}
=&\,
\widetilde{\cal V}^{L}\bigg|^{\eta+}
+
\widetilde{\cal V}^{L}_{q\bar{q}g,\textrm{ UV; }\eta+}
=
\widetilde{\cal V}^{L}\bigg|^{\eta-}
+
\widetilde{\cal V}^{L}_{q\bar{q}g,\textrm{ UV; }\eta-}
\\
=&
\widetilde{\cal V}^{L}\bigg|^{\textrm{pure rap. reg.}}
+
\widetilde{\cal V}^{L}_{q\bar{q}g,\textrm{ UV; pure rap. reg.}}
\\
=&\, 
\frac{1}{2} \left[ \log\left(\frac{z}{1\!-\!z}\right)\right]^2
-\frac{\pi^2}{6}
+\frac{5}{2}
  +\order{\epsilon}+\order{\eta f(\epsilon)}
\label{VL_dipole}
 \, .
    \end{split}
\end{equation}
This exactly same result has been  obtained in Ref.~\cite{Beuf:2017bpd} with the cut-off regulator. It is both UV and rapidity finite.

\subsection{Rapidity subtraction for the dipole operator}

After grouping all of the dipole-like terms in the previous subsection, all of the 
UV divergences in the NLO corrections have canceled each other. 
The only divergences left are contained in the $q \bar q g$ contribution, and they have the form of the rapidity divergence $1/\eta$ (see Eqs.~\eqref{sigma_L_qqbarg_a_like_BK_eta_pole_plus_2} and
\eqref{sigma_L_qqbarg_a_like_BK_eta_pole_minus_2}, and their symmetric \eqref{sigma_L_qqbarg_b_like_BK_eta_pole_plus_2} and
\eqref{sigma_L_qqbarg_b_like_BK_eta_pole_minus_2}).
For both the $\eta^+$ and $\eta^-$ regulators (and hence also the pure rapidity regulator), the rapidity divergent contribution in Eq.~\eqref{sigma_L_qqbarg_a_like_BK_eta_pole_plus_2} or
\eqref{sigma_L_qqbarg_a_like_BK_eta_pole_minus_2} is related to the LO cross section for finite $\epsilon$ by the replacement of the operator ${\textrm{Re}}\left[1-{\cal S}_{01}\right]$ by 
\begin{equation}
\, 
\frac{1}{\eta}\, \frac{2\as \cf}{\pi}\,
\Gamma(1\!-\!\epsilon)^2\, \pi^{2\epsilon}
\mu^{2\epsilon}\,
\int \frac{\dd[2-2\epsilon]{\xt_2}}{2\pi}
\left[\frac{\xt_{20}}{(\xt_{20}^2)^{1-\epsilon}}\vdot\left(\frac{\xt_{20}}{(\xt_{20}^2)^{1-\epsilon}}\!-\!\frac{\xt_{21}}{(\xt_{21}^2)^{1-\epsilon}}\right)\right]
{\textrm{Re}}\left[{\cal S}_{01}\!-\!{\cal S}_{012}\right]
\label{a_like_rap_div}
\, .
\end{equation}
The analog rapidity divergent term present in Eq.~\eqref{sigma_L_qqbarg_b_like_BK_eta_pole_plus_2} or
\eqref{sigma_L_qqbarg_b_like_BK_eta_pole_minus_2}
reads:
\begin{equation}
\, 
\frac{1}{\eta}\, \frac{2\as \cf}{\pi}\,
\Gamma(1\!-\!\epsilon)^2\, \pi^{2\epsilon}
\mu^{2\epsilon}\,
\int \frac{\dd[2-2\epsilon]{\xt_2}}{2\pi}
\left[-\frac{\xt_{21}}{(\xt_{21}^2)^{1-\epsilon}}\vdot\left(\frac{\xt_{20}}{(\xt_{20}^2)^{1-\epsilon}}\!-\!\frac{\xt_{21}}{(\xt_{21}^2)^{1-\epsilon}}\right)\right]
{\textrm{Re}}\left[\dipole\!-\!\tripole\right]
\label{b_like_rap_div}
\, .
\end{equation}
These expressions are valid for all three rapidity regulators considered in this work.

The rapidity-divergent terms \eqref{a_like_rap_div} and \eqref{b_like_rap_div} can be removed by a redefinition of the dipole operator that is analogous to the standard UV renormalization of composite operators.
This procedure can be called \textit{rapidity renormalization} or \textit{rapidity subtraction}.
All of the color dipole or tripole operators appearing earlier in the present study should be understood as \emph{rapidity-unsubtracted operators}. 
Then, similarly to the $\msbar$ scheme for UV renormalization, we define the  \emph{rapidity-subtracted dipole operator} (or rapidity-renormalized) from the unsubtracted one as\footnote{Note that we use here the dipole operator $\dipole$ instead of the operator $\textrm{Re}\left[1\!-\!\dipole\right]$ present in the LO contribution, hence the minus sign in the $1/\eta$ pole term relative to Eqs.~\eqref{a_like_rap_div} and \eqref{b_like_rap_div}.}
\begin{equation}
    \begin{split}
\dipole \Big|^{\eta +}_{\textrm{sub.}}
\equiv &\,
\dipole \Big|^{\eta +}_{\textrm{unsub.}} 
-
\frac{1}{\eta}\, \left(\frac{\nu^+}{\nu_B^+}\right)^{\eta} \frac{2\as \cf}{\pi}\,
\Gamma(1\!-\!\epsilon)^2\, \pi^{2\epsilon}
\mu^{2\epsilon}\,
\\
&\times
\int \frac{\dd[2-2\epsilon]{\xt_2}}{2\pi}
\left[\frac{\xt_{20}}{(\xt_{20}^2)^{1-\epsilon}}\!-\!\frac{\xt_{21}}{(\xt_{21}^2)^{1-\epsilon}}\right]^2
\left[\dipole \!-\!\tripole\right]
+\order{\as^2}
\label{rap_renorm_dipole_plus_1}
\, 
    \end{split}
\end{equation}
in the case of the $\eta^+$ regulator. 
Since the definition \eqref{def_etaplus_reg} of this regulator involves the scale $\nu_B^+$ but not $\nu^+$, the unsubtracted dipole operator should depend on $\nu_B^+$ but not on $\nu^+$. Hence, the dependence on $\nu^+$ of the subtracted dipole operator comes entirely from the rapidity counterterm in Eq.~\eqref{rap_renorm_dipole_plus_1}, and one finds 
\begin{equation}
    \begin{split}
\nu^+ \partial_{\nu^+} \dipole \Big|^{\eta +}_{\textrm{sub.}}
= &\,
-
 \frac{2\as \cf}{\pi}\,
\Gamma(1\!-\!\epsilon)^2\, \pi^{2\epsilon}
\mu^{2\epsilon}\,
\int \frac{\dd[2-2\epsilon]{\xt_2}}{2\pi}
\left[\frac{\xt_{20}}{(\xt_{20}^2)^{1-\epsilon}}\!-\!\frac{\xt_{21}}{(\xt_{21}^2)^{1-\epsilon}}\right]^2
\left[\dipole \!-\!\tripole\right]
\\
&
+O(\eta)+\order{\as^2}
\\
= &\,
-
 \frac{2\as \cf}{\pi}\,
\int \frac{\dd[2]{\xt_2}}{2\pi}\,
\frac{\xt_{01}^2}{\xt_{20}^2\, \xt_{21}^2}\,
\left[\dipole \!-\!\tripole\right]
+O(\epsilon)+O(\eta f(\epsilon))+\order{\as^2}
\label{BK_dipole_plus}
\, ,
    \end{split}
\end{equation}
which is the B-JIMWLK evolution equation for the dipole operator in the $\eta^+$ scheme, at finite $\epsilon$ and then at $\epsilon= 0$. Moreover, the subtracted dipole operator is independent of the scale $\nu_B^+$ if and only if the unsubtracted dipole operator obeys the same evolution \eqref{BK_dipole_plus}, but with respect to $\nu_B^+$. 

In the case of the $\eta^-$ regulator, the situation is similar, with the rapidity renormalization of the dipole operator performed as 
\begin{equation}
    \begin{split}
\dipole \Big|^{\eta -}_{\textrm{sub.}}
\equiv &\,
\dipole \Big|^{\eta -}_{\textrm{unsub.}} 
-
\frac{1}{\eta}\, \left(\frac{\nu_B^-}{\nu^-}\right)^{\eta} \frac{2\as \cf}{\pi}\,
\Gamma(1\!-\!\epsilon)^2\, \pi^{2\epsilon}
\mu^{2\epsilon}\,
\\
&\times
\int \frac{\dd[2-2\epsilon]{\xt_2}}{2\pi}
\left[\frac{\xt_{20}}{(\xt_{20}^2)^{1-\epsilon}}\!-\!\frac{\xt_{21}}{(\xt_{21}^2)^{1-\epsilon}}\right]^2
\left[\dipole \!-\!\tripole\right]
+\order{\as^2}
\label{rap_renorm_dipole_minus_1}
\, ,
    \end{split}
\end{equation}
leading to the evolution equation 
\begin{equation}
    \begin{split}
\nu^- \partial_{\nu^-} \dipole \Big|^{\eta -}_{\textrm{sub.}}
= &\,
 \frac{2\as \cf}{\pi}\,
\Gamma(1\!-\!\epsilon)^2\, \pi^{2\epsilon}
\mu^{2\epsilon}\,
\int \frac{\dd[2-2\epsilon]{\xt_2}}{2\pi}
\left[\frac{\xt_{20}}{(\xt_{20}^2)^{1-\epsilon}}\!-\!\frac{\xt_{21}}{(\xt_{21}^2)^{1-\epsilon}}\right]^2
\left[\dipole \!-\!\tripole\right]
\\
&
+O(\eta)+\order{\as^2}
\\
= &\,
 \frac{2\as \cf}{\pi}\,
\int \frac{\dd[2]{\xt_2}}{2\pi}
\frac{\xt_{01}^2}{\xt_{20}^2\, \xt_{21}^2}\,
\left[\dipole \!-\!\tripole\right]
+O(\epsilon)+O(\eta f(\epsilon))+\order{\as^2}
\label{BK_dipole_minus}
\, 
    \end{split}
\end{equation}
in that scheme.

Finally, in the case of the pure rapidity regulator, the rapidity subtraction is performed as
\begin{equation}
    \begin{split}
\dipole \Big|^{\textrm{pure rap. reg.}}_{\textrm{sub.}}
\equiv &\,
\dipole \Big|^{\textrm{pure rap. reg.}}_{\textrm{unsub.}} 
-
\frac{1}{\eta}\, \left(\frac{\nu_B^-\, \nu^+}{\nu_B^+\, \nu^-}\right)^{\frac{\eta}{2}} \frac{2\as \cf}{\pi}\,
\Gamma(1\!-\!\epsilon)^2\, \pi^{2\epsilon}
\mu^{2\epsilon}\,
\\
&\times
\int \frac{\dd[2-2\epsilon]{\xt_2}}{2\pi}
\left[\frac{\xt_{20}}{(\xt_{20}^2)^{1-\epsilon}}\!-\!\frac{\xt_{21}}{(\xt_{21}^2)^{1-\epsilon}}\right]^2
\left[\dipole \!-\!\tripole\right]
+\order{\as^2}
\label{rap_renorm_dipole_pure_rap_1}
\, .
    \end{split}
\end{equation}
Introducing the rapidity $y_{(\nu)}$ associated with $\nu^+$ and $\nu^-$, 
\begin{equation}
y_{(\nu)} \equiv \frac{1}{2}\, \log\left(\frac{\nu^+}{\nu^-}\right) 
\, ,
\end{equation}
one thus finds
\begin{equation}
    \begin{split}
\partial_{y_{(\nu)}} \dipole \Big|^{\textrm{pure rap. reg.}}_{\textrm{sub.}}
= &\,
 -\frac{2\as \cf}{\pi}\,
\Gamma(1\!-\!\epsilon)^2\, \pi^{2\epsilon}
\mu^{2\epsilon}\,
\int \frac{\dd[2-2\epsilon]{\xt_2}}{2\pi}
\\
&\times
\left[\frac{\xt_{20}}{(\xt_{20}^2)^{1-\epsilon}}\!-\!\frac{\xt_{21}}{(\xt_{21}^2)^{1-\epsilon}}\right]^2
\left[\dipole \!-\!\tripole\right]
+O(\eta)+\order{\as^2}
\\
= &\,
-
 \frac{2\as \cf}{\pi}\,
\int \frac{\dd[2]{\xt_2}}{2\pi}\,
\frac{\xt_{01}^2}{\xt_{20}^2\, \xt_{21}^2}\,
\left[\dipole \!-\!\tripole\right]
+O(\epsilon)+O(\eta f(\epsilon))+\order{\as^2}
\label{BK_dipole_pure_rap}
\, 
    \end{split}
\end{equation}
in this scheme for the evolution.

In principle, one should be able to obtain the same results \eqref{BK_dipole_plus}, \eqref{BK_dipole_minus} and  \eqref{BK_dipole_pure_rap} by calculating the NLO contribution to the dipole operator from its operator definition, using the three considered rapidity regulators. 
This important cross-check is however beyond the scope of the present paper, and left for further study.

A similar rapidity subtraction is in principle performed for the tripole operator $\tripole$. However, since $\tripole$ does not appear in the LO cross section, but only at NLO and beyond, its rapidity subtraction term corresponds to a NNLO contribution to the cross section which is beyond the perturbative order considered in this work.


\section{Final results}

From now on, we will always use rapidity-subtracted operators to write our results. At this stage, no divergence of any kind is left, so that it is safe to take both the $\eta\rightarrow 0$ and the $\epsilon\rightarrow 0$ limits. Then, the longitudinal photon cross section at NLO, can be written in the form
\begin{align}
\sigma^{\gamma^*}_{L} 
=&\, \sigma^{\gamma^*}_{L} \Big|_{\textrm{dipole}}
+\sigma^{\gamma^*}_{L} \Big|_{q\rightarrow g}
+\sigma^{\gamma^*}_{L} \Big|_{\bar{q}\rightarrow g}
+\order{\aem \as^2}
\, ,
\label{sigma_L_total_final_1}
\end{align}
with the first term containing the LO contribution as well as the  
total dipole-like NLO corrections
\begin{equation}
    \begin{split}
\sigma^{\gamma^*}_{L} \Big|_{\textrm{dipole}}
=&\, 4 N_c \aem
\sum_{f}  e_f^2
\int_{0}^{1} \!\!
\dd{z}
\int\frac{\dd[2]{\xt_0}}{2\pi}
\int\frac{\dd[2]{\xt_1}}{2\pi}
\;
4 Q^2\, z^2 (1\!-\!z)^2\, 
\left[\Kb_{0}\qty(\overline{Q}|\xt_{01}|)\right]^2
\\
&\, \times\,
{\textrm{Re}}\left[1-\dipole\right]\Big|_{\textrm{sub.}}\,
\left[1+\frac{\as \cf}{\pi}\,
\widetilde{\cal V}^{L}_{\textrm{dipole}} 
\right]\:
+\order{\aem \as^2}
\, ,
\label{sigma_L_dipole_final}
    \end{split}
\end{equation}
with the finite and scheme-independent loop factor given in Eq.~\eqref{VL_dipole}.

The second and third terms in Eq.~\eqref{sigma_L_total_final_1} contain the NLO contributions with an extra gluon resolved by the target, which can be considered to be emitted from the quark or from the antiquark respectively. These two contributions are symmetric to each other by exchange of the quark and the antiquark; in particular, their integrands are related to each other by simultaneous exchange of $\xt_0$ and $\xt_1$, and of $z$ and $1\!-\!z$. 
Because of this, these two terms are equal at the integrated level, and one could actually write 
\begin{align}
\sigma^{\gamma^*}_{L} 
=&\,
\sigma^{\gamma^*}_{L} \Big|_{\textrm{dipole}}
+2\sigma^{\gamma^*}_{L} \Big|_{q\rightarrow g}
+\order{\aem \as^2}
\, .
\label{sigma_L_total_final_2}
\end{align}
The contributions with a gluon emitted from the quark or the antiquark can be decomposed as 
\begin{align}
\sigma^{\gamma^*}_{L} \Big|_{q\rightarrow g}
=&\, 
\sigma^{\gamma^*}_{L} \Big|_{q\rightarrow g}^{\textrm{reg.}}
+\sigma^{\gamma^*}_{L} \Big|_{q\rightarrow g}^{\textrm{+ distr.}}
+\sigma^{\gamma^*}_{L} \Big|_{q\rightarrow g}^{\textrm{scheme dep.}}
\,  
\label{sigma_L_q2g_split}
\end{align}
and 
\begin{align}
\sigma^{\gamma^*}_{L} \Big|_{\bar q\rightarrow g}
=&\, 
\sigma^{\gamma^*}_{L} \Big|_{\bar q\rightarrow g}^{\textrm{reg.}}
+\sigma^{\gamma^*}_{L} \Big|_{\bar q\rightarrow g}^{\textrm{+ distr.}}
+\sigma^{\gamma^*}_{L} \Big|_{\bar q\rightarrow g}^{\textrm{scheme dep.}}
\, . 
\label{sigma_L_qbar2g_split}
\end{align}
In each of these two equations, the first two terms containing the rapidity-safe and plus-distribution contributions are independent of the choice of rapidity regulator. 
They are given in Eqs.~\eqref{sigma_L_qqbarg_a_like_rap_safe_finite_leftover_3}, \eqref{sigma_L_qqbarg_a_like_plus_prescr_4},
\eqref{sigma_L_qqbarg_b_like_rap_safe_finite_leftover} and \eqref{sigma_L_qqbarg_b_like_plus_prescr_4}.
The third term in each expression depends on the rapidity regulator and the rapidity renormalization scheme.

\subsection{\texorpdfstring{With $\eta^+$ regulator}
{With eta+ regulator
}
}

For the $\eta^+$ regulator, the scheme-dependent contribution in Eq.~\eqref{sigma_L_q2g_split} can be obtained from
Eq.~\eqref{sigma_L_qqbarg_a_like_BK_eta_pole_plus_2} by adding the corresponding piece of the rapidity counterterm from Eq.~\eqref{rap_renorm_dipole_plus_1}. It amounts to subtracting the $1/\eta$ term in Eq.~\eqref{sigma_L_qqbarg_a_like_BK_eta_pole_plus_2} and replacing $\nu_B^+$ by $\nu^+$ in the leftover logarithm.
Finally, taking
the $\eta\rightarrow 0$ and  $\epsilon\rightarrow 0$ limits, we obtain:
\begin{equation}
    \begin{split}
\sigma^{\gamma^*}_{L} \Big|_{q\rightarrow g}^{\textrm{scheme dep.};\, \eta+}
=&\, 4 N_c \aem
\sum_{f}  e_f^2\;
\int_{0}^{1}
\dd{z}
\int \frac{\dd[2]{\xt_0}}{2\pi} \int \frac{\dd[2]{\xt_1}}{2\pi}\,
4 Q^2\, z^2(1\!-\!z)^2\,
\left[\Kb_{0}\qty(\overline{Q}|\xt_{01}|)\right]^2\,
\\
&\, \times\,
\frac{2\as \cf}{\pi}
\log\left(\frac{z q^+}{\nu^+}\right)
\int \frac{\dd[2]{\xt_2}}{2\pi}
\left[\frac{\xt_{20}}{\xt_{20}^2}\vdot\left(\frac{\xt_{20}}{\xt_{20}^2}\!-\!\frac{\xt_{21}}{\xt_{21}^2}\right)\right]
{\textrm{Re}}\left[ \dipole \!-\! \tripole\right]
\label{sigma_L_q2g_scheme_dep_eta_plus_1}
\, .
    \end{split}
\end{equation}
Similarly, the scheme dependent term in Eq.~\eqref{sigma_L_qbar2g_split} is obtained as the leftover  
from Eq.~\eqref{sigma_L_qqbarg_b_like_BK_eta_pole_plus_2}
after rapidity renormalization, as
\begin{equation}
    \begin{split}
&\sigma^{\gamma^*}_{L} \Big|_{\bar{q}\rightarrow g}^{\textrm{scheme dep.};\, \eta+}
=\, 4 N_c \aem
\sum_{f}  e_f^2\;
\int_{0}^{1}
\dd{z}
\int \frac{\dd[2]{\xt_0}}{2\pi} \int \frac{\dd[2]{\xt_1}}{2\pi}\,
4 Q^2\, z^2(1\!-\!z)^2\,
\left[\Kb_{0}\qty(\overline{Q}|\xt_{01}|)\right]^2\,
\\
&\, \times\,
\frac{2\as \cf}{\pi}
\log\left(\frac{(1\!-\!z) q^+}{\nu^+}\right)
\int \frac{d^{2} \xt_2}{2\pi}
\left[-\frac{\xt_{21}}{\xt_{21}^2}\vdot\left(\frac{\xt_{20}}{\xt_{20}^2}\!-\!\frac{\xt_{21}}{\xt_{21}^2}\right)\right]
{\textrm{Re}}\left[ \dipole \!-\! \tripole\right]
\label{sigma_L_qbar2g_scheme_dep_eta_plus_1}
\, .
    \end{split}
\end{equation}
The total scheme-dependent piece for $\eta^+$ regulator is then
\begin{equation}
    \begin{split}
&\sigma^{\gamma^*}_{L} \Big|_{{q}\rightarrow g}^{\textrm{scheme dep.};\, \eta+}
+
\sigma^{\gamma^*}_{L} \Big|_{\bar{q}\rightarrow g}^{\textrm{scheme dep.};\, \eta+}
\\
=&\, 4 N_c \aem
\sum_{f}  e_f^2\;
\int_{0}^{1}
\dd{z}  
\int \frac{\dd[2]{\xt_0}}{2\pi} \int \frac{\dd[2]{\xt_1}}{2\pi}\,
4 Q^2\, z(1\!-\!z)\,
\left[\Kb_{0}\qty(\overline{Q}|\xt_{01}|)\right]^2\,
\\
&\, \times\,
\frac{2\as \cf}{\pi}
\log\left(\frac{q^+\sqrt{z(1\!-\!z)} }{\nu^+}\right)
\int \frac{d^{2} \xt_2}{2\pi}\,
\frac{\xt_{01}^2}{\xt_{20}^2\, \xt_{21}^2}
{\textrm{Re}}\left[\dipole\!-\! \tripole\right]
\label{sigma_L_total_scheme_dep_eta_plus_1}
\, .
    \end{split}
\end{equation}
Note that, we have used the fact that only the part of the integrand symmetric with respect to $z\leftrightarrow (1\!-\!z)$ gives a non-zero contribution to the integral over $z$.

At this stage, we have the longitudinal photon cross section (or equivalently $F_L$) at NLO expressed in terms of subtracted dipoles and tripole operators in the $\eta^+$ scheme. These operators depend on the scale $\nu^+$, and in addition, there is an explicit dependence on $\nu^+$ in the coefficient of the operator in the NLO contribution \eqref{sigma_L_total_scheme_dep_eta_plus_1}. Similarly as the $\mu$ dependence in standard perturbative calculations, the dependence on $\nu^+$ in the NLO coefficient and the dependence on $\nu^+$ in the dipole operator in the LO contribution are compensating each other thanks to the evolution equation \eqref{BK_dipole_plus} for the dipole, so that the cross section is independent of $\nu^+$, up to NNLO contributions (corresponding here to order $\aem \as^2$ in the photon-target cross section, or to order $\as^2$ in the $F_L$ structure function).

Finally, it remains to discuss the choice of scale $\nu^+$ in practical applications. 
But first, let us note that for dimensional and symmetry reasons, the subtracted dipole operator, evaluated in the state of a hadronic or nuclear target of large momentum $P^-$, should depend on $\nu^+$ through a combination
of the type $2P^- \nu^+/Q_0^2$, with $Q_0$ a typical  scale associated with the target, for example its mass or its initial saturation scale (in the case of a dense target). Hence, $Q_0^2/(2P^-)$ is the typical value of $\nu^+$ at which one can set the initial condition for the B-JIMWLK evolution \eqref{BK_dipole_plus} along $\nu^+$.

From Eq.~\eqref{sigma_L_total_scheme_dep_eta_plus_1}, it is clear that a natural choice $\nu_f^+$ for the value of $\nu^+$ scale in our result for cross section at NLO is
\begin{align}
\nu^+ \mapsto \nu_f^+ 
=&\, 
{\cal C}\, q^+
\label{scale_choice_nu_plus}
\, ,
\end{align}
with a constant ${\cal C}$ of order one, in order to prevent a potentially large logarithm in Eq.~\eqref{sigma_L_total_scheme_dep_eta_plus_1}. One could include as well the factor $\sqrt{z(1\!-\!z)}$ in the definition of $\nu_f^+$. However, the phase space at $z\rightarrow 0$ or $z\rightarrow 1$ is suppressed by the factor $z(1\!-\!z)$ in the first line of Eq.~\eqref{sigma_L_total_scheme_dep_eta_plus_1}. Hence, including a factor $\sqrt{z(1\!-\!z)}$ or not in the definition of $\nu_f^+$ should not make a difference in the case of the $F_L$ structure function in terms of large log resummation. It would only complicate numerical studies.  

All in all, with the choice \eqref{scale_choice_nu_plus} for the factorization scale in $\nu^+$, the dipole operator should be evolved with the B-JIMWLK evolution \eqref{BK_dipole_plus} over a logarithmic range 
\begin{align}
Y_f^{(+)} 
=&\, 
\log\left(
\frac{2P^-\nu_f^+}{Q_0^2}
\right)
\sim
\log\left(
\frac{W^2}{Q_0^2}
\right)
\sim
\log\left(
\frac{1}{x_{Bj}}
\right)
+
\log\left(
\frac{Q^2}{Q_0^2}
\right)
\label{evol_range_nu_plus}
\, ,
\end{align}
with $W^2=(P+q)^2$ being the squared center of mass energy of the photon-target process.


\subsection{\texorpdfstring{With $\eta^-$ regulator}
{With eta- regulator
}
}

Similarly as to the $\eta^+$ regulator,
we can obtain the scheme-dependent terms for the $\eta^-$ regulator
 from Eqs.~\eqref{sigma_L_qqbarg_a_like_BK_eta_pole_minus_2}
 and \eqref{sigma_L_qqbarg_b_like_BK_eta_pole_minus_2} by
 adding the corresponding parts of the rapidity counterterm from Eq.~\eqref{rap_renorm_dipole_minus_1}, thus subtracting the $1/\eta$ term and replacing $\nu_B^-$ by $\nu^-$. Then, one finds
\begin{equation}
    \begin{split}
\sigma^{\gamma^*}_{L} \Big|_{q\rightarrow g}^{\textrm{scheme dep.};\, \eta-}
=&\, 4 N_c \aem
\sum_{f}  e_f^2\;
\int_{0}^{1}
\dd{z}  
\int \frac{\dd[2]{\xt_0}}{2\pi} \int \frac{\dd[2]{\xt_1}}{2\pi}\,
4 Q^2\, z^2(1\!-\!z)^2\,
\left[\Kb_{0}\qty(\overline{Q}|\xt_{01}|)\right]^2\,
\\
&\times 
\frac{2\as \cf}{\pi}
\int \frac{d^{2} \xt_2}{2\pi}
{\textrm{Re}}\left[\dipole \!-\! \tripole\right]
\,\\
&\, \times\,
\left[\log\left(\frac{2z q^+\nu^-\, \xt_{20}^2 }{c_0^2}\right)\,
\frac{\xt_{20}}{\xt_{20}^2}\vdot\left(\frac{\xt_{20}}{\xt_{20}^2}\!-\!\frac{\xt_{21}}{\xt_{21}^2}\right)
-\frac{1}{2}\, \log\left(\frac{\xt_{21}^2 }{\xt_{20}^2}\right)
\frac{\xt_{20}\!\vdot\!\xt_{21} }{\xt_{20}^2\, \xt_{21}^2}
\right]
\label{sigma_L_q2g_scheme_dep_eta_minus_1}
\, 
    \end{split}
\end{equation}
and
\begin{equation}
\begin{split}
&\sigma^{\gamma^*}_{L} \Big|_{\bar{q}\rightarrow g}^{\textrm{scheme dep.};\, \eta-}
=\, 4 N_c \aem
\sum_{f}  e_f^2\;
\int_{0}^{1}
\dd{z} 
\int \frac{\dd[2]{\xt_0}}{2\pi} \int \frac{\dd[2]{\xt_1}}{2\pi}\,
4 Q^2\, z^2(1\!-\!z)^2\,
\left[\Kb_{0}\qty(\overline{Q}|\xt_{01}|)\right]^2\,
\\
& \times
\frac{2\as \cf}{\pi}
\int \frac{\dd[2]{\xt_2}}{2\pi}
{\textrm{Re}}\left[\dipole \!-\! \tripole\right]
\,
\\
&\, \times\,
\left[-\log\left(\frac{2(1\!-\!z) q^+\nu^-\, \xt_{21}^2 }{c_0^2}\right)\,
\frac{\xt_{21}}{\xt_{21}^2}\vdot\left(\frac{\xt_{20}}{\xt_{20}^2}\!-\!\frac{\xt_{21}}{\xt_{21}^2}\right)
+\frac{1}{2}\, \log\left(\frac{\xt_{21}^2 }{\xt_{20}^2}\right)
\frac{\xt_{20}\!\vdot\!\xt_{21} }{\xt_{20}^2\, \xt_{21}^2}
\right]
\label{sigma_L_qbar2g_scheme_dep_eta_minus_1}
\, ,
\end{split}
\end{equation}
so that the total scheme-dependent piece for the  $\eta^-$ regulator is
\begin{equation}
    \begin{split}
& \sigma^{\gamma^*}_{L} \Big|_{q\rightarrow g}^{\textrm{scheme dep.};\, \eta-}
+
\sigma^{\gamma^*}_{L} \Big|_{\bar{q}\rightarrow g}^{\textrm{scheme dep.};\, \eta-}
\\
=&\, 4 N_c \aem
\sum_{f}  e_f^2\;
\int_{0}^{1}
\dd{z}
\int \frac{\dd[2]{\xt_0}}{2\pi} \int \frac{\dd[2]{\xt_1}}{2\pi}\,
4 Q^2\, z^2(1\!-\!z)^2\,
\left[\Kb_{0}\qty(\overline{Q}|\xt_{01}|)\right]^2\,
\,
\\
&\, \times\,
\frac{2\as \cf}{\pi}\int \frac{\dd[2]{\xt_2}}{2\pi}
{\textrm{Re}}\left[\dipole\!-\!\tripole\right]\,
\\
&\times 
\left(\frac{\xt_{20}}{\xt_{20}^2}\!-\!\frac{\xt_{21}}{\xt_{21}^2}\right)
\vdot \bigg[\log\left(\frac{2z q^+\nu^-\, \xt_{20}^2 }{c_0^2}\right)\,
\frac{\xt_{20}}{\xt_{20}^2}
-\log\left(\frac{2(1\!-\!z) q^+\nu^-\, \xt_{21}^2 }{c_0^2}\right)\,
\frac{\xt_{21}}{\xt_{21}^2}
\bigg]
\\
=&\, 4 N_c \aem
\sum_{f}  e_f^2\;
\int_{0}^{1}
\dd{z}
\int \frac{\dd[2]{\xt_0}}{2\pi} \int \frac{\dd[2]{\xt_1}}{2\pi}\,
4 Q^2\, z^2(1\!-\!z)^2\,
\left[\Kb_{0}\qty(\overline{Q}|\xt_{01}|)\right]^2\,
\,
\\
&\, \times\,
\frac{2\as \cf}{\pi}\int \frac{\dd[2]{\xt_2}}{2\pi}
{\textrm{Re}}\left[\dipole\!-\!\tripole\right]\,
\Bigg\{ 
\frac{\xt_{01}\cdot \xt_{21}}{\xt_{20}^2\, \xt_{21}^2}
\log\left(\frac{2\sqrt{z(1\!-\!z)} q^+\nu^-\, \xt_{20}^2 }{c_0^2}\right)\,
\\
&
-
\frac{\xt_{01}\cdot \xt_{20}}{\xt_{20}^2\, \xt_{21}^2}
\log\left(\frac{2\sqrt{z(1\!-\!z)} q^+\nu^-\, \xt_{21}^2 }{c_0^2}\right)
\Bigg\}
\label{sigma_L_total_scheme_dep_eta_minus_1}
\, ,
    \end{split}
\end{equation}
where, in the last step, we have dropped antisymmetric terms with respect to $z\leftrightarrow (1\!-\!z)$ in the integrand, which vanish when integrated over $z$. Again, the explicit dependence on $\nu^-$ in the coefficient in Eq.~\eqref{sigma_L_total_scheme_dep_eta_minus_1} compensates the dependence  on $\nu^-$ of the dipole operator in the LO term, leading to a $\nu^-$ independent cross section up to NNLO corrections.  

In the $\eta^-$ scheme, the dependence on $\nu^-$ of the  rapidity-subtracted dipole operator, when evaluated in the state of a  target of large momentum $P^-$, should occur via the combination $\nu^-/P^-$ in order to preserve invariance under longitudinal boosts.

When choosing a natural value $\nu_f^-$ for $\nu^-$ in the cross section, one should remember that the same value has to be taken for the dipole operator in the LO contribution and for the NLO coefficient in Eq.~\eqref{sigma_L_total_scheme_dep_eta_minus_1}. Since the position $\xt_2$ is absent in the LO contribution, $\nu_f^-$ cannot depend on it, so that one cannot fully remove the potentially large logs from Eq.~\eqref{sigma_L_total_scheme_dep_eta_minus_1} by choosing the scale $\nu^-$. Instead, it is convenient to rewrite Eq.~\eqref{sigma_L_total_scheme_dep_eta_minus_1} as
\begin{equation}
    \begin{split}
& \sigma^{\gamma^*}_{L} \Big|_{q\rightarrow g}^{\textrm{scheme dep.};\, \eta-}
+
\sigma^{\gamma^*}_{L} \Big|_{\bar{q}\rightarrow g}^{\textrm{scheme dep.};\, \eta-}
\\
=&\, 4 N_c \aem
\sum_{f}  e_f^2\;
\int_{0}^{1}
\dd{z}
\int \frac{\dd[2]{\xt_0}}{2\pi} \int \frac{\dd[2]{\xt_1}}{2\pi}\,
4 Q^2\, z^2(1\!-\!z)^2\,
\left[\Kb_{0}\qty(\overline{Q}|\xt_{01}|)\right]^2\,
\,
\\
&\, \times\,
\frac{2\as \cf}{\pi}\int \frac{\dd[2]{\xt_2}}{2\pi}
{\textrm{Re}}\left[\dipole\!-\!\tripole\right]\,
\Bigg\{ 
\frac{\xt_{01}^2}{\xt_{20}^2\, \xt_{21}^2}
\log\left(\frac{2\sqrt{z(1\!-\!z)} q^+\nu^-\, }{Q^2}\right)\,
\\
&
\hspace{3cm}
+\frac{\xt_{01}\cdot \xt_{21}}{\xt_{20}^2\, \xt_{21}^2}
\log\left(\frac{\xt_{20}^2 Q^2}{c_0^2}\right)\,
-
\frac{\xt_{01}\cdot \xt_{20}}{\xt_{20}^2\, \xt_{21}^2}
\log\left(\frac{\xt_{21}^2 Q^2}{c_0^2}\right)
\Bigg\}
\label{sigma_L_total_scheme_dep_eta_minus_2}
\, ,
    \end{split}
\end{equation}
and remove the potentially large log in the first term thanks to the scale choice
\begin{align}
\nu^- \mapsto \nu_f^- 
=&\, \frac{Q^2}{2 q^+ }\, \frac{1}{{\cal C}} 
\label{scale_choice_nu_minus}
\, .
\end{align}
Like in the $\eta^+$ scheme case, we have introduced a constant ${\cal C}$ of order one, but not the factor $\sqrt{z(1\!-\!z)}$ which cannot produce large logs for $F_L$. 

The choice \eqref{scale_choice_nu_minus} for the factorization scale in $\nu^-$ amounts to performing the B-JIMWLK evolution \eqref{BK_dipole_minus} of the dipole operator over the logarithmic range
\begin{align}
Y_f^{(-)}
=&\, 
\log\left(
\frac{P^-}{\nu_f^-}
\right)
\sim
\log\left(
\frac{W^2}{Q^2}
\right)
\sim
\log\left(
\frac{1}{x_{Bj}}
\right)
\label{evol_range_nu_minus}
\, .
\end{align}
Note that in the case of the $\eta^-$ regulator, corresponding to an evolution formulated in the $\nu^-$ variable, the logarithmic range of evolution $Y_f^{(-)}$ is more closely approximated by $\log(1/x_{Bj})$ than in the case of the $\eta^+$ regulator, in which the typically large separation between the hard scale $Q^2$ and the soft scale $Q_0^2$ also contributes to  $Y_f^{(+)}$ \eqref{evol_range_nu_plus}. This is consistent with the general expectations for the differences between low $x$ evolutions formulated either along $k^+$ or along $k^-$, as discussed in Ref.~\cite{Beuf:2014uia} for example.

The second and third terms in the bracket in Eq.~\eqref{sigma_L_total_scheme_dep_eta_minus_2} are unaffected by the scale choice \eqref{scale_choice_nu_minus}, but potentially large. 
Let us note that these two terms have the same form as the large log contributions 
\eqref{large_coll_log_plus_dist_q2g} and \eqref{large_coll_log_plus_dist_qbar2g} extracted from the $+$ distribution contribution \eqref{sigma_L_qqbarg_a_like_plus_prescr_3} 
and its symmetric counterpart by exchange of the quark and the antiquark, except that they have an opposite sign, and that no restriction on the integral over $\xt_2$ is imposed in Eq.~\eqref{sigma_L_total_scheme_dep_eta_minus_2}. Hence, in the domain $\xt_{20}^2\sim \xt_{21}^2\gg 1/Q^2 \gtrsim \xt_{01}^2 $, corresponding to the regime collinear to the target, these large logs cancel each other. By contrast, the log in the second term in Eq.~\eqref{sigma_L_total_scheme_dep_eta_minus_2} also becomes large in the domain 
$\xt_{20}^2\ll \xt_{21}^2\sim \xt_{01}^2 \lesssim 1/Q^2 $, and the log in the third term becomes large in the domain 
$\xt_{21}^2\ll \xt_{20}^2\sim \xt_{01}^2 \lesssim 1/Q^2$. These two domains correspond to anticollinear regimes, meaning collinear to the projectile. Hence, one finds that by changing the rapidity regulator from $\eta^+$ to $\eta^-$, the large collinear log contributions coming from the plus distribution NLO terms get cancelled, and replaced by large anticollinear logs. This is exactly the pattern expected when changing the evolution variable from $k^+$ to $k^+$ in low $x$ evolution equations, see the discussion in the introduction and in Refs.~\cite{Salam:1998tj,Beuf:2014uia} for example.    

Finally, we compare our results, in particular Eq.~\eqref{sigma_L_total_scheme_dep_eta_minus_2} in the $\eta^-$ scheme, to the ones obtained in Ref.~\cite{Liu:2022ijp}. In that reference, the authors calculate the NLO correction to forward inclusive single jet production in pA collisions, using SCET but following closely the CGC approach and approximations. In particular, they use the $\eta$ regulator from Refs.~\cite{Chiu:2011qc,Chiu:2012ir}, which reduces to our $\eta^-$ regulator for the modes collinear to the target. Hence, in that calculation, the low $x$ evolution of the target is obtained as an evolution along $k^-$. The final result of Ref.~\cite{Liu:2022ijp}, given in Eq.~(5.51) there, contains the contribution ${\cal H}_{q,BK}$ (see Eq.~(5.52) there), which is equivalent (up to the harmless $z$ dependence) to the first term in the bracket in our result \eqref{sigma_L_total_scheme_dep_eta_minus_2}, and the contribution ${\cal H}_{q,kin}$ (see Eq.~(5.53) in Ref.~\cite{Liu:2022ijp}), which is equivalent to the second and third terms in the bracket in our result \eqref{sigma_L_total_scheme_dep_eta_minus_2}. Of course, the photon virtuality $Q$ in our case is replaced by the jet transverse momentum in Ref.~\cite{Liu:2022ijp} as the hard scale of the problem. This comparison gives us confidence that the contributions in the bracket in Eq.~\eqref{sigma_L_total_scheme_dep_eta_minus_2} should appear in the NLO correction  to any low-$x$ observable which involve only the dipole operator at LO, 
when calculated in the $\eta^-$ scheme.

\subsection{With pure rapidity regulator}

In the pure rapidity regulator case, the scheme-dependent part of the cross section can be obtained in a similar way as with the other regulators, thanks to the rapidity counterterm from Eq.~\eqref{rap_renorm_dipole_pure_rap_1}. The final result is once again the average of the results obtained with the $\eta^+$ and $\eta^-$ regulators, which is  
\begin{equation}
    \begin{split}
&\,
\sigma^{\gamma^*}_{L} \Big|_{q\rightarrow g}^{\textrm{scheme dep.};\, \textrm{pure rap. reg.}}
+
\sigma^{\gamma^*}_{L} \Big|_{\bar{q}\rightarrow g}^{\textrm{scheme dep.};\, \textrm{pure rap. reg.}}
\\
=&\, 4 N_c \aem
\sum_{f}  e_f^2\;
\int_{0}^{1}
\dd{z}
\int \frac{\dd[2]{\xt_0}}{2\pi} \int \frac{\dd[2]{\xt_1}}{2\pi}\,
4 Q^2\, z^2(1\!-\!z)^2\,
\left[\Kb_{0}\qty(\overline{Q}|\xt_{01}|)\right]^2\,
\,
\\
&\, \times\,
\frac{2\as \cf}{\pi}\int \frac{\dd[2]{\xt_2}}{2\pi}
{\textrm{Re}}\left[\dipole\!-\!\tripole\right]\,
 \bigg[\frac{1}{2}\log\left(\frac{2z(1\!-\!z)(q^+)^2\nu^-\, \xt_{20}^2 }{c_0^2\, \nu^+}\right)\,
\frac{\xt_{01}\cdot \xt_{21}}{\xt_{20}^2\, \xt_{21}^2}
\,
\\
&\, 
\hspace{3cm}
-
\frac{1}{2}\log\left(\frac{2z(1\!-\!z) (q^+)^2\nu^-\, \xt_{21}^2 }{c_0^2\, \nu^+}\right)\,
\frac{\xt_{01}\cdot \xt_{20}}{\xt_{20}^2\, \xt_{21}^2}
\bigg]
\\
=&\, 4 N_c \aem
\sum_{f}  e_f^2\;
\int_{0}^{1}
\dd{z}
\int \frac{\dd[2]{\xt_0}}{2\pi} \int \frac{\dd[2]{\xt_1}}{2\pi}\,
4 Q^2\, z^2(1\!-\!z)^2\,
\left[\Kb_{0}\qty(\overline{Q}|\xt_{01}|)\right]^2\,
\,
\\
&\, \times\,
\frac{2\as \cf}{\pi}\int \frac{\dd[2]{\xt_2}}{2\pi}
{\textrm{Re}}\left[\dipole\!-\!\tripole\right]\,
 \bigg[\frac{1}{2}\log\left(\frac{2z(1\!-\!z)(q^+)^2\nu^-}{Q^2\, \nu^+}\right)\,
\frac{\xt_{01}^2}{\xt_{20}^2\, \xt_{21}^2}
\,
\\
&\, 
\hspace{3cm}
+
\frac{1}{2}\log\left(\frac{\xt_{20}^2 Q^2 }{c_0^2}\right)\,
\frac{\xt_{01}\cdot \xt_{21}}{\xt_{20}^2\, \xt_{21}^2}
-
\frac{1}{2}\log\left(\frac{\xt_{21}^2 Q^2 }{c_0^2}\right)\,
\frac{\xt_{01}\cdot \xt_{20}}{\xt_{20}^2\, \xt_{21}^2}
\bigg]
\label{sigma_L_total_scheme_dep_pure_rap_reg_1}
\, ,
    \end{split}
\end{equation}
dropping the terms antisymmetric in $z\leftrightarrow (1\!-\!z)$, which vanish when integrated over $z$, and organizing the terms in a similar way as in Eq.~\eqref{sigma_L_total_scheme_dep_eta_minus_2}.

In this case of the pure rapidity regulator, the dipole operator evaluated in the target state should depend on $\nu^+/\nu^-$ via $2(P^-)^2 \nu^+/Q_0^2 \nu^-$, following our discussion of the $\eta^+$ and $\eta^-$ regulator cases.
By choosing the same values $\nu_f^+$ \eqref{scale_choice_nu_plus} and $\nu_f^-$ \eqref{scale_choice_nu_minus} as earlier for $\nu^+$ and $\nu^-$, one removes the potentially large logarithm in the first term in Eq.~\eqref{sigma_L_total_scheme_dep_pure_rap_reg_1}. Moreover, one finds the logarithmic range for evolution in the pure rapidity regularization scheme
\begin{align}
Y_f
=&\, 
\frac{1}{2}
\log\left(
\frac{2(P^-)^2\nu_f^+}{Q_0^2\, \nu_f^-}
\right)
\sim
\frac{1}{2}\log\left(
\frac{W^4}{Q^2 Q_0^2}
\right)
\sim
\log\left(
\frac{1}{x_{Bj}}
\right)
+
\frac{1}{2}\log\left(
\frac{Q^2}{Q_0^2}
\right)
\label{evol_range_pure_rap}
\, ,
\end{align}
which is, as expected, the average between the ranges $Y_f^{(+)}$ \eqref{evol_range_nu_plus} and  $Y_f^{(-)}$ \eqref{evol_range_nu_minus} obtained in the $\eta^+$ and $\eta^-$ schemes.

Moreover, the second and third terms in the bracket in Eq.~\eqref{sigma_L_total_scheme_dep_pure_rap_reg_1} are half of the ones obtained in Eq.~\eqref{sigma_L_total_scheme_dep_eta_minus_2} in the $\eta^-$ regulator case. Hence, the cancellation of the collinear logs coming from the plus distribution terms is incomplete, so that in the pure rapidity regulator case, one obtains large logs both in the collinear and in the anticollinear regimes, but with factors $1/2$ compared to the ones obtained with either the $\eta^+$ scheme or the $\eta^-$ scheme, as expected.

\subsection{With cut-off in $k^+$}

The results of Ref.~\cite{Beuf:2017bpd} for $F_L$ at NLO, derived with a cut-off in $k^+$ as rapidity regulator, can also be written in the form given in Eqs.~\eqref{sigma_L_total_final_1}, \eqref{sigma_L_q2g_split} and \eqref{sigma_L_qbar2g_split}. The only difference is the expression of the scheme dependent contributions. They can be read-off from Ref.~\cite{Beuf:2017bpd}, as
\begin{equation}
    \begin{split}
\sigma^{\gamma^*}_{L} \Big|_{q\rightarrow g}^{\textrm{scheme dep.\: cut-off}}
=&\, 4 N_c \aem
\sum_{f}  e_f^2\;
\int_{0}^{1}
\dd{z}
\int \frac{\dd[2]{\xt_0}}{2\pi} \int \frac{\dd[2]{\xt_1}}{2\pi}\,
4 Q^2\, z^2(1\!-\!z)^2\,
\left[\Kb_{0}\qty(\overline{Q}|\xt_{01}|)\right]^2\,
\\
&\, \times\,
\frac{2\as \cf}{\pi}
\int^{1}_{\frac{k^+_{\min}}{z q^+}} \frac{d\xi}{\xi}
\int \frac{\dd[2]{\xt_2}}{2\pi}
\left[\frac{\xt_{20}}{\xt_{20}^2}\vdot\left(\frac{\xt_{20}}{\xt_{20}^2}\!-\!\frac{\xt_{21}}{\xt_{21}^2}\right)\right]
{\textrm{Re}}\left[ \dipole \!-\! \tripole\right]
\label{sigma_L_q2g_scheme_dep_cut-off_1}
\, ,
    \end{split}
\end{equation}
and
\begin{equation}
    \begin{split}
\sigma^{\gamma^*}_{L} \Big|_{\bar{q}\rightarrow g}^{\textrm{scheme dep.\: cut-off}}
=&\, 4 N_c \aem
\sum_{f}  e_f^2\;
\int_{0}^{1}
\dd{z}
\int \frac{\dd[2]{\xt_0}}{2\pi} \int \frac{\dd[2]{\xt_1}}{2\pi}\,
4 Q^2\, z^2(1\!-\!z)^2\,
\left[\Kb_{0}\qty(\overline{Q}|\xt_{01}|)\right]^2\,
\\
&\, \times\,
\frac{2\as \cf}{\pi}
\int^{1}_{\frac{k^+_{\min}}{(1\!-\!z) q^+}} \frac{d\xi}{\xi}
\int \frac{d^{2} \xt_2}{2\pi}
\left[-\frac{\xt_{21}}{\xt_{21}^2}\vdot\left(\frac{\xt_{20}}{\xt_{20}^2}\!-\!\frac{\xt_{21}}{\xt_{21}^2}\right)\right]
{\textrm{Re}}\left[ \dipole \!-\! \tripole\right]
\label{sigma_L_qbar2g_scheme_dep_cut-off_1}
\, .
    \end{split}
\end{equation}
In Ref.~\cite{Beuf:2017bpd} and its applications, the operator on the right hand side of Eqs.~\eqref{sigma_L_q2g_scheme_dep_cut-off_1} 
and \eqref{sigma_L_qbar2g_scheme_dep_cut-off_1} was taken at a rapidity factorization scale dependent on $\xi$. However, such dependence is formally a higher order effect. If instead, the operator does not depend on $\xi$, the integral over $\xi$ can be performed explicitly, leading to results identical to Eqs.~\eqref{sigma_L_q2g_scheme_dep_eta_plus_1} and \eqref{sigma_L_qbar2g_scheme_dep_eta_plus_1} in the $\eta^+$ regulator scheme, up to the replacement of the scale $\nu^+$ by the cut-off $k^+_{\min}$.


\section{Conclusions\label{sec:conclu}}

In summary, we have proposed three possible rapidity regulators,  \eqref{def_etaplus_reg}, \eqref{def_etaminus_reg} and  
\eqref{def_pure_rap_reg} for higher order calculations in QCD at low $x$ including gluon saturation, as alternatives to the usual $k^+$ cut-off \eqref{k_plus_cutoff}. We have validated their implementation by using them to revisit the calculation from Refs.~\cite{Beuf:2016wdz,Beuf:2017bpd} of the $F_L$ inclusive DIS structure function at NLO in the dipole factorization approach at low $x$. 

On the one hand, the new regulators allow us to cleanly separate rapidity divergences from the other ones, in particular soft divergences, due to the opposite limit chosen for the removal of the rapidity and dimensional regulators, by comparison to the $k^+$ cut-off \eqref{k_plus_cutoff}. Our calculation of $F_L$ at NLO show the consistency of the new regulators, by maintaining the cancellation between diagrams of all divergences except for rapidity ones, despite different looking results for individual diagrams (with for example the appearance of $1/\epsilon^2$ terms instead of double logs of the $k^+$ cut-off). 

On the other hand, the three proposed rapidity regulators are each formulated in terms of a different kinematical variable: $k^+$, $k^-$ or rapidity. Then, our calculation confirms that by choosing one of these regulators, one also chooses the corresponding quantity as the evolution variable for the low-$x$ evolution equations. This is in particular confirmed by the observed  large collinear or anticollinear logs observed in $F_L$ at NLO, and their dependence on the choice of the rapidity regularization scheme.
Hence, in particular, our $\eta^-$ regulator allows one to choose the BK or B-JIMWLK to be formulated from the start of the calculation in terms of the $k^-$ variable, suitable in the case of DIS observables, and not by a posteriori manipulation of the results.

More work is however necessary in order to fully understand these new rapidity regulators, and notably the numerical implementation of the corresponding results. 
In particular, the scales $\nu^{\pm}$ at which the dipole and other multipole operators in the NLO corrections are evaluated cannot depend on the $k^+$ of the new gluon, or its momentum fraction $\xi$ in our notation. This property might also require us to rethink the implementation of the collinear improvement of evolution equation by  kinematical constraint.

As an immediate follow-up of the present study, we plan to revisit the calculation of the $F_T$ DIS structure function at NLO with the rapidity regulators from Sec.~\ref{sec:regulators} for completeness.
Moreover, we plan to study the impact of these rapidity regulators in the calculation of NLO corrections to less inclusive observables, in which soft and collinear physics play a more important role, such as dijet production in DIS.


\begin{acknowledgments}

We would like to thank Enrico Herrmann, Zhong-Bo Kang, Tuomas Lappi, and Farid Salazar for discussions.
TA is supported in part by the National Science Centre (Poland) under the research Grant No. 2023/50/E/ST2/00133 (SONATA BIS 13). 
GB is supported in part by the National Science Centre (Poland) under the research grant no 2020/38/E/ST2/00122 (SONATA BIS 10).
J.P. is supported by the National Science Foundation, under grant No.~PHY-2515057, and by the U.S. Department of Energy, Office of Science, Office of Nuclear Physics, within the framework of the Saturated Glue (SURGE) Topical Theory Collaboration.

\end{acknowledgments}


\appendix

\section{Integrals}
\label{app:details}

In this Appendix, we will give the detailed steps for evaluating some of the more difficult integrals in Sec.~\ref{sec:vertex_corrections}. 

To evaluate Eq.~\eqref{V1L_B0_div_eta_plus_eta_pole}, we first integrate in $\zeta$ and expand at small $\eta$ (at finite $\epsilon$) as:
\begin{equation}
\begin{split}
&{\cal V}_{1}^{L}\bigg|_{\mathcal{B}_0\textrm{;}\eta\textrm{ pole}}^{\eta+}
\\
=&\,
\frac{1}{\eta}\left[\frac{z q^+}{\nu_B^+}\right]^{\eta}\,
\Gamma\left(1\!+\!\epsilon\right)\,
\left[4\pi\,\mu^2\right]^{\epsilon}\,
\int_{0}^{1}\!\! \dd{y}y^{-1-\epsilon+\eta}\;
\frac{\Pt^2  +  \overline{Q}^2}{
\left[\left(1\!-\!y\right)\, \Pt^2
  +  \overline{Q}^2 \right]^{1+\epsilon}}
 \\
=&\,
\left[\frac{1}{\eta}
+\log\left(\frac{z q^+}{\nu_B^+}\right)
\right]\,
\Gamma\left(1\!+\!\epsilon\right)\,
\left[
\frac{
4\pi\,\mu^2
}{
\Pt^2  + \overline Q^2
}\right]^{\epsilon}\,
\int_{0}^{1}\!\! \dd{y}y^{-1-\epsilon}
\qty[
1 -y
\frac{\Pt^2}{\Pt^2 + \overline Q^2}
]^{-1-\epsilon}
 \\
&\, +
\Gamma\left(1\!+\!\epsilon\right)\,
\left[
\frac{
4\pi\,\mu^2
}{
\Pt^2  + \overline Q^2
}\right]^{\epsilon}\,
\int_{0}^{1}\!\! \dd{y}y^{-1-\epsilon}
 \log y 
\qty[
1 -y
\frac{\Pt^2}{\Pt^2 + \overline Q^2}
]^{-1-\epsilon}
+O(\eta)
\label{V1L_B0_div_eta_plus_eta_pole_computed}
 \, .
\end{split}
\end{equation}
The contribution in the first line, including the $\eta$ pole, can be written in terms of the incomplete beta function by using Eq.~\eqref{eq:inc_beta}.
For the contribution in the second line of Eq.~\eqref{V1L_B0_div_eta_plus_eta_pole_computed},which is independent of $\eta$, we can do the following:
\begin{equation}
    \begin{split}
&\int_{0}^{1}\!\! \dd{y}
\qty[
y 
(
1 -y
x
)
]^{-1-\epsilon}
\log y 
= 
\int_{0}^{1}\!\! \dd{y}
y^{-1-\epsilon}
\log y 
\sum_{n=0}^\infty
\frac{\Gamma(-\epsilon)}{
\Gamma(n+1)
\Gamma(-n-\epsilon)
}
(-xy)^{n}
\\
=&
-
\sum_{n=0}^\infty
\frac{\Gamma(-\epsilon)}{
\Gamma(n+1)
\Gamma(-n-\epsilon)
}
\frac{1}{(n-\epsilon)^2}
   (-x)^n
=
\sum_{n=0}^\infty
\frac{\Gamma(-\epsilon) \Gamma(1+n+\epsilon)}{
\Gamma(n+1)
\Gamma(\epsilon)
\Gamma(1-\epsilon)
}
\frac{1}{(n-\epsilon)^2}
   x^n
   \\
   =&
   -\frac{1}{\epsilon^2}
   - \sum_{n=0}^\infty
   \frac{1}{n^2}x^n
   + \order{\epsilon}
   =
   -\frac{1}{\epsilon^2}
   - \li(x)
   + \order{\epsilon},
    \end{split}
\end{equation}
where we have used the properties of the gamma function and the series representation of the dilogarithm function.
This gives us
\begin{equation}
\begin{split}
{\cal V}_{1}^{L}\bigg|_{\mathcal{B}_0\textrm{;}\eta\textrm{ pole}}^{\eta+}
=&\,
\Gamma\left(1\!+\!\epsilon\right)\,
\left[
\frac{
4\pi\,\mu^2
}{
\Pt^2  + \overline Q^2
}\right]^{\epsilon}\,
\Bigg\{
\left[\frac{1}{\eta}
+\log\left(\frac{z q^+}{\nu_B^+}\right)
\right]
\qty( \frac{\Pt^2}{\Pt^2 + \overline Q^2} )^{\epsilon}
\text{B}\qty( \frac{\Pt^2}{\Pt^2 + \overline Q^2} ; -\epsilon,-\epsilon )
\\
&\,
- \frac{1}{\epsilon^2}
- \li \qty(
\frac{\Pt^2}{\Pt^2 + \overline Q^2}
)
\Bigg\}
+ \order{\epsilon}+ \order{\eta f(\epsilon) }
\label{V1L_B0_div_eta_plus_eta_pole_computed2}
 \, ,
\end{split}
\end{equation}
which can be rewritten as Eq.~\eqref{V1L_B0_div_eta_plus_eta_pole}. Note that in Eq.~\eqref{V1L_B0_div_eta_plus_eta_pole_computed2}, the full dependence in $\epsilon$ is kept in the coefficient of $1/\eta$, whereas any contribution proportional to a positive power of $\eta $ is dropped, whatever is its dependence on $\epsilon$, following the prescription of taking the $\eta \rightarrow 0$ limit before the $\epsilon \rightarrow 0$ limit.

For Eq.~\eqref{V1L_B0_div_eta_plus_plus_prescr}, we have:
\begin{equation}
    \begin{split}
&{\cal V}_{1}^{L}\bigg|_{\mathcal{B}_0\textrm{;+ distr.}}^{\eta+}
\\
=&\, 
\Gamma\left(1\!+\!\epsilon\right)\,
\left[4\pi\,\mu^2\right]^{\epsilon}\,
\int_{0}^{1}\!\! \frac{\dd{\zeta}}{(\zeta)_+}\; 
\int_{0}^{1}\!\! \dd{y} y^{-1-\epsilon} \left[1\!+\!\frac{z\zeta}{(1\!-\!z)}\right]^{-1-\epsilon}
\\
&\times
\left[\left(1\!-\!y\right)\, \Pt^2+ (1\!-\!y\zeta)\, \overline{Q}^2 \right]^{-1-\epsilon}
\\
&\, \times\;
\left\{\left[\left(1\!-\!y\right)\, \Pt^2+ (1\!-\!y\zeta)\, \overline{Q}^2 \right] + y\Pt^2 \left(1\!+\!\frac{z\zeta}{(1\!-\!z)}\right)
\right\}
+\order{\eta}
\\
=&
\Gamma\left(1\!+\!\epsilon\right)\,
\left[4\pi\,\mu^2\right]^{\epsilon}\,
\int_{0}^{1}\!\! \frac{\dd{\zeta}}{\zeta}\; 
\int_{0}^{1}\!\! \dd{y} y^{-1-\epsilon} 
\\
&\times
\Biggl\{
\qty[\Pt^2 + \overline Q^2]^{-\epsilon}
\qty{
\left[
1-y
\frac{\Pt^2 + \zeta \overline Q^2}{\Pt^2 + \overline Q^2}
\right]^{-\epsilon}
\left[1\!+\!\frac{z\zeta}{(1\!-\!z)}\right]^{-1-\epsilon}
- 
\left[
1-y
\frac{\Pt^2 }{\Pt^2 + \overline Q^2}
\right]^{-\epsilon}
}
\\
&
+y 
\qty{
\frac{\Pt^2}{\qty[\Pt^2+\overline Q^2 - y \qty(\Pt^2 +\zeta \overline Q^2)]^{1+\epsilon}}
\left[1\!+\!\frac{z\zeta}{(1\!-\!z)}\right]^{-\epsilon}
-
\frac{\Pt^2}{\qty[\Pt^2+\overline Q^2 - y\Pt^2 ]^{1+\epsilon}}
}
\Biggr\}
\label{V1L_B0_div_eta_plus_plus_prescr_computed}
 \, .
    \end{split}
\end{equation}
As the second term in the brackets is proportional to $y$, there is no UV divergence corresponding to $y \to 0$ and we are allowed to take $\epsilon =0$.
For the first term, we are allowed to replace expressions like $[1-ay]^{-\epsilon} \mapsto 1
$
because after  the expansion
\begin{equation}
    [1- a y]^{-\epsilon} = 1 - \epsilon \log(1-a y) + \order{\epsilon}
\end{equation}
we can set $\epsilon \to 0$ for the term $y^{-1-\epsilon} \log(1-ay)$ which gives a convergent integral of $y$.
We can then rewrite Eq.~\eqref{V1L_B0_div_eta_plus_plus_prescr_computed} as:
\begin{equation}
    \begin{split}
&{\cal V}_{1}^{L}\bigg|_{\mathcal{B}_0\textrm{;+ distr.}}^{\eta+}
\\
=&
\Gamma\left(1\!+\!\epsilon\right)\,
\left[
\frac{4\pi\,\mu^2}{\Pt^2 + \overline Q^2}\right]^{\epsilon}\,
\int_{0}^{1}\!\! \frac{\dd{\zeta}}{\zeta}\; 
\int_{0}^{1}\!\! \dd{y} \,
y^{-1-\epsilon}
\qty{
\left[1\!+\!\frac{z\zeta}{(1\!-\!z)}\right]^{-1-\epsilon}
- 
1
}
\\
&
+
\frac{\Pt^2}{\Pt^2 + \overline{Q}^2}
\int_{0}^{1}\!\! \frac{\dd{\zeta}}{\zeta}\; 
\int_{0}^{1}\!\! \dd{y} 
\qty{
\qty[1 - y \frac{\Pt^2 + \zeta \overline Q^2}{\Pt^2 + \overline{Q}^2}]^{-1}
-
\qty[1 - y \frac{\Pt^2 }{\Pt^2 + \overline{Q}^2}]^{-1}
}
+\order{\epsilon}
\\
=&
-\frac{\Gamma\left(1\!+\!\epsilon\right)}{\epsilon}\,
\left[
\frac{4\pi\,\mu^2}{\Pt^2 + \overline Q^2}\right]^{\epsilon}
\int_{0}^{1}\!\! \frac{\dd{\zeta}}{\zeta}
\qty{
\left[1\!+\!\frac{z\zeta}{(1\!-\!z)}\right]^{-1-\epsilon}
- 
1
}
\\
&
+\int_{0}^{1}\!\! \frac{\dd{\zeta}}{\zeta}
\Biggr\{-
\frac{\Pt^2}{\Pt^2 + \zeta \overline{Q}^2}
\log(
\frac{
(1-\zeta)
\overline Q^2}{\Pt^2 + \overline Q^2}
)
+
\log(
\frac{\overline Q^2}{\Pt^2 + \overline Q^2}
)
\Biggr\}
+\order{\epsilon}
\label{V1L_B0_div_eta_plus_plus_prescr_computed2}
 \, .
    \end{split}
\end{equation}
For the first term, we can write the integral as:
\begin{equation}
\begin{split}
    &
    \int_0^1 \frac{\dd{\zeta}}{\zeta} 
\qty{
\qty(1\!+\!\frac{z\zeta}{(1\!-\!z)})^{-\epsilon}
\qty[
\qty(1\!+\!\frac{z\zeta}{(1\!-\!z)})^{-1}
-1
]
+
\qty[
\qty(1\!+\!\frac{z\zeta}{(1\!-\!z)})^{-\epsilon}
- 
1
]
}
\\
=&
\int_0^1 \dd{\zeta}
\qty{
-\qty(1\!+\!\frac{z\zeta}{(1\!-\!z)})^{-1-\epsilon}
\frac{z}{1-z}
- \frac{\epsilon}{\zeta}
\log(1\!+\!\frac{z\zeta}{(1\!-\!z)})
}
+
\order{\epsilon^2}
\\
=&
\frac{1}{\epsilon}
\qty[
(1-z)^\epsilon -1
]
+\epsilon
\li \qty( - \frac{z}{1-z} )
+
\order{\epsilon^2},
\end{split}
\end{equation}
and for the terms on the second line we get:
\begin{equation}
 \begin{split}
&    \int_0^1 \frac{\dd{\zeta}}{\zeta}
    \qty[
-
\frac{\Pt^2}{\Pt^2 + \zeta \overline{Q}^2}
\log(
\frac{
(1-\zeta)
\overline Q^2}{\Pt^2 + \overline Q^2}
)
+
\log(
\frac{\overline Q^2}{\Pt^2 + \overline Q^2}
)]
\\
=&
    \int_0^1 \dd{\zeta}
    \qty[
-
\frac{\Pt^2}{\Pt^2 + \zeta \overline{Q}^2}
\frac{\log(1-\zeta)}{\zeta}
+
\frac{\overline Q^2}{\Pt^2 + \zeta \overline Q^2}
\log(
\frac{\overline Q^2}{\Pt^2 + \overline Q^2}
)]
\\
=&
\frac{\pi^2}{6}
- \li\qty(
\frac{\overline Q^2}{\Pt^2 + \overline Q^2}
)
-
\log(
\frac{\Pt^2}{\Pt^2 + \overline Q^2}
)
\log(
\frac{\overline Q^2}{\Pt^2 + \overline Q^2}
)
=
 \li\qty(
\frac{\Pt^2}{\Pt^2 + \overline Q^2}
)
.
 \end{split}  
\end{equation}
Using these in Eq.~\eqref{V1L_B0_div_eta_plus_plus_prescr_computed2}, we finally get
\begin{equation}
    \begin{split}
{\cal V}_{1}^{L}\bigg|_{\mathcal{B}_0\textrm{;+ distr.}}^{\eta+}
=&
\Gamma\left(1\!+\!\epsilon\right)\,
\left[
\frac{4\pi\,\mu^2}{\Pt^2 + \overline Q^2}\right]^{\epsilon}
\Biggl\{
-
\frac{1}{\epsilon}
\log(1-z)
-\frac{1}{2} \qty\Big[\log(1-z)]^2
-
\li \qty( - \frac{z}{1-z} )
\\
&
+
 \li\qty(
\frac{\Pt^2}{\Pt^2 +\overline Q^2}
)
\Biggr\}
+\order{\epsilon}
\label{V1L_B0_div_eta_plus_plus_prescr_computed3}
 \, .
    \end{split}
\end{equation}
This can then be rewritten as Eq.~\eqref{V1L_B0_div_eta_plus_plus_prescr}.

For the $\eta^-$ regulator, we need to calculate Eq.~\eqref{V1L_B0_div_eta_minus_eta_pole}:
\begin{equation}
    \begin{split}
{\cal V}_{1}^{L}\bigg|_{\mathcal{B}_0\textrm{;}\eta\textrm{ pole}}^{\eta-}
=&\,
4\pi\, \mu^{2\epsilon}\left[2z q^+\nu_B^-\right]^{\eta}\,
\int_{0}^{1}\!\! \dd{y} y^{\eta}
\int_{0}^{1}\!\! \dd{\zeta} \zeta^{\eta-1} \;
\\
&\times
\int \frac{\dd[2-2\epsilon]{\Kt'}}{(2\pi)^{2-2\epsilon}}\;
\frac{\left[ \Pt^2+  \overline{Q}^2 \right] 
}{\left[{\Kt'}^2 
+ y \left(\left(1\!-\!y\right)\, \Pt^2+  \overline{Q}^2 \right) \right]^2
\left[\left(\Kt'+y\Pt\right)^2\right]^{\eta} }
 \\
=& \,
\frac{1}{\eta}
4\pi\, \mu^{2\epsilon}\left[2z q^+\nu_B^-\right]^{\eta}\,
\int_{0}^{1}\!\! \dd{y} y^{\eta}
\\
&\times
\int_0^1 \dd{x} 
\frac{1}{(4\pi)^{1-\epsilon}} \frac{\Gamma(1+\eta +\epsilon)}{\Gamma(\eta)}
\frac{x (1-x)^{\eta-1} \qty[\Pt^2 + \overline Q^2] }{
\qty[
xy (\Pt^2 +\overline{Q}^2)
-x^2 y^2 \Pt^2
]^{1+\epsilon+\eta}
}
\\
=& \,
 \frac{\Gamma(1+\eta +\epsilon)}{\Gamma(1+\eta)}
\qty[
\frac{4\pi \mu^2}{\Pt^2 + \overline Q^2}
]^{\epsilon}
\left[
\frac{2z q^+\nu_B^-}{\Pt^2 + \overline Q^2}
\right]^{\eta}\,
\\
&\times
\int_{0}^{1}\!\! \dd{y}
\int_0^1 \dd{x} 
y^{-1-\epsilon}
x^{-\epsilon-\eta} (1-x)^{\eta-1} 
\qty[
1
-x y 
\frac{\Pt^2}{\Pt^2 + \overline Q^2}
]^{-1-\epsilon-\eta}
\label{V1L_B0_div_eta_minus_eta_pole_computed_app}
 \,,
    \end{split}
\end{equation}
where we combined the denominators using the Feynman parametrization and then integrated over the momentum $\Kt'$.
We are then left to compute the integrals in the last line.
To proceed further, we note that the $\eta$ pole corresponds to the limit $x \to 1$ in the integrand.
It is then useful to isolate this part by writing the last line as:
\begin{equation}
\label{eq:xy_integral_split}
    \begin{split}
&\int_{0}^{1}\!\! \dd{y}
\int_0^1 \dd{x} 
y^{-1-\epsilon}
x^{-\epsilon-\eta} (1-x)^{\eta-1} 
\\
&\times
\Biggl\{
\qty[
1
-y 
\frac{\Pt^2}{\Pt^2 + \overline Q^2}
]^{-1-\epsilon-\eta}
+
\qty(
\qty[
1
-x y 
\frac{\Pt^2}{\Pt^2 + \overline Q^2}
]^{-1-\epsilon-\eta}
-\qty[
1
-y 
\frac{\Pt^2}{\Pt^2 + \overline Q^2}
]^{-1-\epsilon-\eta}
)
\Biggr\}.
    \end{split}
\end{equation}
For the first term in Eq.~\eqref{eq:xy_integral_split}, we can integrate over $x$ and do an expansion in $\eta$:
\begin{equation}
    \begin{split}
&\int_{0}^{1}\!\! \dd{y}
\int_0^1 \dd{x} 
y^{-1-\epsilon}
x^{-\epsilon-\eta} (1-x)^{\eta-1} 
\qty[
1
-y 
\frac{\Pt^2}{\Pt^2 + \overline Q^2}
]^{-1-\epsilon-\eta}
\\
=&
\int_{0}^{1}\!\! \dd{y}
y^{-1-\epsilon}
\frac{\Gamma(1-\epsilon-\eta) \Gamma(\eta)}{\Gamma(1-\epsilon)}
\qty[
1
-y 
\frac{\Pt^2}{\Pt^2 + \overline Q^2}
]^{-1-\epsilon-\eta}
\\
=&
\frac{\Gamma(1-\epsilon-\eta) \Gamma(\eta)}{\Gamma(1-\epsilon)}
\int_{0}^{1}\!\! \dd{y}
y^{-1-\epsilon}
\qty[
1
-y 
\frac{\Pt^2}{\Pt^2 + \overline Q^2}
]^{-1-\epsilon}
\\
&
-\int_{0}^{1}\!\! \dd{y}
y^{-1-\epsilon}
\qty[
1
-y 
\frac{\Pt^2}{\Pt^2 + \overline Q^2}
]^{-1-\epsilon}
\log(1
-y 
\frac{\Pt^2}{\Pt^2 + \overline Q^2})
+\order{\eta}.
    \end{split}
\end{equation}
In the first term, one recognizes the integral representation of the incomplete Beta function \eqref{eq:inc_beta}. 
In the second term, the logarithm regulates the potential divergence at $y=0$, so that one can take $\epsilon=0$ in that case. 
We then get:
\begin{equation}
\label{eq:xy_integral_split_pole}
    \begin{split}
    &
\frac{\Gamma(1-\epsilon-\eta) \Gamma(\eta)}{\Gamma(1-\epsilon)}
\qty(
\frac{\Pt^2}{\Pt^2 + \overline Q^2}
)^\epsilon
\text{B}\qty(
\frac{\Pt^2}{\Pt^2 + \overline Q^2};-\epsilon,-\epsilon)
\\
&-
\int_0^1 \dd{y}
y^{-1}
\qty[
1
-y 
\frac{\Pt^2}{\Pt^2 + \overline Q^2}
]^{-1}
\log(1
-y 
\frac{\Pt^2}{\Pt^2 + \overline Q^2})
+\order{\epsilon}
+\order{\eta f(\epsilon)}
.
\end{split}
\end{equation}
While the final $y$-integral could be expressed in terms of logarithms and dilogarithms, we leave it as is for now.
For the second term in Eq.~\eqref{eq:xy_integral_split}, we can set both $\eta =0$ and $\epsilon =0$ as there are no associated poles.
The integral is then given by:
\begin{equation}
    \begin{split}
&\int_{0}^{1}\!\! \dd{y}
\int_0^1 \dd{x} 
y^{-1} (1\!-\!x)^{-1} 
\qty(
\qty[
1
\!-\!x y 
\frac{\Pt^2}{\Pt^2 \!+\! \overline Q^2}
]^{-1}
\!-\!\qty[
1
\!-\!y 
\frac{\Pt^2}{\Pt^2 \!+\! \overline Q^2}
]^{-1}
)
+\order{\epsilon}
+\order{\eta f(\epsilon)}
\\
=&-\int_{0}^{1}\!\! \dd{y}
\int_0^1 \dd{x} 
\frac{\Pt^2}{\Pt^2 + \overline Q^2}
\qty[
1
-x y 
\frac{\Pt^2}{\Pt^2 + \overline Q^2}
]^{-1}
\qty[
1
-y 
\frac{\Pt^2}{\Pt^2 + \overline Q^2}
]^{-1}
+\order{\epsilon}
+\order{\eta f(\epsilon)}
\\
=&\int_{0}^{1}\!\! \dd{y}
y^{-1}
\qty[
1
-y 
\frac{\Pt^2}{\Pt^2 + \overline Q^2}
]^{-1}
\log(1
-y 
\frac{\Pt^2}{\Pt^2 + \overline Q^2}
)
+\order{\epsilon}
+\order{\eta f(\epsilon)}.
    \end{split}
\end{equation}
We see that this cancels exactly the leftover integral in Eq.~\eqref{eq:xy_integral_split_pole}, so that we finally have:
\begin{equation}
    \begin{split}
{\cal V}_{1}^{L}\bigg|_{\mathcal{B}_0\textrm{;}\eta\textrm{ pole}}^{\eta-}
=& \,
 \frac{\Gamma(1+\eta +\epsilon)}{\Gamma(1+\eta)}
\qty[
\frac{4\pi \mu^2}{\Pt^2 + \overline Q^2}
]^{\epsilon}
\left[
\frac{2z q^+\nu_B^-}{\Pt^2 + \overline Q^2}
\right]^{\eta}\,
\\
&\times
\frac{\Gamma(1-\epsilon-\eta) \Gamma(\eta)}{\Gamma(1-\epsilon)}
\qty(
\frac{\Pt^2}{\Pt^2 + \overline Q^2}
)^\epsilon
\text{B}\qty(
\frac{\Pt^2}{\Pt^2 + \overline Q^2};-\epsilon,-\epsilon)
+\order{\epsilon}
+\order{\eta f(\epsilon)}
\\
=& \,
\Gamma(1 +\epsilon)
\qty[
\frac{4\pi \mu^2}{\Pt^2 + \overline Q^2}
]^{\epsilon}
\qty(
\frac{\Pt^2}{\Pt^2 + \overline Q^2}
)^\epsilon
\text{B}\qty(
\frac{\Pt^2}{\Pt^2 + \overline Q^2};-\epsilon,-\epsilon)
\\
&
\times\,
\Bigg\{ \frac{1}{\eta}\,\left[
\frac{2z q^+\nu_B^-}{\Pt^2 + \overline Q^2}
\right]^{\eta}\,
+ \Psi(1+\epsilon)-\Psi(1-\epsilon)
\Bigg\}
+\order{\epsilon}
+\order{\eta f(\epsilon)}
\\
=& \,\frac{\Gamma(1 +\epsilon)}{\eta}\,\left[
\frac{2z q^+\nu_B^-}{\Pt^2 + \overline Q^2}
\right]^{\eta}\,
\qty[
\frac{4\pi \mu^2}{\Pt^2 + \overline Q^2}
]^{\epsilon}\,
\qty(
\frac{\Pt^2}{\Pt^2 + \overline Q^2}
)^\epsilon
\text{B}\qty(
\frac{\Pt^2}{\Pt^2 + \overline Q^2};-\epsilon,-\epsilon)
\\
&
- \frac{\pi^2}{3}
+\order{\epsilon}
+\order{\eta f(\epsilon)}
\label{V1L_B0_div_eta_minus_eta_pole_computed2}
 \,,
    \end{split}
\end{equation}
using the expansion \eqref{eq:inc_beta} of the incomplete Beta function, and the value $\Psi'(1) =\pi^2/6$ of the trigamma function at $1$. The result indeed corresponds to 
 Eq.~\eqref{V1L_B0_div_eta_minus_eta_pole}.

Finally, we need to consider Eq.~\eqref{V1L_B0_div_eta_minus_plus_prescr}:
\begin{equation}
    \begin{split}
{\cal V}_{1}^{L}\bigg|_{\mathcal{B}_0\textrm{;+ distr.}}^{\eta-}
=&\,
4\pi\, \mu^{2\epsilon}\,
\int_{0}^{1}\!\! \dd{y} 
\int_{0}^{1}\!\! \frac{\dd{\zeta}}{(\zeta)_+} 
\int \frac{\dd[2-2\epsilon]{\Kt'}}{(2\pi)^{2-2\epsilon}}\;
\\
&\times
\frac{\left\{\left[\left(1\!-\!y\right)\, \Pt^2+ (1\!-\!y\zeta)\, \overline{Q}^2 \right] + y\Pt^2 \left(1\!+\!\frac{z\zeta}{(1\!-\!z)}\right)
\right\}}{\left[{\Kt'}^2 
+ y \left(1\!+\!\frac{z\zeta}{(1\!-\!z)}\right) \left(\left(1\!-\!y\right)\, \Pt^2+ (1\!-\!y\zeta)\, \overline{Q}^2 \right) \right]^2}
\\
=&\,
4\pi\, \mu^{2\epsilon}\,
\int_{0}^{1}\!\! \dd{y} 
\int_{0}^{1}\!\! \frac{\dd{\zeta}}{(\zeta)_+}\, 
\frac{\Gamma( 1+\epsilon )}{(4\pi)^{1-\epsilon}}
\\
&\times
\frac{\left[\left(1\!-\!y\right)\, \Pt^2+ (1\!-\!y\zeta)\, \overline{Q}^2 \right] + y\Pt^2 \left(1\!+\!\frac{z\zeta}{(1\!-\!z)}\right)
}{
\qty[
y \left(1\!+\!\frac{z\zeta}{(1\!-\!z)}\right) \left(\left(1\!-\!y\right)\, \Pt^2+ (1\!-\!y\zeta)\, \overline{Q}^2 \right) 
]^{1+\epsilon}
}
\\
=&
{\cal V}_{1}^{L}\bigg|_{\mathcal{B}_0\textrm{;+ distr.}}^{\eta+} 
\label{V1L_B0_div_eta_minus_plus_prescr_computed}
 \, ,
    \end{split}
\end{equation}
where we noticed that this agrees with Eq.~\eqref{V1L_B0_div_eta_plus_plus_prescr_computed}, as expected.


\section{Fourier transforms}
\label{app:FT}

In order to compute the LFWF in the mixed space, Eq.~\eqref{FT_of_qqbar_LFWF_L}, we need to Fourier transform the momentum-space expression.
The following identities turn out to be useful~\cite{Beuf:2016wdz}:
\begin{align}
&\int \frac{\dd[2-2\epsilon]{\Pt}}{(2\pi)^{2-2\epsilon}}\;
\frac{\Pt^m
}{ \big[\Pt^2\big]^{1+\frac{\eta}{2}} }\:   e^{i\Pt\vdot \rt}
=\,
\frac{i}{2\pi}\,
\frac{\rt^m}{\rt^2}\,
\frac{\Gamma\left(1\!-\!\epsilon\!-\!\frac{\eta}{2}\right)}{\Gamma\left(1\!+\!\frac{\eta}{2}\right)}
\left[\pi\, \rt^2\right]^{\epsilon}
\left[\frac{\rt^2}{4}\right]^{\frac{\eta}{2}}
\, ,
\label{eq:FT_power}
\\
 &  \int \frac{\dd[2-2\epsilon]{\Pt}}{(2\pi)^{2-2\epsilon}}
    \frac{e^{i \Pt \vdot \rt}}{\qty[\Pt^2 + \overline Q^2]^{1+n} }
    =
    \frac{1}{2\pi}
    \frac{1}{(4\pi)^n}
    \frac{1}{\Gamma(1+n)}
    \qty(
\frac{\overline Q}{2\pi \abs{\rt}}
)^{-\epsilon-n}
\Kb_{-\epsilon-n}
\qty(\abs{\rt} \overline Q),
\label{eq:FT_LO}
\\
   \begin{split}      
   &\int \frac{\dd[2-2\epsilon]{\Pt}}{(2\pi)^{2-2\epsilon}}
    \frac{e^{i \Pt \vdot \rt}}{\left[\Pt^2 + \overline Q^2\right]}
    \log(
    \frac{\Pt^2 + \overline Q^2}{
    \overline Q^2
    }
    )
    = \frac{1}{2\pi}
    \qty(
\frac{\overline Q}{2\pi \abs{\rt}}
)^{-\epsilon}
\Kb_{-\epsilon}
\qty(\abs{\rt} \overline Q)\:  \frac{(-1)}{2}
\log(
\frac{\rt^2 \overline Q^2}{c_0^2})
+ \order{\epsilon},
   \end{split}
   \label{eq:FT_log}
\end{align}
where $c_0 = 2 e^{-\gamma_E}$. All of these can be derived by using the standard Schwinger trick
\begin{align}
\frac{1}{D^{\alpha}}
= &\,
\frac{1}{\Gamma(\alpha)}
\int_{0}^{+\infty} \dd \tau\, 
\tau^{\alpha-1}\, e^{-\tau D}
\end{align}
to transform them into Gaussian integrals.

We will also encounter more complicated Fourier transforms that, to our knowledge, have not been encountered before in NLO calculations in QCD at low $x$.
These involve the incomplete beta function:
\begin{equation}
\begin{split}
   & \int \frac{\dd[2-2\epsilon]{\Pt}}{(2\pi)^{2-2\epsilon}}
    \frac{e^{i \Pt \vdot \rt}}{\qty[\Pt^2 + \overline Q^2]^{1+\epsilon}}
    \times 
\qty( \frac{\Pt^2}{\Pt^2 + \overline Q^2} )^{\epsilon}
\text{B}\qty( \frac{\Pt^2}{\Pt^2 + \overline Q^2} ; -\epsilon,-\epsilon )
 \\
    =&
    \frac{1}{2\pi} 
    \frac{
    \Gamma(-\epsilon)}{
\Gamma(1+\epsilon)}
\qty( \frac{ \pi \abs{\rt}^3}{2 \overline Q} )^{\epsilon}
\Kb_{-\epsilon} \qty( \abs{\rt}\overline Q),
\end{split}
\label{eq:FT_beta}
\end{equation}
and a combination of the dilogarithm and the squared logarithm:
\begin{equation}
    \begin{split}
&\int\frac{\dd[2]{\Pt}}{(2\pi)^{2}}\,
\frac{e^{i\Pt\vdot\rt}}{\Pt^2+\overline{Q}^2}
\left\{2\, \textrm{Li}_2\left(\frac{\Pt^2}{\Pt^2\!+\!\overline{Q}^2}\right)
-3 \left[ \log\left(\frac{\Pt^2\!+\!\overline{Q}^2}{\overline{Q}^2}\right)\right]^2
\right\}
\\
=&\,
\frac{1}{2\pi}\, \Kb_{0}\left(|\rt|\overline{Q}\right)\,
\left\{
\frac{\pi^2}{3}
-\left[\log\left(\frac{\rt^2 \overline{Q}^2}{c_0^2}\right)\right]^2
\right\}.
\label{FT_dilog_and_log2}
    \end{split}
\end{equation}
Note that the specific linear combination of dilogarithm and squared logarithm in Eq.~\eqref{FT_dilog_and_log2} has a remarkably simple Fourier transform, but the Fourier transforms of each of the two terms individually cannot be written in terms of elementary functions and reasonably standard special functions. 
Let us now provide proofs of Eqs.~\eqref{eq:FT_beta} and \eqref{FT_dilog_and_log2}.

The Fourier transform involving the incomplete beta function, Eq.~\eqref{eq:FT_beta}, can be done by a direct computation.
To see this, it is easier to first write the incomplete beta function in its integral representation~\eqref{eq:inc_beta}.
The Fourier transform can then be written as
\begin{equation}
\begin{split}
   & \int \frac{\dd[2-2\epsilon]{\Pt}}{(2\pi)^{2-2\epsilon}}
    \frac{e^{i \Pt \vdot \rt}}{\qty[\Pt^2 + \overline Q^2]^{1+\epsilon}}
    \times 
\qty( \frac{\Pt^2}{\Pt^2 + \overline Q^2} )^{\epsilon}
\text{B}\qty( \frac{\Pt^2}{\Pt^2 + \overline Q^2} ; -\epsilon,-\epsilon )
 \\
    =&
   \int \frac{\dd[2-2\epsilon]{\Pt}}{(2\pi)^{2-2\epsilon}}
\int_{0}^{1}\dd{y} 
\frac{e^{i \Pt \vdot \rt}}{\qty[ y \qty( (1-y) \Pt^2 + \overline Q^2 ) ]^{1+\epsilon}}.
\end{split}
\end{equation}
We can integrate over $\Pt$ using Eq.~\eqref{eq:FT_LO}, giving us:
\begin{equation}
\begin{split}
   & 
   \int \frac{\dd[2-2\epsilon]{\Pt}}{(2\pi)^{2-2\epsilon}}
\int_{0}^{1}\dd{y} 
\frac{e^{i \Pt \vdot \rt}}{\qty[ y \qty( (1-y) \Pt^2 + \overline Q^2 ) ]^{1+\epsilon}}
\\
=&
\int_{0}^{1} \dd{y}
\frac{1}{2\pi} 
\qty( \frac{\overline Q^2}{ \pi \rt^2} )^{-\epsilon}
\frac{y^{-1-\epsilon}}{ \Gamma(1+\epsilon) (1-y)} \Kb_{-2\epsilon} \qty(\frac{\abs{\rt}\overline Q}{\sqrt{1-y}})
.
\end{split}
\end{equation}
The final integral over $y$ can be done by changing to a new intergration variable $w \equiv 1/(1-y)$ and the following identity (see Eq.~(6.592-12) of Ref.~\cite{MR2360010}):
\begin{equation}
    \int_1^\infty \dd{x} x^{ -\frac{1}{2} \nu} (x  -1)^{\mu-1} \Kb_\nu(a \sqrt{x})
    = \Gamma(\mu) 2^\mu a^{-\mu} \Kb_{\nu -\mu}(a).
\end{equation}
This allows us to write the integral as
\begin{equation}
\begin{split}
    &
\int_{0}^{1} \dd{y}
\frac{1}{2\pi} 
\qty( \frac{\overline Q^2}{ \pi \rt^2} )^{-\epsilon}
\frac{y^{-1-\epsilon}}{ \Gamma(1+\epsilon) (1-y)} \Kb_{-2\epsilon} \qty(\frac{\abs{\rt}\overline Q}{\sqrt{1-y}})\\
    =&
    \frac{1}{2\pi} \frac{1}{
\Gamma(1+\epsilon)}
\qty( \frac{ \pi \rt^2}{\overline Q^2} )^{\epsilon}
\int_{1}^{\infty} \dd{w}
\frac{w^{\epsilon}}{
\qty[w-1]^{1+\epsilon}
}
 \Kb_{-2\epsilon} \qty(\sqrt{w} \abs{\rt}\overline Q)\\
    =&
    \frac{1}{2\pi} \frac{\Gamma(-\epsilon)}{
\Gamma(1+\epsilon)}
\qty( \frac{ \pi  \abs{\rt}^3}{2\overline Q} )^{\epsilon}
\Kb_{-\epsilon} \qty( \abs{\rt}\overline Q)
\end{split}
\end{equation}
which is Eq.~\eqref{eq:FT_beta}.

The Fourier transform~\eqref{FT_dilog_and_log2} is more complicated to derive.
It turns out that it is easier to consider the inverse Fourier transform instead:
\begin{equation}
   \label{eq:dilog_FT_inv2}
   \begin{split}
   & \int \dd[2]{\rt} e^{-i \Pt \vdot \rt}
   \frac{1}{2\pi}
   \Kb_0 \qty( r \overline Q ) 
    \log^2 \qty(\frac{\rt^2 \overline Q^2}{c_0^2})
    \\
    =&\, \frac{1}{\left[\overline Q^2 + \Pt^2\right]} \qty[  -2 \li \qty(\frac{\Pt^2}{\overline Q^2 + \Pt^2})
    + 3 \log^2 \qty(\frac{\overline Q^2}{\overline Q^2 + \Pt^2}) + \frac{\pi^2}{3}]
   \end{split}
\end{equation}
from which Eq.~\eqref{FT_dilog_and_log2} directly follows.
To show this, we first note that
\begin{equation}
   \partial_\nu^2 \qty[ \qty(\frac{\rt^2 \overline Q^2}{c_0^2})^\nu ]_{\nu = 0} 
   =
\log^2 \qty(\frac{\rt^2 \overline Q^2}{c_0^2}),
\end{equation}
so that we can consider the following Fourier transform:
\begin{equation}
\begin{split}
    &\int \dd[2]{\rt} e^{-i \Pt \vdot \rt} \frac{1}{2\pi} \Kb_0 \qty( r \overline Q ) 
    \qty(\frac{\rt^2 \overline Q^2}{c_0^2})^\nu \\
    =&
    \frac{ e^{2\nu \gamma_E}}{2\pi}  \int \dd[2]{\rt} e^{-i \Pt \vdot \rt} 
    \times \frac{1}{2}\int_0^\infty \frac{\dd{t}}{t} \exp( - t \frac{\rt^2 \overline{Q}^2}{4} - \frac{1}{t} )
    \times
    \int_0^\infty \frac{\dd{u}}{u} \frac{1}{\Gamma(-\nu) u^\nu} \exp(- u \frac{\rt^2 \overline{Q}^2}{4} ) .
\end{split}
\end{equation}
Here,
we have rewritten the Bessel function and the power $ (\rt^{2})^\nu$ in terms of the following integrals:
\begin{align}
    \Kb_0 \qty( r \overline Q ) &=  \frac{1}{2}\int_0^\infty \frac{\dd{t}}{t} \exp( - t \frac{\rt^2 \overline{Q}^2}{4} - \frac{1}{t} ),
    \\
    (\rt^{2})^\nu &=  \int_0^\infty \frac{\dd{u}}{u} \frac{1}{\Gamma(-\nu) } 
    \qty(\frac{4}{\overline Q^2 u})^\nu
    \exp(- u \frac{\rt^2 \overline{Q}^2}{4} ) .
\end{align}
This turns out to be useful because now the integral over $\rt$ is a simple Gaussian integral:
\begin{equation}
\begin{split}
    &\int \dd[2]{\rt} e^{-i \Pt \vdot \rt} \frac{1}{2\pi} \Kb_0 \qty( r \overline Q ) 
    \qty(\frac{\rt^2 \overline Q^2}{c_0^2})^\nu \\
    =&
    \frac{ e^{2\nu \gamma_E}}{2\pi}  \int \dd[2]{\rt} e^{-i \Pt \vdot \rt} 
    \times \frac{1}{2}\int_0^\infty \frac{\dd{t}}{t} \exp( - t \frac{\rt^2 \overline{Q}^2}{4} - \frac{1}{t} )
    \times
    \int_0^\infty \frac{\dd{u}}{u} \frac{1}{\Gamma(-\nu) u^\nu} \exp(- u \frac{\rt^2 \overline{Q}^2}{4} ) \\
    =&  \frac{ e^{2\nu \gamma_E}}{\Gamma(-\nu) \overline Q^2} 
        \int_0^\infty \frac{\dd{t}}{t} \int_0^\infty \frac{\dd{u}}{u} \frac{1}{(t+u) u^\nu}
        \exp[ -\frac{1}{t} - \frac{1}{t+u} \frac{\Pt^2}{\overline Q^2} ].
\end{split}
\end{equation}
We can proceed further by writing $u \equiv tw$ and using $w$ as an integration variable:
\begin{equation}
\begin{split}
&\frac{ e^{2\nu \gamma_E}}{\Gamma(-\nu) \overline Q^2} 
        \int_0^\infty \frac{\dd{t}}{t} \int_0^\infty \frac{\dd{u}}{u} \frac{1}{(t+u) u^\nu}
        \exp[ -\frac{1}{t} - \frac{1}{t+u} \frac{\Pt^2}{\overline Q^2} ]\\
   =&
    \frac{ e^{2\nu \gamma_E}}{\Gamma(-\nu) \overline Q^2} 
        \int_0^\infty \frac{\dd{t}}{t} \int_0^\infty \frac{\dd{w}}{w} \frac{1}{ t^{\nu + 1} (w+1) w^\nu}
        \exp[ -\frac{1}{t} \qty(1 +  \frac{1}{1+w} \frac{\Pt^2}{\overline Q^2}) ] \\
    =&   \frac{ e^{2\nu \gamma_E}}{ \overline Q^2}  \frac{\Gamma(1+\nu)}{\Gamma(-\nu)}
    \int_0^\infty \frac{\dd{w}}{w \qty[w + 1 /\lambda ]}  \qty(\frac{1+w}{w \qty[w +1 /\lambda]})^\nu,
\end{split}
\end{equation}
where we have denoted $\lambda = \overline Q^2 /\qty[\overline Q^2 + \Pt^2]$.
At this point, it would be tempting to use
\begin{equation}
    \frac{\Gamma(1+\nu)}{\Gamma(-\nu)} \qty( a e^{2\gamma_E} )^\nu
    = - \nu -  \log(a) \nu^2 + \mathcal{O}(\nu^3)
\end{equation}
to expand in $\nu$ and get the $  \log(a) \nu^2$ term.
However, if we do this it turns out that the resulting integral is divergent when $w \to 0 $ and thus this expansion does not commute with integration.
To deal with this, we write
\begin{equation}
     \int_0^\infty \frac{\dd{w}}{w \qty[w + 1/\lambda ]}  \qty(\frac{1}{w \qty[w +1/\lambda]})^\nu 
  \qty\Big{ \qty[ (1+w)^\nu - 1] +1 }
\end{equation}
and divide the integral into two parts.
For the first part, we note that we can now expand in $\nu$ and get
\begin{equation}
\begin{split}
   & \partial_\nu^2 \qty[ 
    \frac{\Gamma(1+\nu)}{\Gamma(-\nu)} e^{2\nu \gamma_E}
    \int_0^\infty \frac{\dd{w}}{w \qty[w +1/ \lambda ]}  \qty(\frac{1}{w \qty[w +1/\lambda]})^\nu 
\qty[ (1+w)^\nu - 1]  ]_{\nu = 0}
\\
=& -2   \int_0^\infty \frac{\dd{w}}{w \qty[w + 1/\lambda ]}  \log\qty(1+w) \\
=& -2\lambda \qty[ \frac{\pi^2}{3} + \frac{1}{2} \log^2 \qty(\frac{1-\lambda}{\lambda} ) + \li \qty( -\frac{\lambda}{1-\lambda} ) ].
\end{split}
\end{equation}
For the second part, we first integrate and then take the derivative:
\begin{equation}
    \frac{\Gamma(1+\nu)}{\Gamma(-\nu)} e^{2\nu \gamma_E}   \int_0^\infty \frac{\dd{w}}{w \qty[w + 1/ \lambda ]}  \qty(\frac{1}{w \qty[w +1/\lambda]})^\nu 
    = \frac{ (4 e^{2\gamma_E})^\nu \Gamma( \frac{1}{2} +\nu ) \Gamma(1+\nu) }{\sqrt{\nu} }\lambda^{1+2\nu},
\end{equation}
and thus
\begin{equation}
    \partial_\nu^2 \qty[\frac{ (4 e^{2\gamma_E})^\nu \Gamma( \frac{1}{2} +\nu ) \Gamma(1+\nu) }{\sqrt{\nu}}\lambda^{1+2\nu}]_{\nu =0} 
    = 2\lambda \qty[ \frac{\pi^2}{3} +2 \log^2( \lambda) ].
\end{equation}
Putting everything together, we now have:
\begin{equation}
    \int \dd[2]{\rt} e^{-i \Pt \vdot \rt} \frac{1}{2\pi} \Kb_0 \qty( r \overline Q ) 
    \log^2\qty(\frac{\rt^2 \overline Q^2}{c_0^2})
    = \frac{2  \lambda}{\overline Q^2}
    \qty[ 2 \log^2(\lambda) - \frac{1}{2} \log^2 \! \qty(\frac{1-\lambda}{\lambda}) -\li \qty( - \frac{\lambda}{1-\lambda } )  ].
\end{equation}
Finally, using the identity~\eqref{eq:dilog_identity}
we get
\begin{equation}
\begin{split}
    &\int \dd[2]{\rt} e^{-i \Pt \vdot \rt} \frac{1}{2\pi} \Kb_0 \qty( r \overline Q ) 
    \log^2\qty(\frac{\rt^2 \overline Q^2}{c_0^2})
    =\frac{2  \lambda}{\overline Q^2}
    \qty[ \frac{3}{2}\log^2(\lambda) - \li(1-\lambda) + \frac{\pi^2}{6}] 
    \\
    =&\, \frac{1}{\left[\overline Q^2 + \Pt^2\right]} \qty[ 3 \log^2\qty(\frac{\overline Q^2}{\overline Q^2 + \Pt^2}) - 2 \li \qty( \frac{\Pt^2}{\overline Q^2 + \Pt^2} ) + \frac{\pi^2}{3} ].
\end{split}
\end{equation}
This proves Eq.~\eqref{eq:dilog_FT_inv2}.


\bibliographystyle{JHEP-2modlong.bst}
\bibliography{mybib}

\end{document}